\documentclass[letterpaper,journal]{IEEEtran}
\usepackage{amsfonts}
\usepackage[cmex10]{amsmath}
\usepackage{algorithmic}
\usepackage{algorithm}
\usepackage{amssymb}
\usepackage{array}
\newcolumntype{L}[1]{>{\raggedright\arraybackslash}m{#1}} 
\usepackage{bm}

\ifCLASSOPTIONcompsoc
  \usepackage[caption=false,font=normalsize,labelfont=sf,textfont=sf]{subfig}
\else
  \usepackage[caption=false,font=footnotesize]{subfig}
\fi

\usepackage{textcomp}
\usepackage{stfloats}
\usepackage{url}
\usepackage{verbatim}
\usepackage{graphicx}
\usepackage{cite}
\usepackage[table,xcdraw]{xcolor}
\usepackage{diagbox}
\usepackage{makecell}
\usepackage{multirow}

\begin{document}

\title{Holographic Surface Enabled Integrated Sensing and Communications}

\author{Haobo~Zhang,~\IEEEmembership{Member,~IEEE,}
        Shuhao~Zeng,~\IEEEmembership{Member,~IEEE,}
        Xinyuan~Hu,~\IEEEmembership{Graduate Student Member,~IEEE,}
        Shupei~Zhang,~\IEEEmembership{Graduate Student Member,~IEEE,}
        Boya~Di,~\IEEEmembership{Senior Member,~IEEE,}
        Hongliang~Zhang,~\IEEEmembership{Member,~IEEE,}
        Lianlin~Li,~\IEEEmembership{Senior Member,~IEEE,}
        Lin~Chang,
        Weixiang~Jiang,
        Shaohui~Sun,~\IEEEmembership{Member,~IEEE,}
        Zhu~Han,~\IEEEmembership{Fellow,~IEEE,}
        Yonina~Eldar,~\IEEEmembership{Fellow,~IEEE,}
        
        Rui~Zhang,~\IEEEmembership{Fellow,~IEEE,}
        M\'erouane~Debbah,~\IEEEmembership{Fellow,~IEEE,}
        and Lingyang~Song,~\IEEEmembership{Fellow,~IEEE}\\
\thanks{Haobo Zhang is with School of Electronic and Computer Engineering, Peking University Shenzhen Graduate School, Shenzhen 518055, China, and also with Department of Engineering, University of Cambridge, Cambridge CB2 1PZ, UK (email: haobo.zh97@gmail.com).}
\thanks{Shuhao Zeng is with School of Electronic and Computer Engineering, Peking University Shenzhen Graduate School, Shenzhen 518055, China, and also with Department of Electrical and Computer Engineering, Princeton University, NJ 08544, USA (email: shuhao.zeng96@gmail.com).}
\thanks{Xinyuan Hu, Shupei Zhang, Boya Di, Hongliang Zhang, Lianlin Li, and Lin Chang are with School of Electronics, Peking University, Beijing 100871, China (email: huxiny@pku.edu.cn; zhangshupei@pku.edu.cn; boya.di@pku.edu.cn; hongliang.zhang@pku.edu.cn; lianlin.li@pku.edu.cn; linchang@pku.edu.cn).}
\thanks{Weixiang Jiang is with the State Key Laboratory of Millimeter Waves, School of Information Science and Engineering, Southeast University, Nanjing 210096, China (email: wxjiang81@seu.edu.cn).}
\thanks{Shaohui Sun is with the State Key Laboratory of Wireless Mobile Communication, Datang Mobile Communications Equipment Co., Ltd, Beijing 10083, China (email: sunshaohui@catt.cn).}
\thanks{Zhu Han is with Electrical and Computer Engineering Department, University of Houston, Houston, TX 77004, USA, and also with the Department of Computer Science and Engineering, Kyung Hee University, Seoul 446-701, South Korea (email: hanzhu22@gmail.com).}
\thanks{Yonina C. Eldar is with the Faculty of Mathematics and Computer Science, Weizmann Institute of Science, Rehovot 7610001, Israel (e-mail: yonina.eldar@weizmann.ac.il).}
\thanks{Rui Zhang is with the Department of Electrical and Computer Engineering, National University of Singapore, Singapore 117583 (email: elezhang@nus.edu.sg).}
\thanks{M\'erouane Debbah is with Khalifa University of Science and Technology, PO Box 127788, Abu Dhabi, UAE, and also with Korea University, Seoul 02841, South Korea (email: merouane.debbah@ku.ac.ae).}
\thanks{Lingyang Song is with the School of Electronics, Peking University, Beijing 100871, China, and also with the School of Electronic and Computer Engineering, Peking University Shenzhen Graduate School, Shenzhen 518055, China (email: lingyang.song@pku.edu.cn).}
}

\markboth{IEEE Transactions on Communications,~Vol.~14, No.~8, August~2025}%
{Shell \MakeLowercase{\textit{et al.}}: A Sample Article Using IEEEtran.cls for IEEE Journals}

\maketitle

\begin{abstract}
Integrated sensing and communications (ISAC) is an essential 6G capability for joint data transmission and environmental sensing.
To support 6G scenarios with stringent ISAC performance requirements, existing massive-MIMO-based systems are expected to scale toward ultra-massive MIMO. However, this scaling incurs prohibitive cost and power consumption when realized using widely adopted phased arrays with complex phase shifters and feeding networks. 
Recently, holographic integrated sensing and communications (HISAC) has emerged as a promising paradigm to address this issue. 
It employs reconfigurable holographic surfaces (RHSs), a type of leaky-wave antenna, as a cost- and energy-efficient implementation of ultra-massive MIMO-based ISAC, and offers enhanced flexibility for ISAC beam synthesis through holographic beamforming. 
In this paper, we provide a comprehensive tutorial on HISAC, focusing on how RHS-enabled holographic beamforming can be exploited to jointly support communication and sensing under practical hardware constraints. We first introduce the fundamentals of RHSs and discuss the unique leakage power constraint of holographic beamforming. We then present a general optimization framework for HISAC and show how HISAC enhances joint communication and sensing, sensing-assisted communication, and communication-assisted sensing. We further present HISAC system implementations and experimental results.
Finally, we outline promising research directions for HISAC, highlighting the potential of HISAC in advancing efficient, flexible, and high-performance ISAC networks.
\end{abstract}

\begin{IEEEkeywords}
Integrated sensing and communications (ISAC), reconfigurable holographic surface (RHS), holographic beamforming, massive MIMO, 6G.
\end{IEEEkeywords}

\section{Introduction}

\subsection{Motivation}

\IEEEPARstart{T}{he} sixth generation (6G) of wireless networks is envisioned to extend the capabilities of 5G by providing not only seamless broadband connectivity but also native support for sensing services~\cite{liu2022a, huang2024challenges}. This paradigm shift is driven by emerging applications such as autonomous driving~\cite{zhao2025a} and the low-altitude economy~\cite{jiang2025integrated}, which demand both efficient data transmission and accurate environmental perception. At the same time, it also has the potential to alleviate spectrum congestion arising from the rapid growth of data traffic, as radar spectrum resources can be reused for communication purposes~\cite{wei2023integrated}. As a result, integrated sensing and communications (ISAC) has been recognized as a key feature of 6G, where hardware and spectral resources are efficiently shared to enable the joint operation of communication and sensing, allowing wireless infrastructure to evolve into a unified platform with dual-functional abilities~\cite{luo2025isac, liu2023integrated_book}.

To support ISAC scenarios with stringent communication and sensing performance requirements, ultra-massive multiple-input multiple-output (MIMO) has emerged as an important enabling technology. Evolving from massive MIMO, a cornerstone of 5G, ultra-massive MIMO scales the array dimension to hundreds or even thousands of antenna elements~\cite{feng2025recent}. Such ultra-large arrays provide orders-of-magnitude improvements in spatial resolution and array gain, thereby enabling highly directional beams for high-throughput multi-user communications~\cite{zeng2024ris} and fine-grained sensing with high accuracy~\cite{gao2025integrated}. Moreover, ultra-massive MIMO offers richer spatial degrees of freedom~(DoFs) for dual-functional waveform design and interference suppression, which are essential to meet the stringent ISAC requirements~\cite{yue2025directivity}.

However, realizing ultra-massive MIMO-based ISAC in practice is hindered by the issue of prohibitively high hardware cost and power consumption. This is because existing implementations largely rely on phased-array architectures, which require numerous phase shifters, power amplifiers, and complex feeding networks. This motivates the development of new transceiver architectures that can simultaneously deliver high ISAC performance and high energy and cost efficiency, thereby paving the way for scalable and adaptive ISAC in future 6G networks.

\subsection{Holographic Integrated Sensing and Communications}

To overcome the limitations mentioned above, holographic integrated sensing and communications (HISAC) has recently been introduced as a promising solution~\cite{zhang2022holographic}, whose core lies in the implementation of reconfigurable holographic surface (RHS)-enabled holographic beamforming for sensing and communication functions~\cite{zhang2023holographic_i}. An RHS is a type of leaky-wave antenna that integrates a large number of sub-wavelength radiating elements. By varying the bias voltages of its onboard diodes, the radiation amplitudes of the low-power elements can be independently adjusted to encode the desired \emph{holographic pattern}~\cite{di2025holographic}, thereby realizing holographic beamforming for dynamic beam manipulation.
Unlike the parallel feeding network in phased arrays, the RHS adopts a simplified structure, where embedded feeds excite the elements through serial feeding.
Moreover, the entire RHS, including the feeding network and the radiating elements, can be implemented using printed-circuit-board~(PCB) fabrication, resulting in an ultra-thin profile and low cost.
Consequently, the cost and power efficiencies of RHSs are substantially boosted~\cite{deng2022reconfigurable}, which addresses the scalability challenge of ultra-massive MIMO.

The concept of holographic antennas/surfaces~\cite{checcacci1970holographic} can be traced back to the 1960s, in which the original idea was to apply the holographic principle from optics to antenna design~\cite{checcacci1968holographic}. Specifically, metallic patches of different sizes and arrangements were used to record the holographic pattern. When excited by the reference wave from the feed, the holographic pattern redirected the energy to form a desired beam. However, the generated beam direction was fixed once the hologram was fabricated, which severely limited the flexibility and applicability of conventional holographic antennas. In recent years, the rapid progress of metamaterial technologies has enabled the reconfigurability of holographic surfaces~\cite{sleasman2016waveguide}. By loading active components such as positive-intrinsic-negative~(PIN) diodes onto the metamaterial antenna element and adjusting its external bias voltage, each element can switch between different states with varied radiation amplitudes. In this way, the surface can generate reconfigurable holographic patterns and flexibly steer beams toward desired directions, thereby supporting mobile communication and target sensing. 

One of the key enablers that fundamentally distinguishes HISAC from conventional ISAC architectures lies in the holographic beamforming enabled by RHSs. Due to the differences in hardware architecture and working mechanism, the holographic beamforming possesses the following {\em unique characteristics} that are typically not applicable in traditional analog beamforming enabled by phased arrays:
\begin{itemize}
    \item \emph{Dense elements:} The element spacing in RHSs is less than half-wavelength, indicating a denser deployment than that of phased arrays, which typically employ half-wavelength spacing. As a result, holographic beamforming offers a larger number of controllable parameters than conventional analog beamforming for a given aperture size.
    \item \emph{Adjustable array aperture:} The effective aperture for holographic beamforming can be adjusted by activating or deactivating subsets of radiation elements, allowing flexible switching between narrow and wide beams. However, the aperture size for traditional analog beamforming is typically fixed.
    \item \emph{Reconfigurable aperture position:} The aperture position in holographic beamforming can also be optimized since electronic antenna movement can be realized through the selective activation of RHS subarray at different locations. In contrast, phased arrays have to rely on extra components such as mechanical structures or switch circuits for movement. 
    \item \emph{Additivity of holographic patterns:} To support multi-user and multi-target holographic beamforming, multiple holographic patterns corresponding to different directions can be superposed and simultaneously encoded on a single RHS. This is not feasible for conventional analog beamforming as it has to satisfy the constant modulus constraint.
\end{itemize}

The above characteristics of RHS-enabled holographic beamforming give rise to several key advantages of HISAC over conventional ISAC architectures. Representative advantages include low cost and power consumption, tunable coverage, channel adaptation, and low complexity.
Specifically, the low cost and power consumption of HISAC stem from the use of low-power metamaterial elements and simple serial feeding structures in RHSs, which can be readily fabricated using mature PCB manufacturing techniques. 
Moreover, by flexibly rescaling the effective array aperture, gain-flat beams with adjustable beamwidths can be generated, allowing HISAC to adapt to diverse coverage requirements. 
In addition, the ability to reposition the effective aperture provides a practical realization of movable antennas~\cite{zhang2025fluid, zhang2025movable}, offering an additional mechanism to adapt to varying channel conditions in ISAC scenarios. Furthermore, the additivity of holographic patterns enables low-complexity generation of multiple beams towards different users and targets. Together, these advantages highlight the potential of HISAC to meet the stringent performance requirements of 6G ISAC, while offering enhanced flexibility and scalability across a wide range of deployment scenarios.

HISAC is essentially different from reconfigurable intelligent surface~(RIS)/intelligent reflecting surface~(IRS)-based ISAC in both channel model and operation mechanism~\cite{di2025reconfigurable}. Specifically, RISs/IRSs represent a class of reflective metasurfaces that control wireless propagation by imposing adjustable phase shifts on incident signals. Since the feeding antennas of RISs/IRSs are apart from the surface, the transmitted signals have to undergo an additional reflection before being redirected by the RIS/IRS. In contrast, RHSs integrate the feed and the metasurface into a unified structure, thus removing the extra reflection path. Moreover, RISs/IRSs rely on tuning the phase shifts of their elements to steer reflected signals, whereas RHSs adjust the radiation amplitudes of their elements to construct reconfigurable holographic patterns. 

It is also worth emphasizing that HISAC is not a straightforward extension of RHS-aided communication. In the latter, the holographic pattern is designed solely to optimize communication-oriented metrics such as achievable rate and outage probability, and the receiver only performs data decoding. By contrast, HISAC must simultaneously satisfy both communication and sensing requirements. This calls for new models that characterize the propagation of both communication and sensing signals, as well as holographic pattern designs that enhance and balance both the channel gain of communication users and the illumination power of sensing targets. In addition, the processing of echo signals received by RHSs must be carefully investigated to enable reliable target detection and parameter estimation.

\subsection{Use Cases}

\begin{figure*}[!t]
\centering
\includegraphics[width=5.5in]{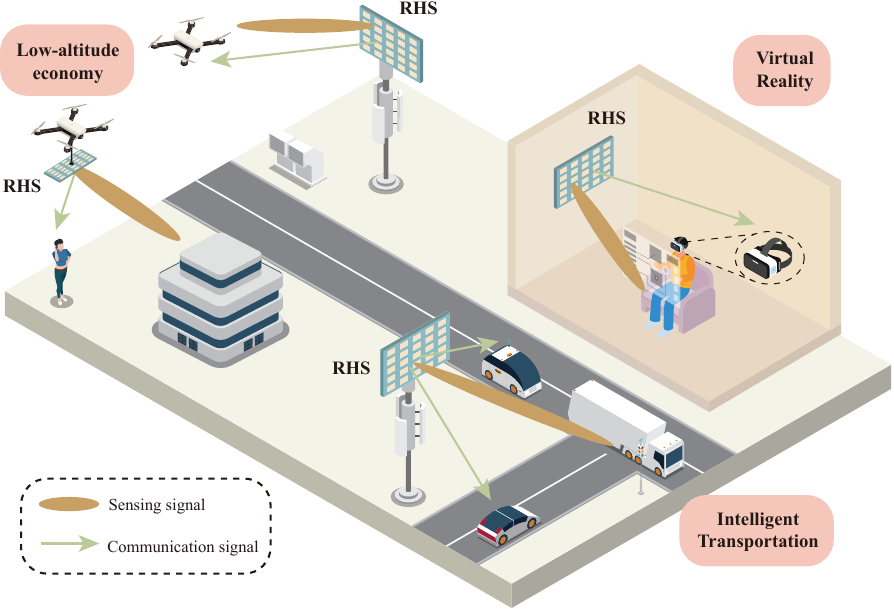}
\caption{Use cases of HISAC.}
\label{f_use_cases}
\end{figure*}

The distinctive properties of HISAC open up a wide range of potential applications in future wireless systems. Several representative use cases shown in Fig.~\ref{f_use_cases} are highlighted below.
\begin{itemize}
    \item \emph{Low-Altitude Economy:} Different from conventional radar dedicated for UAV sensing, base stations~(BSs) empowered by HISAC can not only monitor UAV trajectories but also maintain robust air-to-ground communication links, supporting command transmission and payload data delivery. Owing to the low-power and lightweight design of RHSs, HISAC is also particularly attractive for wide-area deployment and seamless integration with unmanned traffic management systems.
    \item \emph{Virtual Reality:} In virtual reality (VR) applications, precise awareness of the user’s surroundings is essential~\cite{wang2024integration}. Due to the large antenna arrays of RHSs with ultra-dense elements, fine-grained sensing of the environment can be performed by HISAC. Furthermore, the compact and cost-efficient nature of RHSs allows them to be densely embedded into indoor walls or ceilings, thus guaranteeing seamless connectivity and stable throughput for VR systems.
    \item \emph{Intelligent Transportation:} Vehicular networks rely on both real-time data exchange and reliable environmental perception to support autonomous driving and cooperative mobility. HISAC with large-scale aperture enables vehicles and roadside units to establish high-throughput communication links while simultaneously detecting pedestrians, vehicles, and road hazards with fine spatial resolution.
\end{itemize}

\begin{figure}[!t]
\centering
\includegraphics[width=3in]{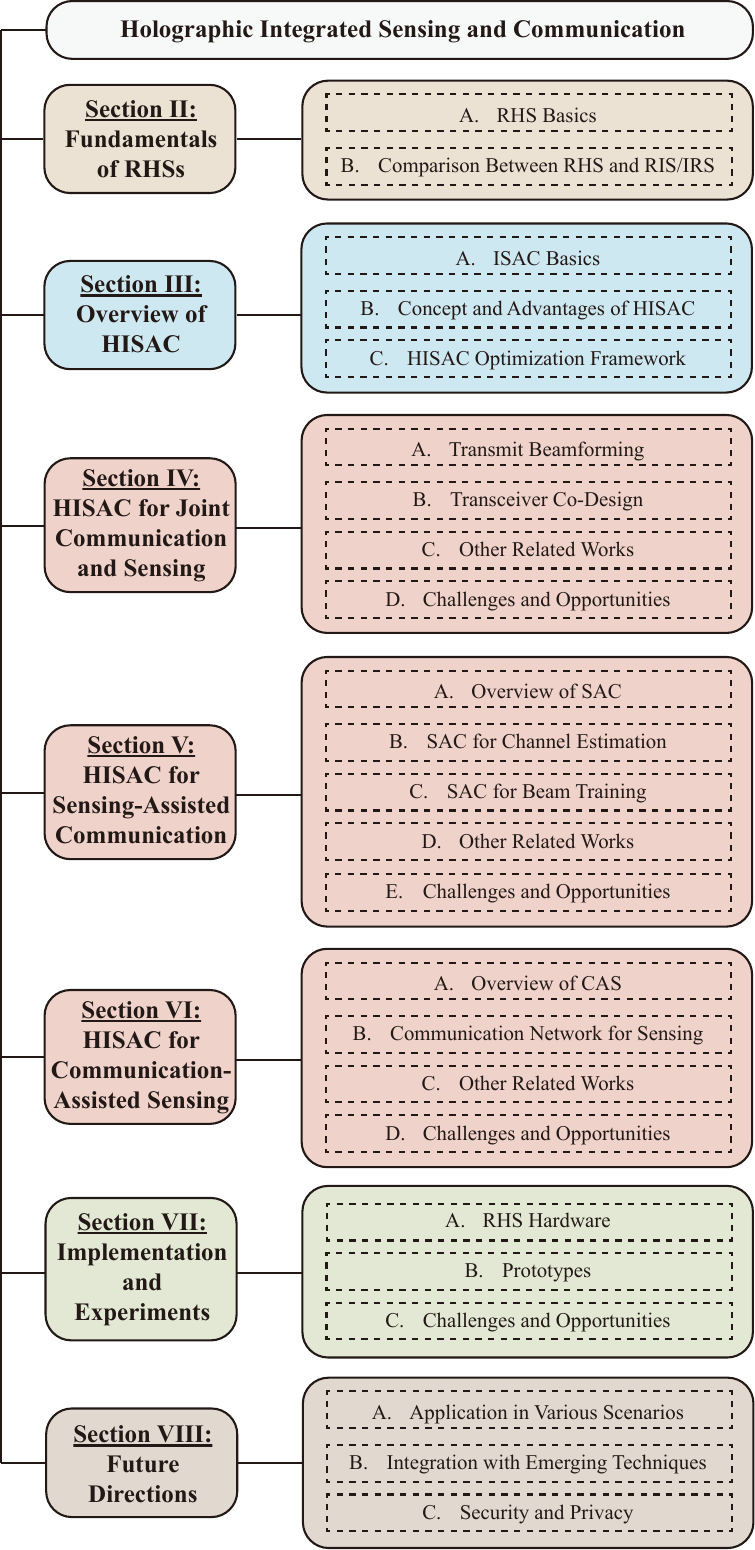}
\caption{Outline of this paper.}
\label{f_outline}
\end{figure}

\subsection{Contribution and Organization}

In the literature, several survey and tutorial papers have investigated ISAC systems enabled by ultra-massive MIMO and metasurface technologies~\cite{ding2024advancements, lu2024a, dai2025tutorial, hakimi2024roadmap, liu2024near, galappaththige2025cell, shi2024ris, han2025network, osorio2025the}. Specifically, the authors in~\cite{lu2024a, dai2025tutorial, hakimi2024roadmap, liu2024near} focus on ISAC channel modeling and signal processing methodologies tailored to the near-field region of ultra-massive MIMO systems. In addition, several surveys have examined cell-free massive MIMO-empowered ISAC, where multiple distributed access points~(APs) collaboratively provide joint sensing and communication services~\cite{galappaththige2025cell, shi2024ris, han2025network, osorio2025the}. These studies primarily rely on conventional ultra-massive MIMO implementations, such as phased arrays or distributed antenna systems. In parallel, metasurface-enabled ISAC has also attracted increasing attention, and a number of surveys have reviewed the associated ISAC frameworks~\cite{magbool2025a, li2025ris, zhu2025enabling, sur2024a, chopra2025ris, tishchenko2025the}. However, the emphasis of these works has largely been placed on reflective metasurfaces such as RIS/IRS\footnote{It is worth noting that HISAC and RIS/IRS-enabled ISAC are suited to different application scenarios and system requirements, as will be discussed in Sec.~\ref{ss_cbrr}.}, while RHSs, which operate as compact radiating apertures with embedded feeding structures, have remained largely unexplored in existing surveys and tutorials.

As discussed earlier, HISAC exhibits a number of distinctive advantages such as low cost, low power consumption, and tunable coverage, which endow it with promising potential for future ISAC systems. As a result, an increasing number of works have investigated HISAC-related topics from various perspectives, such as beamforming design and channel estimation. However, these studies typically focus on specific problems or scenarios, and a comprehensive and unified understanding of HISAC as a new ISAC paradigm is still lacking. Although the concept of HISAC has been mentioned in the survey paper~\cite{magbool2025a}, the fundamental hardware architectures and operating principles of RHSs, which form the physical basis of HISAC, have not been systematically introduced. Moreover, the key advantages of HISAC enabled by holographic beamforming, as well as the works related to sensing-assisted communication~(SAC) and communication-assisted sensing~(CAS), have not been reviewed. Therefore, a comprehensive tutorial on HISAC is essential to understand its hardware foundations and to clarify its capability boundaries and design insights.

Against this background, this paper aims to deliver a tutorial on HISAC that addresses the limitations of existing works by incorporating the description of RHS hardware, outlining the key advantages of HISAC, developing a general HISAC optimization framework, and providing a more comprehensive review of existing works. The main contributions are summarized as follows.

\begin{itemize}
  \item We introduce the fundamentals of RHSs, covering the operating principles, hardware architectures, and signal modeling in the literature. In particular, we explain how the holographic principle is exploited to reconstruct radiation patterns, and discuss the unique leakage power constraint of RHSs. This lays the foundation for understanding the role of RHSs in enabling HISAC.
  \item Then, we summarize the key advantages of HISAC, which help clarify its performance potential. We further present a general HISAC optimization framework synthesized from existing studies, highlighting the key design variables, objective functions, and constraints involved. In addition, two new case studies on HISAC performance tradeoffs are developed to illustrate how the framework can be applied in practice.
  \item To further offer insights into the design and advantages of HISAC, we explore three architectures of HISAC and review the related works in the literature. The first is joint communication and sensing (JCAS), where a common ISAC beam is generated via holographic beamforming to serve both functions simultaneously. The second is SAC, where sensing results are leveraged to support communication tasks such as beam training. The third is CAS, in which communication waveforms or networks are exploited to improve sensing performance.
  \item We also examine the implementation aspects of holographic beamforming-enabled systems. The design of RHS elements and array configurations are first provided. Experimental evaluations in existing works are also reviewed to demonstrate the feasibility and performance of HISAC in practical deployments.
  \item Finally, we outline the future research directions of HISAC in diverse scenarios, including satellite ISAC, autonomous driving, and robotics. We also discuss its potential integration with emerging technologies such as deep learning, wireless power transfer, and non-orthogonal multiple access (NOMA). In addition, we highlight the security and privacy issues associated with HISAC.
\end{itemize}

The rest of this paper is organized as illustrated in Fig.~\ref{f_outline}. Section~\ref{s_fr} introduces the fundamentals of RHSs. Section~\ref{s_ohics} provides an overview of HISAC principles and advantages. Sections~\ref{s_hjcas}, \ref{s_hsac}, and \ref{s_hcas} discuss the JCAS, SAC, and CAS frameworks of HISAC, respectively. Section~\ref{s_ie} describes the implementation aspects of holographic beamforming-enabled systems, followed by experimental results that demonstrate their feasibility. Section~\ref{s_e} explores future research directions of HISAC. Finally, Section~\ref{s_c} concludes the paper.

For clarity, the following notations are adopted throughout the paper. The holographic pattern is denoted by $\bm{\psi}$. The resulting RHS beamformer is $\bm{M}$ (or $\bm{m}$ in the 1D case). The digital beamformers are represented by $\mathbf{B}$. Unless otherwise specified, bold lowercase letters denote vectors and bold uppercase letters denote matrices.

\section{Fundamentals of RHSs}
\label{s_fr}

In this section, we introduce the fundamentals of RHSs. We start by describing their working principles and hardware architectures, followed by the holographic beamforming model and the corresponding leakage power constraint. Building on these fundamentals, we further compare RHSs with RISs/IRSs to clarify the distinctive characteristics of RHS-enabled systems.

\subsection{RHS Basics}
\label{ss_rb}

\subsubsection{Working Principle} 
\label{sss_wp}

\begin{figure}[!t]
\centering
\includegraphics[width=3.3in]{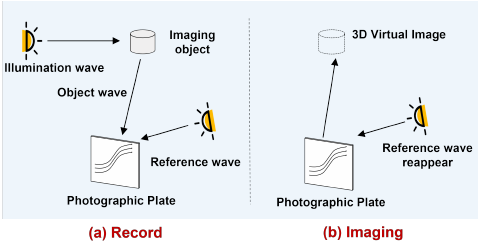}
\caption{Illustration of optical holographic principle.}
\label{f_principle}
\end{figure}

The operation of an RHS relies on the holographic principle, which was first introduced in optics. 
The optical holographic principle records the complete optical wavefront, including both amplitude and phase, thereby enabling the reconstruction of the object image~\cite{pedrotti2017introduction}. In this process, two coherent waves are involved: an \emph{object wave} carrying information of the object and a \emph{reference wave} serving as a phase reference. As shown in Fig.~\ref{f_principle}(a), to record the full wavefront, one beam serves as the reference wave and is directly projected onto the recording medium (i.e., the photographic plate). The reference wave at position $\bm{r}$ can be modeled as
\begin{equation}
    \xi_r(\bm{r}) = A_r(\bm{r}) e^{j\phi_r(\bm{r})},
\end{equation}
where $A_r(\bm{r})$ and $\phi_r(\bm{r})$ denote the amplitude and phase of the reference wave, respectively. Another beam, coherent with the reference wave, illuminates the object and is then reflected to form the object wave and reach the recording medium. According to Huygens’ principle, the object wave can be regarded as the superposition of a large number of spherical waves originating from every point on the object. Thus, the object wave at position $\bm{r}$ is expressed as
\begin{equation}
    \xi_o(\bm{r}) = \sum_{\bm{r'}} \rho_{\bm{r'}}(\bm{r}) = A_o(\bm{r}) e^{j\phi_o(\bm{r})},\label{e_U_o}
\end{equation}
where $\bm{r'}$ is a point on the object. $\rho_{\bm{r'}}$ is the spherical wave from $\bm{r'}$. $A_o(\bm{r})$ and $\phi_o(\bm{r})$ denote the amplitude and phase of the object wave, respectively.

Owing to the coherence between the reference and object waves, they interfere with each other, and the resulting intensity at position $\bm{r}$, which is also referred to as a holographic pattern, is given by
\begin{align}
    I(\bm{r}) =& |\xi_o(\bm{r}) + \xi_r(\bm{r})|^2\notag\\
    =& |\xi_o(\bm{r})|^2\! +\! |\xi_r(\bm{r})|^2\! +\! \xi_o(\bm{r})\xi_r^*(\bm{r})\! +\! \xi_o^*(\bm{r})\xi_r(\bm{r}).
\end{align}
The first two terms correspond to the intensities of the object and reference waves, respectively. The other two terms encode both amplitude and phase information of the object wave. This means that the holographic pattern preserves the complete wavefront information of the object wave.

Next, to reconstruct the virtual image of the object, the recording medium with the holographic pattern is illuminated by a beam that has the same wavelength and propagation direction as the original reference wave. Typically, the pattern recorded by the photographic plate is linearly proportional to the holographic pattern $I(\bm{r})$, which is given by
\begin{equation}
    T(\bm{r}) = T_0 + \beta I(\bm{r}),
\end{equation}
where $T_0$ is the average response of the plate, and $\beta$ denotes the modulation coefficient. As a result, when illuminated by the same reference wave, the wave reconstructed by the holographic pattern is given by
\begin{align}
    \xi_T(\bm{r}) =~& \xi_r(\bm{r}) T(\bm{r})\notag\\
    =~& \xi_r(\bm{r})(T_0+ \beta|\xi_o(\bm{r})|^2 + \beta|\xi_r(\bm{r})|^2)\notag\\
    &+ \beta \xi^*_o(\bm{r})\xi_r(\bm{r})\xi_r(\bm{r}) + \beta \xi_o(\bm{r}) |\xi_r(\bm{r})|^2.\label{e_U_T}
\end{align}
In optical systems, the different terms in~(\ref{e_U_T}) can be spatially separated.
Among them, the last term is of particular interest as it can directly reconstruct the object wave. Specifically, when the reference wave is a plane wave, its amplitude $A_r(\bm{r})$ remains constant across the photographic plate, and the third term reduces to $\beta |A_r|^2 \xi_o(\bm{r})$, which is a scaled replica of the object wave $\xi_o(\bm{r})$. Consequently, a virtual three-dimensional image of the object appears at its original position.

The concept of holography was later translated from optics to microwave engineering~\cite{checcacci1968holographic}. The underlying principle remains the same: interference patterns are encoded into a surface structure, which is then used to synthesize desired far-field beams. The key challenge lies in realizing such surface structures at microwave frequencies, leading to the notion of the holographic antenna~\cite{deschamps1967some}. In~\cite{checcacci1968holographic}, the holographic patterns were implemented on a holographic antenna based on a paraffin plate with variations in surface thickness, metallic strip placement and width. Following this, \cite{elsherbiny2004holographic} integrated the feed that emits the reference wave directly into the surface structure of the antenna to improve compactness and enhance radiation efficiency. 

\subsubsection{RHS Architecture}
\label{sss_ra}

\begin{figure}[!t]
\centering
\subfloat[]{\includegraphics[width=2.5in]{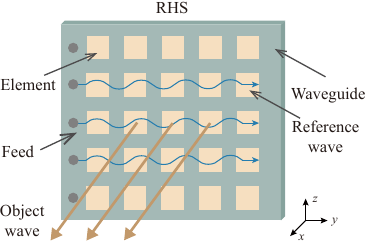}
\label{f_RHS_a}}
\hfil
\subfloat[]{\includegraphics[width=2.2in]{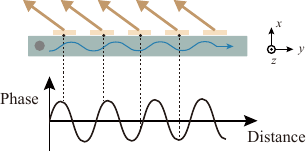}
\label{f_RHS_b}}
\caption{Diagram of an RHS: (a) perspective view; (b) bottom view.}
\label{f_RHS}
\end{figure}

In traditional holographic antennas, the holographic pattern was fixed once fabricated, and the antenna could only radiate beams in predetermined directions. This lack of reconfigurability severely limited the flexibility of holographic antennas for modern wireless systems, where dynamic beam adaptation is crucial. To enable the beam-steering of holographic antennas, metamaterial elements have been incorporated into holographic antenna designs, which leads to the development of RHS.

An RHS can be regarded as a special class of leaky-wave antenna that converts a reference wave into leaky radiation modulated by holographic patterns~\cite{nannetti2007leaky, lang2025leaky}. As shown in Fig.~\ref{f_RHS}, its structure typically integrates three key components: the feed, the waveguide, and the metamaterial elements.

\begin{itemize}
    \item \textbf{Feed:} It is embedded either at the edge or the bottom of the RHS and is directly connected to the RF chain. This feed converts the input RF signal into an electromagnetic wave, referred to as the reference wave.
    \item \textbf{Waveguide:} It acts as the propagation medium for the reference wave. As the wave travels along the waveguide, it sequentially excites the metasurface elements distributed across the RHS. Each element extracts a portion of the guided energy and converts it into a leaky wave, which is then radiated into free space. In contrast to the complex parallel feeding networks of phased arrays, the serial feeding structure adopted by the RHS is more compact and  simpler, making it well suited for scaling to thousands of antenna elements or beyond.
    \item \textbf{Metamaterial elements:} These elements are sub-wavelength in size and are fabricated using engineered metamaterial structures with tailored electromagnetic properties. 
    
    In particular, the inter-element spacing of RHSs can be smaller than half a wavelength, i.e.,
    \begin{equation}
        d_e \le \frac{\lambda}{2}, \label{e_d_e}
    \end{equation}
    where $d_e$ denotes the spacing between adjacent elements and $\lambda$ is the wavelength. 
    
    To enable dynamic control, active devices such as PIN diodes or varactors are embedded into the elements, allowing their radiation amplitude to be tuned in real time\footnote{There also exists another class of metasurface-based antennas, namely dynamic metasurface antennas (DMAs), whose element responses typically satisfy a Lorentzian constraint with coupled amplitude and phase~\cite{shlezinger2021dynamic}. In this paper, however, we focus on RHSs with independently tunable amplitudes.}. By coordinating the radiation amplitudes of all elements, the RHS can effectively encode a holographic pattern to form the desired object wave without relying on expensive phase shifters.
\end{itemize}

As for the object wave reconstruction, consider an RHS comprising $L$ feeds and $N$ metamaterial elements, and each feed excites the RHS elements in the same row. Assume the desired object wave is a plane wave toward direction $(\theta_o, \phi_o)$, and the object wave at the location of the $n$-th element, denoted by $\bm{r}_n$, is $\Psi_o(\bm{r}_n, \theta_o, \phi_o)$.
Besides, the reference wave at the location of the $n$-th element in the same row is denoted by $\Psi_r(\bm{r}'_n)$, where $\bm{r}'_n$ is the displacement from the $n$-th element to the feed in the same row.
According to the holographic principle introduced in~\ref{sss_wp}, reconstructing the object wave requires recording the holographic pattern
\begin{equation}
    \Psi_i(n, \theta_o, \phi_o) = \Psi_o(\bm{r}_n, \theta_0, \phi_0) \Psi^*_r(\bm{r}'_n).
\end{equation}
When the holographic pattern $\Psi_i(\bm{r}_n, \theta_o, \phi_o)$ is excited by the reference wave, the reconstructed wave is obtained as
\begin{equation}
    \Psi'_o(\bm{r}_n)\! =\! \Psi_i(n, \theta_o, \phi_o) \Psi_r(\bm{r}'_n)\! =\! \Psi_o(\bm{r}_n, \theta_o, \phi_o) |\Psi_r(\bm{r}'_n)|^2.
\end{equation}
Therefore, the object wave $\Psi_o(\bm{r}_n, \theta_o, \phi_o)$ is faithfully recovered.

The direct recording of both the amplitude and phase of $\Psi_i(n, \theta_o, \phi_o)$ is challenging owing to the limitation in material and cost. To this end, an amplitude-controlled holographic beamforming method is adopted by the RHS to record the holographic pattern $\Psi_i$ and recover the object wave. Specifically, the phase of the reference wave continuously varies along its propagation over the waveguide. If the phase of the reference wave at a certain element is close to that of the object wave, the element can be tuned into a resonant state with a relatively high radiation amplitude, allowing most of the reference wave energy to be radiated into free space. Conversely, when the phase of the reference wave differs significantly from that of the object wave, the element will be set to a non-resonant state with a low radiation amplitude, so that little energy from the reference wave is radiated into free space. 

This implies that to realize a mapping between the element radiation amplitude and the interference holographic pattern $\Psi_i$, the amplitude of each element should be negatively correlated with the phase difference between the reference and object waves at that location. It should be noted that the real part of the interference wave, i.e., $\text{Re}[\Psi_i]$, corresponds to the cosine of the phase difference between the reference and object waves. Its value varies with the phase difference, either increasing or decreasing, thereby satisfying the requirement of amplitude control. Therefore, by adopting $\text{Re}[\Psi_i]$ to characterize the radiation amplitude of each element, the normalized amplitude of the $n$-th element for generating the object wave in direction $(\theta_o,\phi_o)$ can be expressed as
\begin{equation}
    \psi_n(\theta_o,\phi_o) = \frac{\text{Re}\big[\Psi_i(n,\theta_o,\phi_o)\big] + 1}{2}.\label{e_m_n}
\end{equation}
Here, the constant 1 in the numerator shifts $\text{Re}[\Psi_i(n,\theta_o,\phi_o)]$ from $[-1,1]$ to $[0,2]$, so that division by $2$ yields a normalized amplitude in the range $[0,1]$.

\subsubsection{Holographic Beamforming Model}
\label{sss_hbm}

In this part, a detailed mathematical model for RHS-enabled holographic beamforming is provided. Similar to the above section, we consider an RHS that forms a uniform planar array with $N = N_y \times N_z$ radiating elements and $L = N_y$ excitation feeds. 
The EM wave radiated by the $(n_y, n_z)$-th element is
\begin{equation}
    y_n = \psi_{n_y, n_z} w_{n_y, n_z, l} e^{-j\bm{k}_s \bm{r}_{n_y, n_z, l}} x_l,
\end{equation}
where $n = (n_z-1)N_y + n_y$. $\psi_{n_y,n_z} \in [0, 1]$ is the normalized radiation amplitude of the $(n_y, n_z)$-th element and can be tuned by applying different bias voltages. Here, $w_{n_y,n_z, l}$ is the energy coefficient given by~\cite{zhang2025fluid}
\begin{equation}
    w_{n_y, n_z, l} = \sqrt{\chi_{n_y, l}} e^{-\alpha|\bm{r}_{n_y, n_z, l}|}, 
\end{equation}
where $\chi_{n_y, l}$ is the ratio of the power received by an RHS element in the $n_y$-th row to the total power of the reference wave emitted by the $l$-th feed, and we have $\chi_{n_y, l} = 0$ when $n_y \ne l$. Here, $\alpha$ denotes the propagation loss factor with a typical range $[1, 10]$~\cite{zhang2024target}, $\bm{k}_s$ is the wavenumber vector of the RHS, $\bm{r}_{n_y, n_z, l}$ is the location vector from the $l$-th feed to the $(n_y, n_z)$-th element, and $x_l$ is the signal emitted by the $l$-th feed. 

Thus, the relationship between the signals fed into the RHS and radiated by the RHS can be written as
\begin{equation}
    \bm{y} = \bm{M}\bm{x}.
\end{equation}
Here, $\bm{x} = (x_1, \cdots, x_L)^T$ is the signal emitted by the feeds, $\bm{M}$ is the holographic beamformer given by
\begin{equation}
\bm{M} = \operatorname{diag}(\psi_1,\ldots,\psi_N)\bm{F},
\label{eq:Mmatrix}
\end{equation}
where the $(n, l)$-th element of $\bm{F}$ is
$w_{n_y, n_z, l} e^{-j \bm{k}_s\cdot \bm{r}_{n_y, n_z, l}}$. 

Note that the above model is narrowband. In wideband scenarios, both the propagation of the reference wave and the response of RHS elements can become frequency-dependent. Specifically, under serial feeding, the reference wave experiences a frequency-dependent phase shift as it propagates along the surface. In addition, when the bandwidth is sufficiently large, the radiation amplitude of an RHS element may vary significantly with frequency due to their resonant nature, leading to non-uniform beamforming behavior across subcarriers. As a result, the beam direction becomes frequency-dependent, which needs to be carefully addressed~\cite{di2021reconfigurable}.

\subsubsection{Leakage Power Constraint}

Given the amplitude-controlled design and serial feeding structure of RHS, an inherent limitation of leakage power arises~\cite{di2025holographic}. Specifically, in the RHS, the reference wave launched by the feed propagates along the waveguide and sequentially excites the metasurface elements. 
Since the power of reference wave is non-negative, the total radiated power of the serially-fed elements in each row cannot exceed the incident power of the feed. Thus, the leakage power constraint for the $n_y$-th row is
\begin{equation}
    \sum_{n_z} \psi^2_{n_y, n_z} \eta_{n_y, n_z} \le 1,\label{e_sum}
\end{equation}
where $\eta_{n_y, n_z}$ equals $w^2_{n_y, n_z, n_y}$ and represents the ratio of the power accepted by the $(n_y, n_z)$-th element to the incident power\footnote{The coefficient $\eta_{n_y, n_z}$ is an exponential attenuation term along the propagation direction, which captures the serial-feeding behavior in an averaged manner~\cite{smith2017analysis}. As the wave propagates across the RHS, a portion of the power is extracted by each element, leading to a cumulative attenuation that is represented by this exponential factor. Thus, the leakage power constraint accounts for the effect of power attenuation induced by serial-feeding structure. Some recent works have also explored more detailed models to describe such effect~\cite{hu2025reconfigurable_tcom, yang2026reconfigurable}. However, these models inevitably incur higher complexity and are not considered in this tutorial for brevity.}. 

The leakage power constraint differs from conventional power constraints applied to phased arrays. Since each element extracts energy from the propagating reference wave, mutual coupling in the power domain across the RHS aperture are created, which does not arise in traditional phased arrays. Moreover, the leakage power constraint introduces a new perspective on antenna design, i.e., the effective aperture size. Since the radiation amplitudes of RHS elements can be flexibly adjusted, the effective aperture can be dynamically reconfigured to meet diverse communication and sensing requirements. Detailed case studies will be provided in the following sections.

\subsection{Comparison Between RHS and RIS/IRS}
\label{ss_cbrr}

\begin{table}[!t]
\caption{Difference between RHS and RIS/IRS}
\centering
\renewcommand{\arraystretch}{2}
\begin{tabular}{|c|c|c|}
\hline
\textbf{Type} & \textbf{RHS} & \textbf{RIS/IRS}\\
\hline
\textbf{Structure} & Embedded feed & External feed\\
\hline
\makecell{\textbf{Feeding}\\\textbf{Network}} & Serial & Parallel\\
\hline
\makecell{\textbf{Working}\\\textbf{Mechanism}} & Wave Leakage & Wave Reflection\\
\hline
\makecell{\textbf{Typical}\\\textbf{Application}} & Transmit/receive antenna & Passive relays\\
\hline
\end{tabular}
\label{t_RHS_RIS_diff}
\end{table}

In this subsection, we first describe the differences between RHS and RIS/IRS, and then compare their performances for communication and sensing.

\subsubsection{Differences Between RHS and RIS/IRS}

\begin{figure}[!t]
\centering
\subfloat[]{\includegraphics[height=1.8in]{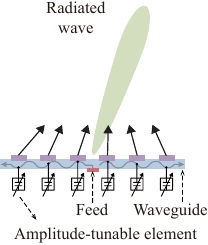}
}
\hfil
\subfloat[]{\includegraphics[height=1.8in]{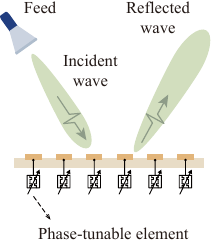}
}
\caption{Comparison of (a) RHS structure and (b) RIS/IRS structure.}
\label{f_compare}
\end{figure}

The differences between RHS and RIS/IRS in hardware structures, radiation mechanisms, and applications are summarized in Table~\ref{t_RHS_RIS_diff}.
Specifically, the feed and the waveguide of the RHS are tightly integrated into the same surface, as shown in Fig.~\ref{f_compare}. The feed injects a guided reference wave into the waveguide, which propagates along the surface and sequentially excites the metamaterial elements. This corresponds to a serial-feeding architecture, where each element extracts part of the guided energy and radiates it into free space. This embedded feed design not only reduces the profile of the antenna but also eliminates the need for external feeding networks. By contrast, RIS/IRS do not embed their own feeds~\cite{zhang2025large}. Instead, the feed antennas are deployed externally. This leads to a parallel feeding approach, in which all elements are illuminated simultaneously by the external antenna. While parallel feeding simplifies the surface structure itself, it requires dedicated transmitters in the environment and typically results in a bulkier hardware size.

As for the radiation mechanism, the RHS elements act as radiating units that convert the guided reference wave into leaky waves. Their radiation is achieved by coupling a fraction of the guided energy into free space, and the radiation amplitudes of the elements are tuned to encode holographic patterns across the surface. On the other hand, RIS/IRS elements operate passively by reflecting the incident signals. Beam steering is accomplished by controlling the phase shifts of the reflected waves, while the amplitudes remain largely fixed~\cite{yue2023intelligent}.

The functional roles and applications of RHSs and RISs/IRSs in wireless systems also vary~\cite{zeng2021reconfigurable2}. 
Specifically, RHSs, with integrated feeds and leaky-wave radiation, function as compact active transceivers that directly generate and receive electromagnetic waves. They can be deployed on various platforms such as UAVs, satellites, and vehicles, where both communication and sensing functionalities are required under strict size and power constraints~\cite{zeng2018uav}. In contrast, RISs/IRSs operate as passive relays that reshape the propagation environment without actively transmitting signals, making them well suited for network-side deployment~\cite{zeng2022intelligent2, zeng2022reconfigurable2}. For example, it can be used to extend coverage at cell edges or improve connectivity in non-line-of-sight areas~\cite{liu2021reconfigurable, zeng2021reconfigurable}.

\subsubsection{Performance Comparison}

The above differences between RHSs and RISs/IRSs result in distinct performance characteristics.: 
\emph{i}) the aperture gain of RIS/IRS increases quadratically with the number of metamaterial elements for the far-field beamforming case~\cite{tang2021wireless},
whereas the gain of RHS grows approximately linearly with the number of radiating elements~\cite{zhang2022holographic}; \emph{ii}) the RIS/IRS channel is cascaded, and the end-to-end path loss scales with $(d_1 d_2)^2$, where $d_1$ and $d_2$ denote the distances from the feed to the RIS/IRS and from the RIS/IRS to the receiving antenna, respectively. 
Both distances can be substantial in practical deployments.
In contrast, RHS employs an embedded feeding architecture, such that the distances between the feed and the RHS elements are fixed given the RHS hardware, while the path loss of RHS-based system scales with the distance between the RHS and the receiving antenna. 

These characteristics imply that RHS and RIS/IRS may outperform one another under different operating conditions including deployment geometry, link distance, or aperture size. For example, Table~\ref{t_RHS_RIS_comp} shows the optimal metasurface type for maximizing the single-target detection probability under different working frequencies and hardware sizes~\cite{zhang2024target}. Here, “I’’ denotes RIS, “H’’ denotes RHS, and $N_H$ is the number of RHS elements. The physical dimension of the RIS is the same as that of the RHS. Two main trends are revealed in the table. First, for larger apertures, RIS/IRS tends to outperform RHS, which is consistent with the performance characteristics discussed above. Second, at higher frequencies, RHS achieves superior performance even with a fixed number of elements. This is because the smaller wavelength reduces element spacing, thereby shortening the propagation distances between the feed and the RHS elements and alleviating the attenuation of signals propagation in RHSs.

In summary, although both RHSs and RISs/IRSs are based on metamaterials, their feeding architectures and radiation mechanisms are fundamentally different, leading to distinct functional roles and performance characteristics. RISs/IRSs act as passive relays and are therefore well suited for coverage enhancement. In contrast, RHSs function as compact active antennas with embedded feeding structures, making HISAC particularly attractive for ISAC platforms subject to stringent power and size constraints, such as UAVs and satellites. From a system design perspective, these differences also imply distinct modeling and optimization frameworks. Existing RIS-ISAC systems typically aim to optimize the reflection coefficients of a RIS, which is deployed separately from the BS. In contrast, in HISAC systems, the RHS directly serves as the transmit/receive antenna, and the design of holographic pattern is carried out under leakage power and other hardware constraints of the RHS~\cite{liu2023integrated, liu2024snr}. In terms of communication and sensing performance, whether RHSs or RISs/IRSs achieve superior performance depends on the specific operating conditions, such as the link distance and the aperture size.

\begin{table}[!t]
\centering
\caption{The Optimal Metasurface Type for Single-Target Detection Under Various Conditions}
\label{t_RHS_RIS_comp}
\begin{tabular}{|c|c|c|c|c|}
\hline
\diagbox{$f$}{$N_H$} & $16 \times 16$ & $32 \times 32$ & $64 \times 64$ & $128 \times 128$ \\ \hline
1 GHz   & \cellcolor[HTML]{C0C0C0}{I} & \cellcolor[HTML]{C0C0C0}{I} & \cellcolor[HTML]{C0C0C0}{I} & \cellcolor[HTML]{C0C0C0}{I} \\ \hline
2.4 GHz & \cellcolor[HTML]{C0C0C0}{I} & \cellcolor[HTML]{C0C0C0}{I} & \cellcolor[HTML]{C0C0C0}{I} & \cellcolor[HTML]{C0C0C0}{I} \\ \hline
12 GHz  & H & H & \cellcolor[HTML]{C0C0C0}{I} & \cellcolor[HTML]{C0C0C0}{I} \\ \hline
24 GHz  & H & H & H & H \\ \hline
77 GHz  & H & H & H & H \\ \hline
\end{tabular}

\vspace{2mm}
\small \textit{Note: H represents RHS; I represents RIS.}
\end{table}

\section{Overview of Holographic Integrated Communication and Sensing}
\label{s_ohics}

In this section, we provide an overview of HISAC. We begin with a brief review of the fundamentals of ISAC, and then introduce the concept of HISAC together with its key advantages. Building on this discussion, a general optimization framework for HISAC is presented to summarize the key design variables and constraints. Finally, two illustrative examples are developed to demonstrate how the framework can be applied in practice.

\subsection{ISAC Basics}
\label{ss_ib}

ISAC has emerged as a fundamental paradigm for beyond-5G and 6G networks, aiming to unify high-capacity wireless transmission with high-accuracy environmental sensing~\cite{luo2024horus}. Specifically, in conventional wireless systems, communication and sensing are typically implemented in different platforms as separate functions. However, wireless signals inherently possess a dual nature: each transmitting signal not only carries communication data but also interacts with the surrounding environment, embedding rich sensing information in the received signals. This intrinsic property enables the integration of sensing and communication, where both functionalities can be jointly realized through shared spectrum, hardware, and signal processing frameworks. As a result, ISAC improves both spectrum efficiency and functionality, enabling wireless infrastructure to evolve into a dual-purpose platform~\cite{wen2025a}.

Depending on how communication and sensing are integrated, ISAC systems can be categorized into three types~\cite{armeniakos2025stochastic}:
\begin{itemize}
    \item \textbf{Joint Communication and Sensing (JCAS):} In this type, communication and sensing share the same waveform, hardware, and spectrum resources, and the design goal is to simultaneously achieve satisfactory performances of both functions~\cite{zhang2021an}. For example, the transmitting waveforms can be optimized to maximize the achievable rate given the sensing performance constraint.
    \item \textbf{Sensing-Assisted Communication (SAC):} Unlike JCAS, where communication and sensing are conducted simultaneously, SAC performs sensing first and then exploits the sensing outcomes to support subsequent communication operations~\cite{yang2025location,yang2024localization}. For example, environmental perception results, such as the locations of users and scatterers, can assist channel estimation or beam training~\cite{chen2025trends}, thereby enhancing link reliability and reducing signaling overhead.
    \item \textbf{Communication-Assisted Sensing (CAS):} Communication waveforms and networks can also be leveraged to facilitate sensing tasks~\cite{liu2024cooperative}. For instance, multiple BSs can form a cooperative radar network through communication links to improve detection range and coverage.
\end{itemize}
These three paradigms represent different integration mechanisms within the ISAC framework. 
JCAS realizes an \emph{integration gain}, as communication and sensing functions share same resources and operate simultaneously, thus improving spectral efficiency compared with separate systems. 
In contrast, CAS and SAC mainly bring a \emph{coordination gain} as they exploit cross-domain cooperation to strengthen specific functionalities (e.g., using sensing information for communication channel estimation).

To evaluate the performance of ISAC systems, the following three types of metrics are widely used: communication metrics, which reflect the data transmission capability of the system; sensing metrics, which capture the detection and estimation performance of targets; and unified metrics, which explicitly characterize the trade-off and synergy between the two functionalities.

\textbf{Communication Metrics:} The International Mobile Telecommunications~(IMT)-2030 framework defines a broad set of key capabilities that serve as performance indicators for future 6G systems~\cite{ITU-RM2160, zeng2024dual}. Among these, several are directly related to communication performance, such as data rate and reliability~\cite{zeng2024reconfigurable}. The data rate of a MIMO communication system is given by
    \begin{equation}
        C = \log_2 \det\left(\bm{I}_{N_r}+\frac{1}{N_0}\bm{H}\bm{R}\bm{H}^H\right),
    \end{equation}
    where $\bm{I}_{N_r}$ is the identity matrix of size $N_r \times N_r$, $N_0$ is the noise power spectral density, $\bm{H}$ is the channel between the transmit antennas and receive antennas, and $\bm{R} = \mathbb{E}[\bm{x}\bm{x}^H]$ is the covariance matrix of transmit signal $\bm{x}$.
    
    As for reliability, an important metric is the outage probability, defined as the probability that the instantaneous achievable rate $R$ falls below a required threshold $R^{\text{th}}$~\cite{chen2024near}. Mathematically, the outage probability can be given by
    \begin{equation}
        P_{\text{out}} = \Pr [ R < R_{\text{th}} ].
    \end{equation}
    This metric reflects the likelihood that the system fails to meet a minimum quality-of-service requirement under random channel variations or interference.

\textbf{Sensing Metrics:} Besides communication, ISAC performance is also evaluated from the sensing perspective. Typical sensing tasks include target detection and localization~\cite{zhang2021metalocalization}. Unlike communication metrics that focus on throughput or reliability of data delivery, sensing metrics are designed to capture the ability of the system to perceive and reconstruct the surrounding environment with high fidelity. A fundamental indicator in sensing is the probability of detection~\cite{zhang2023holographic}, which characterizes the probability of correctly identifying the presence of a target. Consider a binary hypothesis testing model:
\begin{equation}
    \begin{cases}
        \mathcal{H}_0: \text{no target is present,}\\
        \mathcal{H}_1: \text{a target is present.}
    \end{cases}\label{e_bht}
\end{equation}
Mathematically, the probability of detection is defined as
\begin{equation}
    P_d = \Pr[\mathcal{H}_1 \mid \mathcal{H}_1],
\end{equation}
which denotes the probability of declaring $\mathcal{H}_1$ when it is true. 

Another widely used benchmark is the Cramér-Rao Bound (CRB), which quantifies the accuracy of localization or estimation of other parameters such as radar cross section~(RCS)~\cite{yang2022metaslam, yang2021wireless}. For a vector of unknown target parameters $\bm{\theta} = (\theta_1, \dots, \theta_i, \dots, \theta_I)$, the CRB states that the variance of any unbiased estimator $\hat{\theta}_i$ is lower-bounded by the $i$-th diagonal element of the inverse Fisher information matrix (FIM), which can be expressed as
\begin{equation}
    \mathrm{var}(\hat{\theta}_i) \geq [\bm{J}^{-1}(\bm{\theta})]_{ii},
\end{equation}
where $\bm{J}(\bm{\theta})$ is the FIM of $\bm{\theta}$, quantifying how much information the observation, i.e., the received signals, carries about the unknown parameters $\bm{\theta}$. More specifically, the FIM can be expressed as $\bm{J}(\bm{\theta}) = -\mathbb{E}\!\left[\partial^2 \ln p(\bm{y}; \bm{\theta})/\partial \bm{\theta}\, \partial \bm{\theta}^T \right]$, 
where $p(\mathbf{y}; \boldsymbol{\theta})$ denotes the likelihood function of the received signal $\mathbf{y}$ parameterized by $\bm{\theta}$.

\textbf{Other Metrics:} 
Beyond the above communication- and sensing-only metrics, several unified metrics have been developed to jointly characterize both functionalities under a common theoretical framework.
One representative example is the estimation rate~\cite{dong2023rethinking}.
It quantifies the minimum amount of information required for the receiver to reconstruct or estimate the unknown environmental parameters within a given accuracy constraint, and can be given by

\begin{equation}
    R_E(D_L) = \min_{P_{\hat{\eta}|\eta}:\mathbb{E}[d(\eta,\hat{\eta})]=D_L} I(\eta;\hat{\eta}),
\end{equation}
where $\eta$ is the random parameter, $\hat{\eta}$ is the estimation of $\eta$, $I(\eta;\hat{\eta})$ is the mutual information between $\eta$ and $\hat{\eta}$, $\mathbb{E}[d(\eta,\hat{\eta})]$ is the expected distortion between $\eta$ and $\hat{\eta}$, $P_{\hat{\eta}|\eta}$ is the conditional probability distribution of the estimation $\hat{\eta}$ given the true parameter $\eta$.

In this way, it measures the sensing performance in the same information-theoretic unit as communication rate (bits per second per hertz), thus offering an alternative solution to interpret the interplay between communication and sensing in ISAC systems.

\begin{figure*}[!t]
\centering
\includegraphics[width=4in]{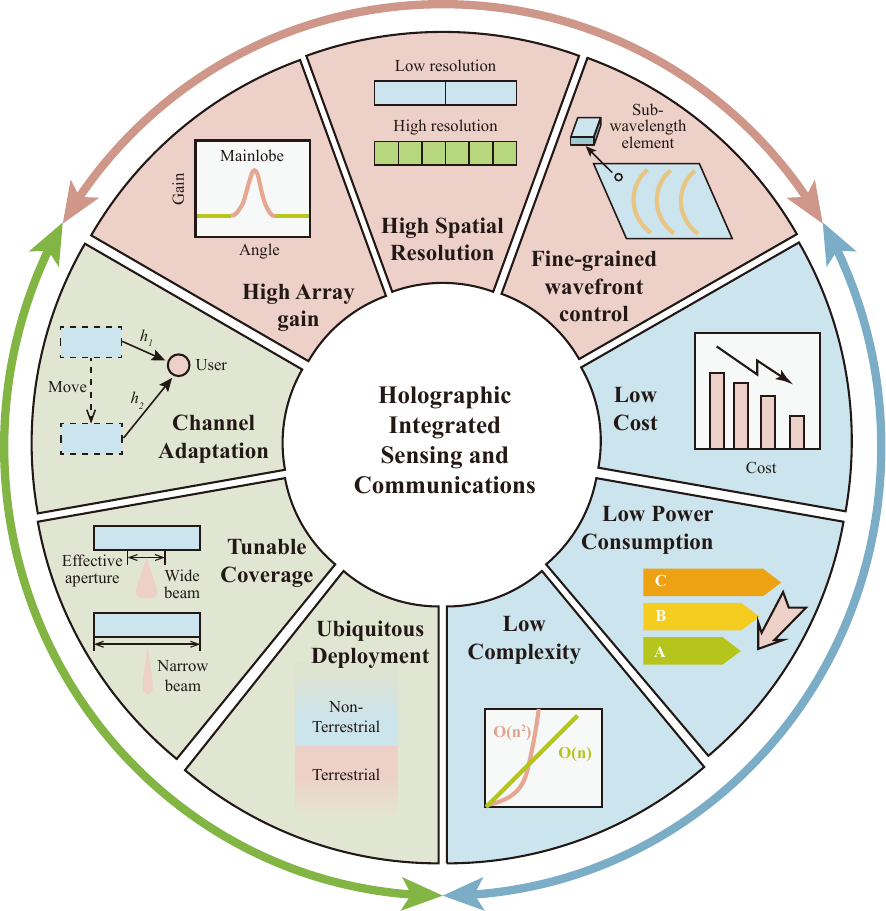}
\caption{Advantages of HISAC.}
\label{f_advantage}
\end{figure*}

In addition to such unified metrics, many other performance measures have also been widely used in existing works. Typical examples include beampattern-based metrics, such as power towards directions of targets and users~\cite{stoica2007on}, sidelobe level~\cite{zhang2024joint}, and beampattern matching error~\cite{hua2023optimal}, which directly characterize the radiation beampattern and are tightly connected to both communication and sensing performance. SNR-based metrics, such as the received SNR or signal-to-interference-plus-noise ratio (SINR) at communication users or sensing receivers, are also commonly adopted. These metrics are typically positively correlated with communication capacity and detection accuracy, while their mathematical expressions are simpler, making them easier to handle in analysis and optimization~\cite{he2022a}.

\subsection{Concept and Advantages of HISAC}
\label{ss_cah}

HISAC extends the principle of ISAC by leveraging RHS-enabled holographic beamforming for communication and sensing. Unlike traditional phased arrays that rely on expensive phase shifters and complex feeding networks, RHSs encode holographic patterns across densely packed sub-wavelength elements to synthesize different radiation patterns. The radiation amplitudes in the holographic pattern can be carefully designed for communication and sensing functions. Owing to the low-cost, low-power, and reconfigurable characteristics of RHSs, HISAC brings various benefits in terms of performance, efficiency, and flexibility. These advantages are illustrated in Fig.~\ref{f_advantage} and elaborated in detail in the following.

\begin{figure}[!t]
\centering
\includegraphics[width=3.3in]{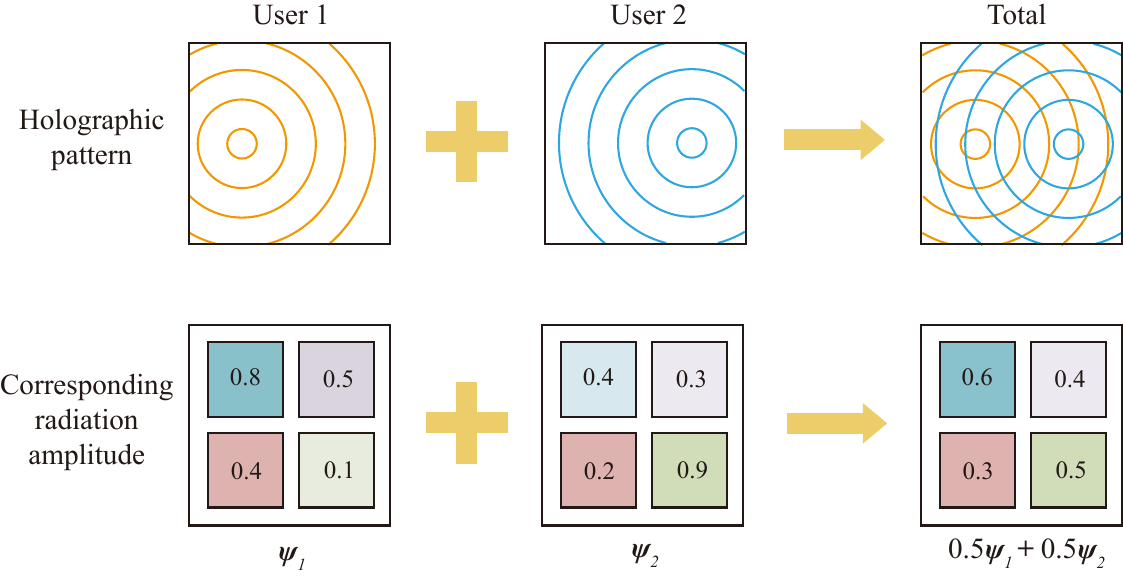}
\caption{Illustration of HDMA.}
\label{f_HDMA}
\end{figure}

\begin{itemize}
  \item \textbf{High Array Gain:} 
  As discussed in the previous section, the array gain of an RHS scales approximately linearly with the number of radiating elements. Moreover, since RHSs do not rely on complex phase shifters or parallel feeding networks, the aperture can be efficiently enlarged to realize ultra-massive MIMO. As a result, HISAC can achieve a high array gain, which enhances both the communication link budget and the sensing performance under the same transmit power constraint.

  \item \textbf{High Spatial Resolution:} The large aperture also leads to a narrower main lobe and finer angular resolution, improving the ability of HISAC to distinguish closely spaced users or targets. 
  Moreover, increasing the aperture size significantly extends the near-field region. Specifically, the Rayleigh distance that separates the near-field and far-field regimes is given by
  \begin{equation}
    R_{\mathrm{NF}} \approx \frac{2D^2}{\lambda},
  \end{equation}
  where $D$ denotes the effective aperture size and $\lambda$ is the wavelength. For link distances $d < R_{\mathrm{NF}}$, near-field effects become pronounced, whereas the conventional far-field approximation holds for $d > R_{\mathrm{NF}}$.
  As a result, the near-field region cannot be ignored for large-aperture RHSs. In this region, beamforming can be performed in the distance-angle domain, enabling 3D beam focusing that further improves distance resolution and target localization accuracy compared with far-field cases~\cite{cao2024unified, zhang2024hybrid}.

  \item \textbf{Fine-Grained Wavefront Control:} HISAC enables fine-grained wavefront control owing to the dense deployment of RHS elements with sub-wavelength spacing, as discussed in Sec.~\ref{sss_ra}. This property provides a larger number of reconfigurable radiation elements within a given physical aperture, thereby allowing more refined manipulation of the radiated electromagnetic field and more flexible beamforming designs. As a result, HISAC facilitates enhanced interference suppression and improved adaptability to complex propagation environments~\cite{iacovelli2026holographic}.
  
  \item \textbf{Low Cost:} 
  Thanks to the metamaterial-based architecture of RHSs, the element responses for holographic beamforming can be adjusted using low-cost PIN diodes or varactors embedded in a PCB structure. This amplitude-based control paradigm eliminates the need for expensive phase shifters and bulky parallel feeding networks commonly used in conventional phased arrays.
  As a result, RHSs are inherently compatible with mature PCB manufacturing processes, enabling scalable and cost-efficient fabrication and facilitating the mass deployment of HISAC transceivers across diverse scenarios.
  \item \textbf{Low Power Consumption:} The absence of active phase shifters also greatly reduces the system’s power requirements, since each element only adjusts its radiation amplitude through low-power electronic control. This property makes HISAC a practical solution for green wireless communication and sensing, aligning with the sustainability objectives of 6G.

  \item \textbf{Low Complexity:} 
  As for implementation complexity, HISAC supports rapid multi-beam generation for multi-user communication and multi-target sensing by employing holographic division multiple access (HDMA) techniques~\cite{deng2022hdma}. The basic idea is to first construct a set of holographic patterns according to the holographic principle, and then obtain the final beamforming matrix through a weighted combination of these patterns. 
  Specifically, based on the holographic principle, we can compute the holographic pattern corresponding to each user/target direction in closed form, which is given in~(\ref{e_m_n}). When superimposing these patterns across the antenna aperture\footnote{Direct superposition of holographic patterns may violate feasibility constraints such as per-element amplitude bounds and the leakage power constraint. To address this issue, a straightforward approach is to apply a re-scaling operation to the superposed pattern. Alternatively, the weights in the superposition can be optimized to satisfy all the constraints while maximizing the ISAC performances.},  multiple directional beams can be produced simultaneously, each aligned to a spatial location of a user or target. An illustrative example with $2$ users is provided in Fig.~\ref{f_HDMA}. 
  Since the weights of patterns rather than the individual element amplitudes need to be optimized, the complexity of beamforming optimization can be significantly reduced.
  Note that such a direct superposition in HDMA is feasible because holographic beamforming on RHSs is realized in an amplitude-controllable manner and does not impose a constant-modulus constraint on the element responses. In contrast, HDMA is not applicable for conventional phased arrays and metasurfaces such as RISs because their beamforming relies on phase shifting and has to follow the constant-modulus constraint~\cite{an2024stacked, shen2023multi}.
  
  \item \textbf{Ubiquitous Deployment:} The planar and feed-integrated architecture of RHSs results in a lightweight and low-profile structure that can be seamlessly embedded into diverse objects such as vehicle surfaces, building facades, and indoor walls. This compact structure facilitates pervasive and dense deployment of RHS panels across heterogeneous environments, thereby extending both communication and sensing coverage of the network. Moreover, the minimal size and weight of RHSs make them ideally suited for resource-constrained platforms such as UAVs~\cite{zeng2021trajectory}, aircraft, and satellites, where traditional large-scale phased arrays are impractical.
\end{itemize}

\begin{figure*}[!t]
\centering
\subfloat[]{\includegraphics[height=2.1in]{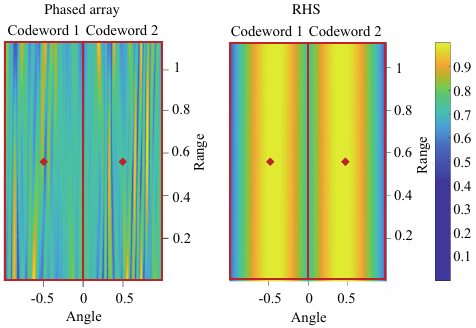}
}
\hfil
\subfloat[]{\includegraphics[height=2.1in]{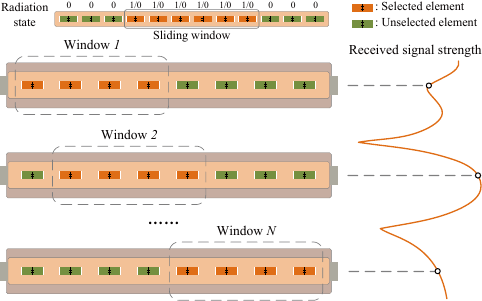}
}
\caption{Illustration of effective aperture with: (a) tunable coverage; (b) movable position.}
\label{f_flex}
\end{figure*}

\begin{itemize}
  \item \textbf{Tunable Coverage:} The flexibility of HISAC in coverage is enabled by the tunable aperture size. Specifically, the effective aperture size for holographic beamforming can be flexibly adjusted by selectively activating or deactivating subsets of RHS elements, leading to different beamwidths given by~\cite{balanis2016antenna}
  \begin{equation}
      B_N = \frac{\lambda}{N_a d_e},\label{e_B_N}
  \end{equation}
  where $N_a$ is the number of activated elements.
  According to~(\ref{e_B_N}), a larger aperture produces narrower beams with higher directivity, improving angular resolution and detection accuracy, whereas a smaller aperture generates wider beams that broaden communication coverage or accelerate initial access. This dynamic aperture reconfiguration enables HISAC to balance the trade-off between sensing resolution and communication coverage in real time. In contrast, PA achieves wide-coverage operation by applying multi-beam design methods or extra hardware components. For example, the schemes proposed in~\cite{cui2023near} design multi-beam codewords by maximizing beam gain at sampled angular points in the near field. However, the finite sampling resolution often causes gain fluctuation or collapse between the sampled points, as illustrated in Fig.~\ref{f_flex}(a). Alternatively, the authors in~\cite{wu2024two} introduce additional switch for each antenna to generate wider beams with flat gains, but they require additional cost and energy consumption. Different from PA-based schemes, the RHS can directly control the size of its equivalent aperture without any auxiliary hardware components. As a result, gain-flat and wide-coverage beams can be achieved through scale-changeable holographic arrays, as shown in Fig.~\ref{f_flex}(a).
  \item \textbf{Channel Adaptation:} 
  Beyond tunable coverage, HISAC can adapt to time-varying channel conditions by emulating antenna movement through reconfigurable aperture positioning~\cite{zhang2025movable}. This capability also stems from the amplitude modulation of RHS elements, which allows subsets of elements to be selectively activated while others are deactivated.
  As illustrated in Fig.~\ref{f_flex}(b), by activating only a contiguous subset of elements, multiple equivalent arrays with identical aperture sizes, which is referred to as sliding windows, can be dynamically formed along the RHS surface. The elements outside the window are deactivated by setting their radiation amplitude to zero, while those within the window remain adjustable. If the number of elements in each equivalent array is $N_a$, up to $(N - N_a + 1)$ possible sliding windows can be generated in holographic beamforming, corresponding to distinct effective antenna positions. Owing to small-scale fading, the channel characteristics vary across these positions, which in turn affects the received signal strength. By exploiting this additional flexibility, HISAC can adaptively select favorable channel regions, thereby maintaining high communication SNR and sensing accuracy even under varying propagation conditions.
\end{itemize}

Building upon the above advantages, HISAC can be flexibly deployed under different paradigms, depending on system requirements and application scenarios. In particular, JCAS is well suited for applications that require simultaneous communication and sensing, such as autonomous driving and UAV networks, where the high array gain and spatial resolution can jointly enhance link reliability and environmental perception. SAC is more appropriate when sensing information is leveraged to improve communication performance, for instance in dynamic environments where beam training, predictive beamforming, and resource allocation can benefit from the flexibility of HISAC. In contrast, CAS is particularly relevant for scenarios where communication signals and network functionalities are directly exploited for sensing. For instance, communication waveforms can be reused as probing signals for environmental sensing, while multi-node coordination enables enhanced coverage and robustness. In such scenarios, the low-cost, low-power, and ubiquitous deployment capability of HISAC further enables scalable and robust sensing performance.

\subsection{HISAC Optimization Framework}
\label{ss_hof}

Existing works have proposed various HISAC schemes to realize the aforementioned benefits. For example, in~\cite{zhang2022holographic}, the digital beamformer $\bm{B}$ at the BS and the holographic beamformer $\bm{M}$ at the RHS are jointly optimized to maximize a radar utility under communication SINR constraints. In~\cite{zhang2025holographic}, a sensing-assisted beam training scheme is designed, where the RHS adapts its holographic beamformer and effective aperture size to scan users in both the near and far fields. In Table~\ref{t_sum}, we compare these HISAC schemes in terms of scope, design variables, objective, and solution method, while the details of several representative schemes will be presented in the following sections.

\begin{table*}[!t]
\centering
\caption{Comparison of HISAC Schemes}
\renewcommand{\arraystretch}{1.3} 
\begin{tabular}{|L{2cm}|L{1.2cm}|m{2cm}|m{4cm}|m{4cm}|m{2cm}|}
\hline
\textbf{Type} & \textbf{Ref.} & \textbf{Scope} & \textbf{Design Variables} & \textbf{Objective} & \textbf{Solution}\\ \hline
\multirow[c]{8}{=}[-16ex]{
Joint \ \ \ com- munication and sensing
} &\cite{zhang2022holographic} & Transmit beamforming & Digital beamformer, holographic beamformer & Maximize the radar utility given communication SINR constraints  & AO, SDR  \\ \cline{2-6}
    & \cite{zhu2025integrated} & Transmit beamforming & Digital beamformer, holographic beamformer & Maximize the sensing mutual information given the user data rate constraint & Charnes-Cooper transform, AO  \\ \cline{2-6}
    & \cite{zeng2025holographic} & Transmit beamforming & Digital beamformer, holographic beamformer & Maximize the minimum data rate under the sensing constraints and mutual coupling model & Fractional program, SDR, Neumann series \\ \cline{2-6}
    & \cite{yu2025joint} & Joint transmit and receive beamforming & Digital beamformer, transmit and receive holographic beamformers, receive filters & Maximize the minimum target SINR given the communication SINR constraints & AO, ZF, SDR, Taylor expansion \\ \cline{2-6}
    & \cite{gao2025covert} & Joint transmit and receive beamforming & Digital beamformer, holographic beamformer, receive filters & Maximize the achievable rate given the sensing and covertness constraint & S-procedure, SCA, Bernstein-type inequality \\ \cline{2-6}
    & \cite{zhang2025rhs} & Integration with RIS & Digital beamformer, holographic beamformer, RIS beamformer & maximize the achievable rate given the minimum sensing SNR threshold & AO, PCCP, SCA \\ \cline{2-6}
    & \cite{nikmaleki2025leveraging} & Integration with RIS & Digital beamformer, holographic beamformer, RIS beamformer, receive filters & maximize the achievable secrecy rate given the sensing SNR constraint & AO, ZF, MRT, Reyleigh quotient \\ \cline{2-6}
    & \cite{zhao2025on, zhao2025performance} & Performance analysis & N/A & Analysis the HISAC performance under spatial correlated channel  & N/A \\ \hline 
\multirow[c]{4}{=}[-7.5ex]{
Sensing-assisted communication
}   & \cite{yue2023channel} & Sensing-assisted channel estimation & N/A & Estimate the CSI for both near and far fields & OMP \\ \cline{2-6}
    & \cite{zhang2024hierarchical, zhang2025holographic} & Sensing-assisted beam training & Holographic beamformer, effective aperture size & Design a codebook covering both near and far fields & Additivity of patterns, Taylor expansion \\ \cline{2-6}
    & \cite{zhang2025fluid, zhang2025movable} & Sensing-assisted beam training & Holographic beamformer, effective aperture size, effective aperture location & Design a codebook for RHS-enabled fluid/movable antenna system & Hierarchical codebook \\ \cline{2-6}
    & \cite{hu2023multi} & Sensing-assisted beamforming & Digital beamformer, holographic beamformer & Minimize the localization CRB used for communication beamforming  & Conjugate gradient, Monte Carlo, AO \\ \hline
\multirow[c]{4}{2cm}[-7.5ex]{
Communication-assisted sensing} & \cite{wei2025wideband} & Wideband OFDM signal design & Digital beamformer, holographic beamformer, data symbols & Minimize the difference between the optimal transmit signal and the signal generated by the RHS  & AO, ADMM \\ \cline{2-6}
    & \cite{wei2024ris} & Wideband OFDM signal design & Digital beamformer, holographic beamformer, RIS beamformer, receive filters & Maximize the worst-case radar SINR given communication SINR constraint & AO, Dinkelbach algorithm, ADMM \\ \cline{2-6}
    & \cite{hu2023holofed} & Multi-band localization & Digital beamformer, holographic beamformer, position estimator, adaptation functions, scheduling probability vectors & Minimize the average MSE of positioning & Stochastic gradient descent, transfer learning \\ \cline{2-6}
    & \cite{li2025reconfigurable} & Distributed sensing & Holographic beamformers of transmit and receive RHSs & Minimize the worst-case SINR over all targets & AO, Gaussian randomization \\\hline

\end{tabular}
\label{t_sum}
\end{table*}

Before delving into the mathematical formulations tailored to specific scenarios, this subsection introduces a general optimization framework that offers a high-level perspective on existing HISAC schemes. It summarizes the key design components of these schemes and offers a starting point for future HISAC designs. Note that this formulation only serves as a general optimization template for HISAC systems. Specific HISAC designs in the literature can be obtained by specifying the objective function, constraints, and design variables according to the considered architecture and application scenario.

Specifically, the general optimization problem is given by
\begin{subequations}
    \begin{align}
        \text{P1:}\max_{\bm{M}, \bm{B}, \bm{r}_1, \bm{r}_2, \bm{\mathcal{R}}} &f_{obj}(\bm{M}, \bm{B}, \bm{r}_1, \bm{r}_2, \bm{\mathcal{R}}),\label{p1_obj}\\
        s.t.~
        &tr(\bm{B}\bm{x}\bm{x}^H\bm{B}^H) = P_t,\label{p_g_tp}\\
        & \sum_{n_z} \psi^2_{n_y, n_z} d_{n_y, n_z} \eta_{n_y, n_z} \le 1,\forall n_y,\label{p_g_lp}\\
        & \psi_{n_y, n_z} \in [0, 1], \forall n_y, n_z,\label{p_g_psi}\\
        & f_i(\bm{M}, \bm{B}, \bm{r}_1, \bm{r}_2, \bm{\mathcal{R}}) \ge g_i, i \in [1, I],\label{p_g_f_i}
    \end{align}
\end{subequations}
where ${\bm{\bm{M}}, \bm{B}, \bm{r}_1, \bm{r}_2, \bm{\mathcal{R}}}$ are the design variables.

Here, $\bm{M}$ denotes the holographic beamformer of the RHS, which serves as a key design parameter for holographic beamforming and is mathematically defined in~(\ref{eq:Mmatrix}). As shown in~(\ref{eq:Mmatrix}), $\bm{M}$ is determined by the holographic pattern $\bm{\psi} = (\psi_1, \dots, \psi_N)$ and the propagation matrix $\bm{F}$.
Since the propagation matrix $\bm{F}$ is fixed once the RHS hardware and operating frequency are given, the holographic beamformer $\bm{M}$ is uniquely specified by the holographic pattern $\bm{\psi}$. Therefore, in the formulated optimization problem, optimizing over $\bm{M}$ is equivalent to optimizing over the element-wise holographic pattern $\bm{\psi}$.
In addition, $\bm{B}$ denotes the digital beamformer at the BS, which determines how data streams and sensing waveforms are precoded before being fed into the holographic aperture. Besides, variables $\bm{r}_1 = (r_{1, 1}, r_{1, 2}, \dots, r_{1, N_y})$ and $\bm{r}_2 = (r_{2, 1}, r_{2, 2}, \dots, r_{2, N_y})$ specify the start and end
element indices of the effective aperture in every row of RHS, respectively. These variables serve as structural design parameters that determine both the size and the position of the effective aperture. To be specific, for the $n_y$-th row, the effective aperture size is given by $r_{2,n_y} - r_{1,n_y} + 1$, while the corresponding aperture center (i.e., position) is given by $(r_{2,n_y} + r_{1,n_y})/2$. Finally, $\bm{\mathcal{R}}$ denotes other design variables such as power and bandwidth allocation. 

Among these variables, the holographic pattern $\bm{\psi}$ corresponds to the physically tunable amplitudes of the RHS elements, whereas the other variables, including $\bm{M}$ and $(\bm{r}_1,\bm{r}_2)$, are introduced as equivalent representations or structural abstractions to facilitate system modeling and optimization, rather than as independent physical control parameters.

Next, we interpret each equation in (P1):
\begin{itemize}
    \item \textbf{Objective function (\ref{p1_obj}):} It characterizes the HISAC performance and can be chosen from the communication and sensing metrics introduced in Sec. III-A according to the requirement of different scenarios. For example, consider a scenario where the main aim of the system is to accurately detect a target by maximizing the radiation power towards the target direction~\cite{stoica2007on}. The objective function can thus be given by
    \begin{equation}
        f_{obj} = |\bm{a}^T(\theta_s)\bm{D}\bm{M}\bm{B}\bm{x}|^2,
    \end{equation}
    where $\bm{a}^T(\theta_s)$ is the steering vector of the RHS towards target direction $\theta_s$. $\bm{x}$ is the signal fed into the RHS. $\bm{D} = \text{diag}(\bm{d})$ is the effective aperture matrix, and $\bm{d} = (d_1, \dots, d_N)$ is the effective aperture vector. Here, $d_n$ represents where the $n$-th element of RHS is activated and can be given by
    \begin{equation}
        d_n = \begin{cases}
        1, \quad n_z \in [r_{1, g(n)}, r_{2, g(n)}],\\
        0, \quad \text{otherwise,}
        \end{cases}
    \end{equation}
    where $g(n)$ is the row index of the $n$-th element.
    \item \textbf{BS Power Constraint~(\ref{p_g_tp}):} $P_t$ is the power budget of the BS. Thus, this constraint ensures that the incident power of all the feeds is $P_t$.
    \item \textbf{Leakage Power Constraint~(\ref{p_g_lp}):} This constraint guarantees that the energy input from each feed of the RHS gradually decreases after passing through each element. Note that the leakage power constraint has a strong impact on the holographic beamformer design for ISAC. First, the optimal amplitude response at each RHS element is coupled with the phase of the reference wave, and this coupling becomes more intricate when the size and position of the effective aperture vary. Even when the effective aperture is fixed, the leakage power constraint also involves quadratic terms in the amplitude variables, which makes the problem non-convex and calls for more advanced optimization methods~\cite{deng2022reconfigurable}.
    \item \textbf{Amplitude Range Constraint~(\ref{p_g_psi}):} This constraint specifies that the radiation amplitude of each RHS element is non-negative and that its maximum value does not exceed 1.
    \item \textbf{Other Constraints (\ref{p_g_f_i}):} Depending on the considered scenario, the optimization problem may have other constraints. For example, when maximizing the radiation power towards the target direction, a SNR constraint can be set for the communication user to ensure the downlink communication quality, which can be given by
    \begin{equation}
        \frac{|\bm{h}\bm{D}\bm{M}\bm{B}\bm{x}|^2}{\sigma^2} \ge \Gamma_c,
    \end{equation}
    where $\bm{h}$ is the communication channel, $\sigma^2$ is the noise power, and $\Gamma_c$ is the SNR threshold.
\end{itemize}

\begin{figure}[!t]
\centering
\includegraphics[width=3.3in]{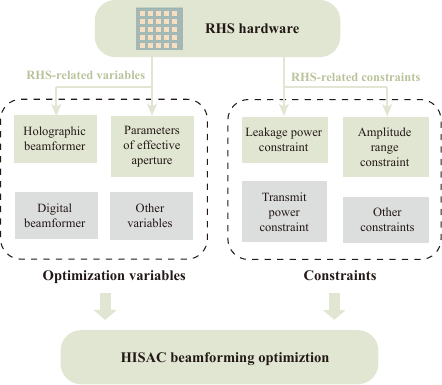}
\caption{Relationship among RHS hardware, variables, constraints, and HISAC beamforming optimization.}
\label{f_relation}
\end{figure}

The relationship among RHS hardware, optimization variables, constraints, and HISAC beamforming optimization is summarized in Fig.~\ref{f_relation}. For a general template of HISAC beamforming optimization, the holographic beamformer, leakage power constraint, and amplitude range constraint are necessary, while other variables or constraints can be included in specific HISAC designs.

To apply this general framework to a specific HISAC scenario, a typical procedure is as follows: (i) determine the design variables $\bm{M}$, $\bm{B}$, $\bm{r}_1$, $\bm{r}_2$, and $\bm{\mathcal{R}}$ according to the system configuration (e.g., the number of RHS radiation elements and RF chains); (ii) specify the incident-power constraint~(\ref{p_g_tp}) and leakage power constraint~(\ref{p_g_lp}) based on the power budget $P_t$; and (iii) select an appropriate objective function and additional constraints according to the design requirements. 

Compared with RIS or phased array-based beamforming with the same aperture, the optimization complexity of HISAC is larger because the RHS contains more tunable elements, resulting in a larger number of variables in holographic beamformer $\bm{M}$. When the RHS aperture becomes extremely large or the available computational resources are limited, the element-wise optimization may become computationally prohibitive. In such cases, as discussed in Sec.~\ref{ss_cah}, HDMA can be adopted to significantly reduce the optimization complexity, which is mainly related to the number of users, targets, and feeds, instead of the total number of RHS elements. More details are provided in Sec.~\ref{ss_tb}.

\begin{figure*}[!t]
\centering
\subfloat[]{\includegraphics[height=2in]{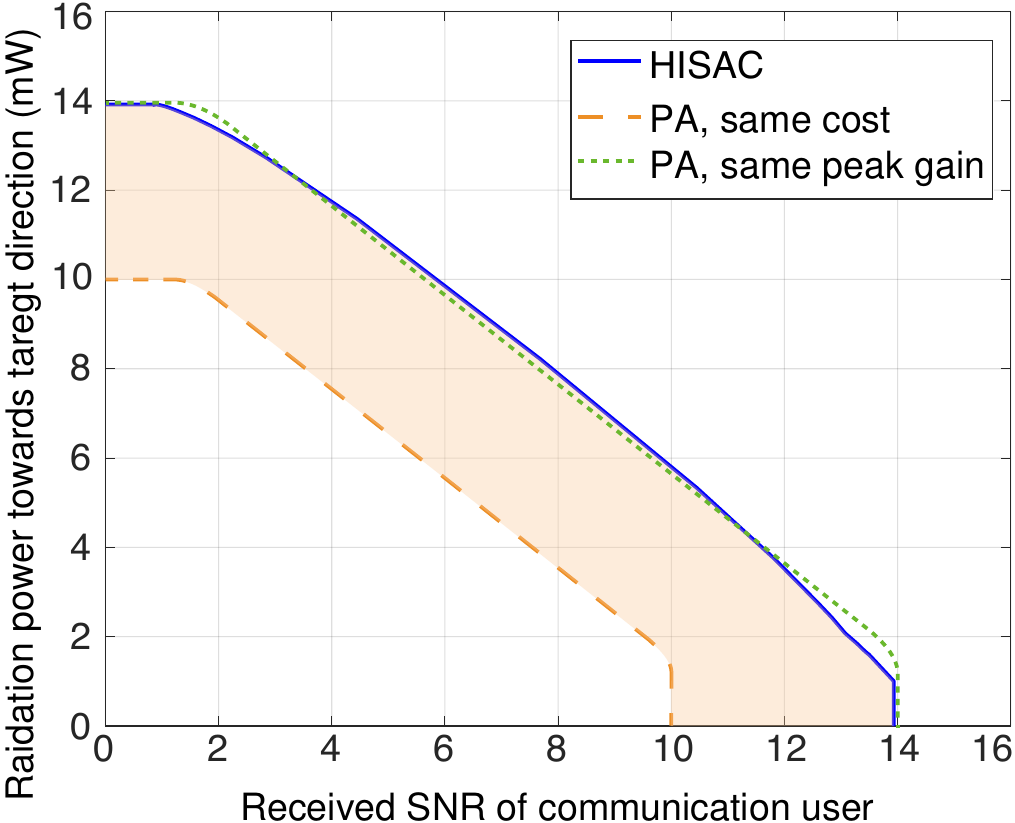}
}
\hfil
\subfloat[]{\includegraphics[height=2in]{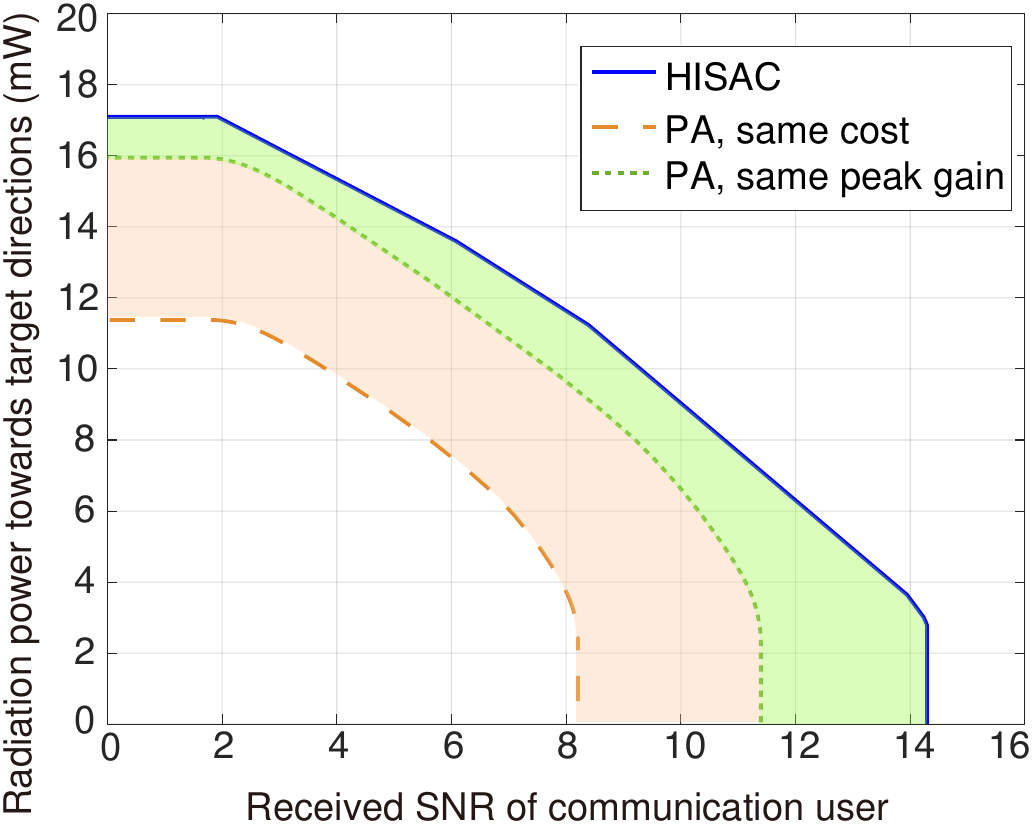}
}
\caption{Communication and sensing performance tradeoff with: (a) single target and LoS communication link; (b) two targets and a two-path communication channel.}
\label{f_compare_single_double}
\end{figure*}

In the following, we present two simple cases to illustrate how this framework can be applied. These two cases can also demonstrate how the high antenna gain and fine-grained wavefront control of HISAC described in Sec.~\ref{ss_cah} benefit both communication and sensing. 

\subsubsection{One-User and One-Target Case} We consider a case with only one communication user and one sensing target. The user direction $\theta_c$ and target direction $\theta_s$ are independently and uniformly distributed over the interval $[-60^\circ, 60^\circ]$. For the communication user, a deterministic LoS channel is employed, and we use the SNR of user as the communication performance, which is positively related to the data rate. The ISAC signals are emitted by an $1$D RHS with one feed, and the whole aperture is employed for beamforming. Thus, only the holographic beamformer $\bm{m} = \text{diag}(\bm{\psi})\bm{f}$ needs to be optimized. 

As for the objective function and additional constraint, we choose to maximize the sensing performance and set a SNR threshold for the communication user. The corresponding optimization problem can thus be written as

\begin{subequations}
    \begin{align}
        \max_{\bm{m}}~& P_s,\\
        s.t.~& |x|^2 = P_t,\label{pc_x}\\
        &\sum_n \psi^2_n \eta_n \le 1,\label{pc_m}\\
        & \psi_n \in [0, 1],\label{pc_01}\\
        & \gamma_c \ge \Gamma_c,
    \end{align}
\end{subequations}
where $P_s = |\bm{a}^T(\theta_s)\bm{m}x|^2$ is the radiation power towards target direction $\theta_s$, and $\eta_n$ is the ratio of the power accepted by the $n$-th element to the incident power.

(\ref{pc_x}) and (\ref{pc_m}) are the constraints of incident power and leakage power, respectively. 
$\gamma_c$ is the received SNR of the user and can be given by
\begin{equation}
    \gamma_c = \dfrac{G |\bm{a}^T(\theta_c)\bm{m}|^2 P_t}{\sigma^2},
\end{equation}
where $G$ is the gain of the LoS channel. Since the formulated problem is a quadratic program, it can be efficiently solved by using existing techniques such as semidefinite relaxation (SDR)~\cite{luo2010semidefinite}.

The Pareto front can be obtained by solving the formulated problem under different communication SNR thresholds $\Gamma_c$. The leakage power constraint in (\ref{pc_m}), together with the amplitude constraint in (\ref{pc_01}), restricts the feasible set of the holographic pattern $\bm{\psi}$, and hence limits the set of realizable holographic beamformers $\bm{m}$. As a result, the achievable beamforming gain under a given input power is bounded. It in turn determines the attainable upper limits of both communication and sensing performance, which is reflected in the resulting Pareto front.

\begin{table}[!t]
\caption{Main Simulation Parameters}
\centering
\renewcommand{\arraystretch}{2}
\begin{tabular}{|c|c|}
\hline
\textbf{Parameters} & \textbf{Values} \\
\hline
Number of RHS elements & $50$\\
\hline
Inter-distance between adjacent RHS elements & $\lambda/3$\\
\hline
Inter-distance between adjacent PA antennas & $\lambda/2$\\
\hline
Number of RF chains & $1$\\
\hline
Maximum transmit power of the RHS & $1$W\\
\hline
\end{tabular}
\label{t_para}
\end{table}

In Fig.~\ref{f_compare_single_double}(a), the Pareto front achieved by the HISAC is illustrated. The main simulation parameters are summarized in Table~\ref{t_para}.
For comparison, we also provide the performance of two PA-based schemes. In the first scheme, the PA has the same cost as the RHS in HISAC scheme. In the second scheme, the aperture size of the PA is enlarged, so that the PA and the RHS have the same peak gain of antenna array, i.e., the maximum gain of the mainlobe. It can be observed that the Pareto front of HISAC lies at the upper-right region of that of the PA-based system with the same cost, and coincides with that of the PA-based system with the same gain. This validates that the performance of HISAC is superior compared with the former scheme, and the performance gain mainly comes from the higher array gain. 

\subsubsection{One-User and Two-Target Case}
Next, we consider a slightly more complex case to gain deeper insights about the fine-grained wavefront control of HISAC. Specifically, suppose there are two sensing targets located at different directions, and the communication link is modeled as a two-path channel comprising a LoS path and a single multipath component. The sum of radiation power towards target directions can be given by
\begin{equation}
    P_s = |\bm{a}^T(\theta_{s, 1})\bm{m}|^2 P_t + |\bm{a}^T(\theta_{s, 2})\bm{m}|^2 P_t,\label{c_p_s}
\end{equation}
where $\theta_{s, 1}$ and $\theta_{s, 2}$ are directions of these targets. 

Besides, the received SNR of the user can be given by
\begin{equation}
    \gamma_c = \dfrac{\left|\left(\sqrt{G_1}\bm{a}^T(\theta_{c, 1})+\sqrt{G_2}\bm{a}^T(\theta_{c, 2})\right)\bm{m}\right|^2 P_t}{\sigma^2},\label{c_gamma_c}
\end{equation}
where $\theta_{c, 1}$ and $\theta_{c, 2}$ are directions of user and scatter, respectively. $G_1$ and $G_2$ are gains of the LoS and scattering paths, respectively. Since both~(\ref{c_p_s}) and (\ref{c_gamma_c}) are quadratic functions, the formulated problem is still a quadratic program.

Fig.~\ref{f_compare_single_double}(b) illustrates the Pareto fronts for this scenario. Different from Fig.~\ref{f_compare_single_double}(a), the Pareto front achieved by HISAC lies in the upper-right region relative to that of the PA-based system with the same peak gain, and its attainable region is notably larger. This indicates that fine-grained wavefront control can also provide improvement for ISAC performance. Although both systems have the same peak gain, when multiple targets are present and the propagation environment becomes more complex, holographic beamforming with denser radiating elements offers finer control over multi-beam synthesis, thereby enabling superior joint communication-sensing performance.

\section{HISAC for Joint Communication and Sensing}
\label{s_hjcas}

As a fundamental form of ISAC, joint communication and sensing (JCAS) employs a common waveform and shared hardware to simultaneously support data transmission and environmental perception.
When combined with holographic beamforming, JCAS can exploit the large aperture and numerous tunable elements of RHSs to generate highly directional beams for both communication and sensing tasks. In this section, we focus on HISAC for JCAS. We begin by reviewing the transmit beamforming architecture for HISAC, where the BS and the transmit RHS jointly enable dual-function operation, while the BS receiver follows a conventional echo processing procedure. We then introduce the transceiver co-design approach in the literature and highlight how joint optimization of the transmit and receive RHSs can further improve communication and sensing performance. The section subsequently reviews other related works on HISAC-enabled JCAS, and concludes with a discussion of the key challenges and opportunities.

\subsection{Transmit Beamforming}
\label{ss_tb}

\subsubsection{System Model}

\begin{figure*}[!t]
\centering
\subfloat[]{\includegraphics[height=2in]{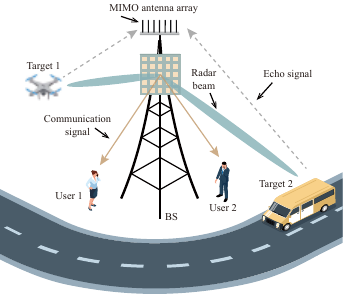}
}
\hspace{2mm}
\subfloat[]{
  \raisebox{6mm}{
    \includegraphics[height=1.6in]{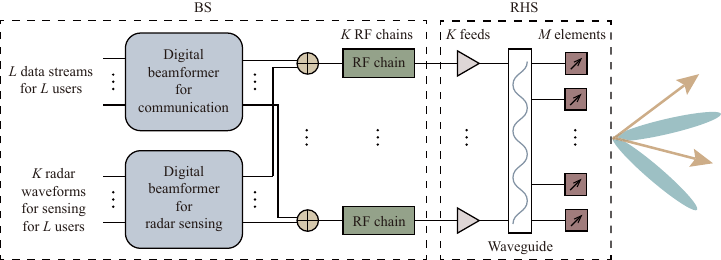}
  }
}
\caption{HISAC-enabled joint communication and sensing: (a) Scenario; (b) Beamforming architecture.}
\label{f_jcas}
\end{figure*}

As shown in Fig.~\ref{f_jcas}(a), the considered system consists of a BS, an $L$-feed RHS serving as a Tx antenna, a MIMO antenna array as a Rx antenna, and $U$ communication users and sensing targets distributed within the coverage area.
To realize downlink data transmission and environmental sensing,
three main procedures are executed sequentially: optimization, transmission, and reception.
In the optimization stage, the transmit ISAC signals generated by the BS and the Tx RHS are designed to balance communication throughput and sensing precision.
Next, in the transmission stage, the optimized ISAC signals are radiated by the RHS to serve users and illuminate targets.
Finally, in the reception stage, the communication users receive their respective data streams, while echoes reflected from the targets are captured by the MIMO antenna array and processed at the BS to determine the presence of targets.

Different from the communication or sensing-only models, the beamforming architecture shown in Fig.~\ref{f_jcas}(b) is adopted to generate the desired ISAC signal.
Specifically, the BS first produces $N_{data}$ data streams and $N_{rad} = L$ radar waveforms, which are processed by digital beamformers $\bm{B}_c$ and $\bm{B}_s$, respectively.
After digital precoding, the generated signal is expressed as
\begin{equation}
    \bm{x} = \bm{B}_c\bm{c} + \bm{B}_s\bm{s},
\end{equation}
where $\bm{c}$ and $\bm{s}$ denote the communication and radar symbols, respectively.
The signal is then up-converted and fed into the RHS through $L$ RF feeds, and the signal transmitted by the RHS can be given by
\begin{equation}
    \bm{z} = \bm{M}(\bm{B}_c\bm{c} + \bm{B}_s\bm{s}).
\end{equation}

For the $u$-th user, the SINR can be given by
\begin{equation}
    \gamma_u =
\frac{|\bm{h}^T_u\bm{M}\bm{b}_{c,u}|^2}
{|\bm{h}^T_u\bm{M}\sum_{u'\ne u}\bm{b}_{c, u'}|^2 + \mathbf{h}^T_u\bm{M}\bm{B}_s\bm{B}^H_s\bm{M}^H\bm{h}^*_u + \sigma^2},
\end{equation}
where $\bm{h}_u$ is the channel from the RHS to the $u$-th single-antenna user, and $\bm{b}_{c,u}$ is the $u$-th column of matrix $\bm{B}_c$. Consequently, the achievable data rate of the $u$-th user is $R_u=\log_2(1+\gamma_u)$.

For sensing performance, the transmit beampattern of the RHS can be optimized to enhance the probability of detection and the accuracy of parameter estimation in the multi-target scenario. According to~\cite{stoica2007on}, maximizing the beampattern gain toward the target directions strengthens the reflected echo signals and thus raises their SINR, resulting in better sensing performance.
Meanwhile, suppressing the inter-beam correlation allows more effective separation of echoes from distinct angles, which also improves the sensing accuracy. Thus, to simultaneously maximize the beamforming gain and minimize the inter-beam correlation, we define the radar utility function as the difference between these two terms, which is given by
\begin{equation}
    f_u = P_a - \alpha_0 RMSC,
\end{equation}
where $P_a$ is the average radiation power towards the target directions and can be given by
\begin{equation}
    P_a = \frac{1}{J} \sum^J_{j=1} P(\theta_j, \phi_j).
\end{equation}
Here, $J$ is the number of target directions, and $P(\theta_j, \phi_j) = \bm{a}^T(\theta_j, \phi_j)\bm{M}\bm{B}\bm{B}^H\bm{M}^H\bm{a}^*(\theta_j, \phi_j)$ is the radiation power towards direction $(\theta_d, \phi_d)$, with the digital beamformer $\bm{B}$ being $(\bm{B}_c, \bm{B}_s)$, $\gamma^l_d$. $RMSC$ is the root mean square cross-correlation among the target directions and can be written as
\begin{equation}
    RMSC = \sqrt{\frac{2}{D(D-1)}\sum^{D-1}_{d=1}\sum^D_{d'=d+1}|P^c(\theta_d,\phi_d,\theta_{d'},\phi_{d'})|^2},
\end{equation}
where $P^c(\theta_j,\phi_j,\theta_{j'},\phi_{j'})$ is the cross-correlation between directions $(\theta_d,\phi_d)$ and $(\theta_{d'},\phi_{d'})$ and is equal to $\bm{a}^T(\theta_j, \phi_j)\bm{M}\bm{B} \bm{B}^H \bm{M}^H \bm{a}^*(\theta_{j'}, \phi_{j'})$. $\alpha_0$ is the weighting factor for $RMSC$.

\subsubsection{Transmit Beamforming Optimization}

Given the above beamforming architecture and performance metrics, the optimization problem can be formulated, aiming to maximize the radar utility $f_u$ given the SINR thresholds for communication users. To enable flexible control of energy allocation across multiple sensing targets, the following constraint is also incorporated:
\begin{equation}
    \gamma^l_j P(\theta_1, \phi_1) \le P(\theta_j, \phi_j) \le \gamma^u_j P(\theta_1, \phi_1), j = 2, \dots, J,
\end{equation}
where $\gamma^u_d$ are two positive coefficients that determine the range of beampattern gains in different directions. As a result, the radiation power toward direction $(\theta_d', \phi_d')$ is restricted to be within the range $[\gamma^{l}_d P(\theta_1, \phi_1), \gamma^u_d P(\theta_1, \phi_1)]$.

To solve the formulated optimization problem, an alternating optimization~(AO)-based holographic beamforming algorithm is proposed, where the digital and holographic beamformers are iteratively optimized. Both the optimization problems for digital and holographic beamformers are quadratic program, and thus can be tackled using SDR or other quadratic programming methods. 

In practice, due to the limitation of active devices such as PIN diodes, the radiation amplitude of an RHS element can only be selected from a discrete set, i.e., $\psi_n \in \{0, \frac{1}{C^s-1}, \dots, 1\}$. To tackle the discrete constraint, we can first relax the discrete constraint to a continuous constraint, and obtain the solution of the continuous optimization problem. Then, the solution can be quantized to the nearest values in $\{0, \frac{1}{C^s-1}, \dots, 1\}$ to satisfy the discrete constraint. 

In addition, due to the large number of radiation elements in RHSs, the complexity of holographic beamformer optimization can be relatively high. As discussed in Sec.~\ref{ss_cah}, the HDMA can be employed to reduce the complexity. According to~\cite{zhang2023energy}, $S = (U + J) \times L$ holographic patterns can be generated to serve $U$ users and detect $J$ targets using an RHS with $L$ feeds, and the final holographic pattern can be obtained as
\begin{equation}
    \bm{\psi} = \sum_{s=1}^{S} \zeta_s \bm{\psi}_s,
\end{equation}
where $\zeta_s$ denotes the weight of the $s$-th holographic pattern. Substituting the resulting $\bm{\psi}$ into the beamforming design framework yields an efficient approach to optimize the beamformers in HISAC. Since the dimension of the weighting factor $\bm{\zeta} = (\zeta_1, \dots, \zeta_S)$ is only related to $U$, $J$, and $L$, the HDMA-based method is particularly suitable for the optimization of large-scale RHS.

\subsubsection{Simulation Results}

\begin{figure}[!t]
\centering
\includegraphics[height=2in]{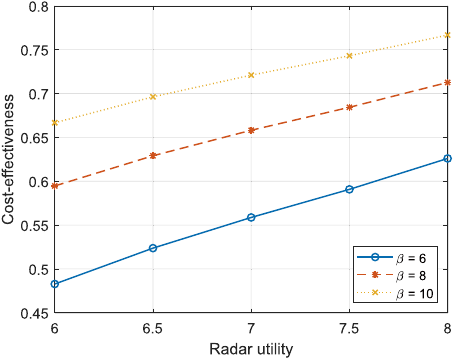}
\caption{Cost-effectiveness of the proposed scheme over the PA-based scheme versus the radar utility $\delta$ for different values of cost ratio $\beta$.}
\label{f_cost_eff}
\end{figure}

The cost-effectiveness of the proposed HISAC-based transmit beamforming scheme is evaluated in comparison with a conventional PA-based scheme. According to~\cite{lu2016wifi}, the cost-effectiveness metric is defined as
\begin{equation}
    \iota(\delta)=1-\dfrac{\alpha_r(\delta)}{\alpha_a(\delta)},
\end{equation}
where $\alpha_r(\delta)$ and $\alpha_a(\delta)$ denote the hardware costs required by the RHS and PA-based systems to reach radar utility $\delta$, respectively.
This metric quantifies the relative hardware cost reduction achieved by using an RHS instead of a PA-based MIMO system. A positive value of $\iota(\delta)$ indicates that the RHS requires lower hardware cost, while a larger $\iota(\delta)$ represents greater cost savings. In Fig.~\ref{f_cost_eff} (see~\cite{zhang2022holographic} for details), we consider a hardware setting with 3-bit amplitude quantization. We can observe that the cost-effectiveness remains positive when cost ratio $\beta \ge 6$, where $\beta$ denotes the ratio of the cost of one antenna in the PA to the cost of an RHS element. Due to the simple structure of an RHS element, the cost ratio can be much greater than 6~\cite{pivotal2019holographic}, thereby confirming that the RHS is more economical than the PA counterpart. The results also indicate that the proposed design remains effective under discrete amplitude constraints, demonstrating its robustness to practical quantization effects. Furthermore, $\iota(\delta)$ increases with the radar utility, implying that the cost advantage of holographic beamforming becomes more significant when stricter sensing performance is required.

\subsection{Transceiver Co-Design}

The above beamforming framework can be generalized to a full transceiver architecture by introducing a receive-side RHS. The coordinated optimization of both transmit and receive radiation amplitudes allows the system to exploit the spatial degrees of freedom on both sides of the RHSs, thereby improving the performance of both communication and sensing.

\subsubsection{System Model}

In the considered scenario, the BS employs two RHSs, one for transmission and one for echo reception. The transmit RHS radiates ISAC signals toward users and targets. At the same time, the receive RHS captures the reflected echoes from the environment.
The echo signal collected by the elements of the receive RHS can be expressed as
\begin{equation}
    \bm{w} = \sum_{j=1}^{J} \beta_j \bm{a}_r(\theta_j,\phi_j) \bm{a}^T_t(\theta_j,\phi_j) \bm{M}^t \bm{x} + \bm{n}_{ext},
\end{equation}
where $\beta_j$ denotes the reflection coefficient of the $j$-th target, and $\bm{a}_t(\theta_j,\phi_j)$ and $\bm{a}_r(\theta_j,\phi_j)$ are the transmit and receive steering vectors, respectively. $\bm{x}$ is the signals sent by the feeds of transmit RHS. $\bm{n}_{ext}$ is the external noise, which comes
from the radio environment.
After analog combining and filtering, the signal extracted by the BS can be written as
\begin{equation}
    \bm{\epsilon}_j = \bm{w}^H_j \sum^J_{j=1} \beta_j \bm{S}_j \bm{x} + \bm{w}^H_j (\bm{\Psi}^r \bm{Q}_r)^T \bm{n}_{ext} + \bm{w}^H_j \bm{n}_{int},
\end{equation}
where $\bm{w}_j$ is the filter corresponding to the $j$-th target, and
$\bm{S}_j = (\bm{M}^r)^T \bm{a}_r(\theta_j,\phi_j) \bm{a}^T_t(\theta_j,\phi_j) \bm{M}^t$
denotes the end-to-end equivalent channel. $\bm{n}_{int}$ is the internal noise coming from
the noisy electronic components in RF chains.

For the transmit RHS, we consider a generalized model where only a portion of the input power is radiated due to the propagation and radiation losses of the reference wave in the RHS. We define the antenna efficiency of transmit RHS $\upsilon_t$ as the ratio between the radiated power and the input power, which satisfies $\upsilon_t < 1$. Similarly, for the receive RHS, only a portion of the incident power collected by the aperture can be delivered to the feed, which is bounded by the maximum antenna efficiency $\upsilon_r$.

\subsubsection{Transceiver Optimization}

The joint optimization problem aims to maximize the minimum sensing SINR while satisfying communication rate and leakage power constraints. Let $\nu_j$ denote the sensing SINR of the $j$-th target, which is given by
\begin{equation}
    \nu_j = \dfrac{|\beta_j \bm{w}^H_j \bm{S}_j \bm{B}|^2}{\left|\bm{w}^H_j\sum^J_{i=1,i\ne j}\beta_i\bm{S}_i\bm{B}\right|^2
+\sigma^2_{ext}|\bm{u}_j|^{2}
+N\sigma^2_{int}}.
\end{equation}
The design variables include the holographic patterns of the transmit RHS $\bm{\psi}_t$ and receive RHS $\bm{\psi}_r$, the digital beamformer $\bm{B}$, and the receive filters $\bm{W} = (\bm{w}_1, \dots, \bm{w}_J)$. This optimization problem is a fractional programming problem with a max-min objective, which can be transformed into an equivalent maximization problem by introducing auxiliary variables. Specifically, let $\tau \in \mathbb{R}$ denote the minimum SINR among all targets, and define $\bm{\Lambda} = (\bm{\lambda}_1,\ldots,\bm{\lambda}_J) \in \mathbb{C}^{(N+U)\times J}$ as a set of complex auxiliary variables according to~\cite{shen2018fractional}. By introducing $\tau$, the max-min structure is converted into a single maximization problem over $\tau$, and the fractional SINR terms are decoupled through $\bm{\Lambda}$, which is given by
\begin{subequations}
    \begin{align}
        \max_{\bm{\psi}^t,\bm{\psi}^r,\bm{B},\bm{W},\bm{\lambda},\tau} &\tau \\
        s.t.~&2\mathrm{Re}\left(\bm{\lambda}^H_j\bm{A}^H_j\right)-|\bm{\lambda}_j|^2 G_j \ge \tau,\ \forall j,\label{c_frac}
    \end{align}
\end{subequations}
where
$\bm{A}_j=\beta_j\bm{w}^H_j\bm{S}_j\bm{B}$ and 
$G_j=\left|\bm{w}^H_j\sum^J_{i=1,i\ne j}\beta_i\bm{S}_i\bm{B}\right|^2
+\sigma^2_{ext}|\bm{u}_j|^{2}
+N\sigma^2_{int}$.
Here, $\bm{A}_j$ and $\bm{G}_j$ correspond to the numerator and denominator of the sensing SINR $\gamma_j$, respectively. The constraint (\ref{c_frac}) is obtained via the quadratic transform technique, which linearizes the fractional SINR terms into a convex form. Notably, when the auxiliary variable is updated as
$\bm{\lambda}^*_j = \bm{A}_j/G_j$,
the left-hand side of (\ref{c_frac}) equals $\gamma_j$, ensuring that the transformed linear constraint is equivalent to the original SINR requirement. This reformulation simplifies the objective function and enables an efficient alternating-update procedure for the transceiver design.

\subsubsection{Theoretical Analysis and Simulation Results}

Theoretical results provide insights into how the leakage power constraint affects the overall ISAC performance. For communication, the achievable SNR of a single user scales linearly with the transmit-side efficiency, which can be given by
\begin{equation}
    \zeta_1 = \frac{\upsilon_{t,1}}{\upsilon_{t,2}} \zeta_2.
\end{equation}
Here, $\zeta_1$ and $\zeta_2$ denote the maximum SNR of the communication user given transmit RHS efficiency $\upsilon_{t, 1}$ and $\upsilon_{t, 2}$, respectively.
This indicates that the downlink communication quality is mainly governed by the efficiency of the transmit RHS. The reason is that the considered scenario corresponds to downlink transmission, in which the receive RHS at the BS does not affect the data transmission to users. For sensing, consider a single-target case, and the relationship between the maximum target SNR and the antenna efficiencies of the two RHSs satisfies
\begin{equation}
    \nu(\upsilon_0,\upsilon_0) > \nu(\upsilon_0,\upsilon_0 - \Delta\upsilon) > \nu(\upsilon_0 - \Delta\upsilon,\upsilon_0),
\end{equation}
where $1 > \upsilon_0 > \Delta\upsilon > 0$, and $\nu(\upsilon_t,\upsilon_r)$ denote the maximum SNR of the sensing target when the maximum antenna efficiencies of the transmit and receive RHSs at the BS are $\upsilon_t$ and $\upsilon_r$, respectively. It shows that both the transmit and receive efficiencies contribute positively to sensing performance, while improvements in the transmit efficiency $\upsilon_t$ yield a larger gain because it directly determines the radiated and reflected signal power, whereas increasing $\upsilon_r$ proportionally amplifies both the signal and noise components.

\begin{figure}[!t]
\centering
\includegraphics[height=2in]{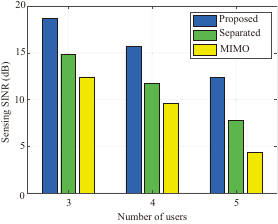}
\caption{Minimum sensing SINR versus the number of users}
\label{f_co-design-comp}
\end{figure}

Fig.~\ref{f_co-design-comp} illustrates the minimum sensing SINR among the sensing targets versus the number of communication users for the co-design of HISAC transceiver (see~\cite{yu2025joint} for details). Besides the PA-based scheme, the performance of the separated scheme is also provided, in which the transmit and receive beamformers are optimized independently. It can be observed that the minimum SINR of the targets decreases as the number of users increases, owing to the stronger interference from communication signals. Under the same hardware cost, the proposed HISAC scheme achieves at least $6.5$dB higher minimum SINR than the PA-based scheme. Furthermore, the SINR achieved by the proposed scheme exceeds that of the separated optimization scheme by at least $3.55$dB, demonstrating the effectiveness of jointly optimizing the transmit and receive RHSs.

\subsection{Other Related Works}

In the literature, extensive research has been conducted on HISAC-enabled JCAS~\cite{zhu2025integrated, nikmaleki2025leveraging, tharaniya2025beamforming, gao2025covert, shao2025robust, zhang2025rhs, adhikary2024age, adhikary2024integrated, xie2025joint}. In particular, this subsection reviews representative studies focusing on two key aspects: spatial correlation~\cite{zhao2025performance, zhao2025on} and mutual coupling~\cite{zeng2025holographic}.

\subsubsection{Spatial Correlation}

In conventional MIMO systems, antennas are usually spaced at approximately half a wavelength, allowing the use of the simplistic independent and identically distributed~(i.i.d.) communication channel model to evaluate the theoretical capacity.
In contrast, the densely packed sub-wavelength antenna elements in HISAC systems lead to significant spatial correlation among channels, rendering the i.i.d. channel model physically inconsistent and thus inapplicable. To accurately characterize the performance bounds of HISAC, the authors in~\cite{zhao2025performance} introduce an approximated Fourier plane-wave series expansion~(FPWSE)-based channel model. The key idea of FPWSE is that the channel response can be represented as the superposition of a series of plane waves. Consider an antenna array with $N = N_x N_y$ antennas and size of $L_x L_y$. The channel response between a transmit point $\bm{s}$ and a receive point $\bm{r}$ is expressed as
\begin{align}
    h&(\bm r,\bm s)=\frac{1}{(2\pi)^2}\!\!\iint\!\!\iint_{\mathcal{D}(\kappa)\times\mathcal{D}(\kappa)}
a_R(k_x, k_y,\bm r)\,\notag\\
&\times H_a(k_x, k_y, \kappa_x, \kappa_y)\,
a_S(\kappa_x, \kappa_y, \bm s)\,
dk_x dk_y d\kappa_x d\kappa_y,
\end{align}
where $\bm \kappa = (\kappa_x, \kappa_y, \kappa_z)^T$ is the wavenumber vector for the transmit array with $\kappa^2_x + \kappa^2_y + \kappa^2_z = k^2_0$, and $\bm k = (k_x, k_y, k_z)^T$ is the wavenumber vector of the receive array with $k^2_x + k^2_y + k^2_z = k^2_0$. $a_S(\kappa_x,\kappa_y,\bm s)$ and $a_R(k_x, k_y, \bm r)$ characterize the aperture response of the transmit and receive arrays at $\bm s$ and $\bm r$, respectively, and $H_a(k_x, k_y, \kappa_x, \kappa_y)$ represents the channel coupling between every pair of transmit direction $\bm \kappa/||\bm \kappa||$ and receive direction $\bm k/||\bm k||$. The integration region is $\mathcal{D}(\kappa) = {(k_x, k_y) \in \mathbb{R}^2: k^2_x + k^2_y \le k^2_0}$. 

When the aperture is discretized at or below half-wavelength spacing, the integral reduces to a discrete Fourier series, and closed-form expressions for several key metrics including communication rate, sensing rate, and outage probability are then derived in~\cite{zhao2025performance}. It is also demonstrated that when spatial correlation is considered, HISAC can achieve higher communication/sensing rate and lower outage probability than conventional MIMO-based ISAC in both downlink and uplink scenarios.

\subsubsection{Mutual Coupling}

Besides the spatial correlation, the deployment of ultra-dense radiation elements may also introduce mutual coupling among them. The former describes the statistical dependence among channel responses, while the latter represents the phenomenon that the current excited on one antenna element perturbs the radiation characteristics of its neighbors~\cite{zeng2025holographic}. Specifically, due to the small spacing between the adjacent sub-wavelength elements in RHSs, the field emitted by one element induces additional currents on nearby elements, which in turn re-radiate fields that feed back to the original element. This effect can alter the effective excitation experienced by every element and consequently modify the overall radiation pattern of the array.

\begin{figure*}[!t]
\centering
\subfloat[]{\includegraphics[height=2in]{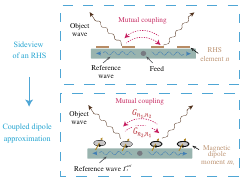}
}
\hspace{16mm}
\subfloat[]{
    \includegraphics[height=2in]{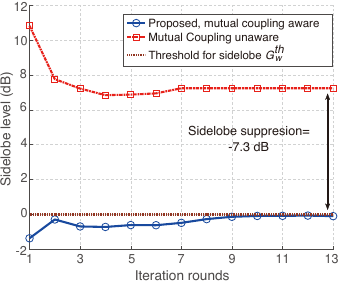}
  }
\caption{Mutual coupling-aware holographic beamforming: (a) Illustration of coupled dipole approximations for RHSs; (b) Sidelobe levels versus iteration rounds.}
\label{f_couple}
\end{figure*}

As shown in Fig.~\ref{f_couple}(a), each radiating element of the RHS can be represented by an equivalent magnetic dipole to model the mutual coupling effect, and the moment of the dipole governs the emitted field of this element. The ensemble of dipoles forms a coupled system governed by
\begin{equation}
    \big((e^{j\tau}\bm{\Theta})^{-1}-G\big)\,m = \bm{f}_l^{ref},
\end{equation}
where $e^{j\tau}\bm{\Theta}=\operatorname{diag}\{\theta_1 e^{j\tau},\dots,\theta_N e^{j\tau}\}$ denotes the polarizability matrix with $\theta_n e^{j\tau}$ being the polarizability of the $n$-th equivalent dipole. $G\in\mathbb{C}^{N\times N}$ records the electromagnetic coupling coefficients between neighboring elements obtained from the Green’s functions. $\bm{m}$ represents the magnetic dipole moments, and $\bm{f}_l^{ref}$ represents the reference wave distribution generated by the $l$-th feed of the RHS. 

After mapping this formulation into the holographic beamforming framework, the beamforming matrix becomes
\begin{equation}
    \bm{B} = k\left((e^{j\tau}\bm{\Theta})^{-1}-G\right)^{-1}\tilde{\bm{F}}^{ref},
\end{equation}
where $\tilde{\bm{F}}^{ref} = [\bm{f}_1^{ref}, \dots, \bm{f}_L^{ref}]$ aggregates the reference-wave distributions of all feeds, and $k$ is a complex scaling constant relating the dipole moments to the radiated field. The presence of the coupling matrix $G$ implies that the radiated field from each element depends not only on its local excitation but also on the collective excitation of its neighbors. 

The ideal model assumes that $G=0$, whereas the collective excitation of the neighboring elements also influence the radiation field and may generate undesired sidelobes. If such mutual coupling effect is not considered in holographic beamforming, the generated sidelobes can interfere with the sensing procedure, thus degrading ISAC performances. To cope with this issue, the authors in~\cite{zeng2025holographic} propose a mutual coupling aware holographic beamforming algorithm for the HISAC system, which can effectively suppress sidelobes. Fig.~\ref{f_couple}(b) illustrates the sidelobe level, defined as the beampattern gain of sidelobes, versus the iteration rounds. It can be observed that with the proposed scheme, which incorporates mutual coupling effects, the sidelobe level is effectively suppressed below the predetermined threshold. In contrast, when mutual coupling is neglected, the sidelobe level remains high as the iteration proceeds, resulting in a final value approximately 7.3 dB higher than that of the proposed scheme. This effective sidelobe suppression arises because the proposed scheme accurately models the coupling-induced variations in the radiation fields, enabling precise control of the sidelobe levels through optimized RHS amplitude patterns.

\subsection{Challenges and Opportunities}

One of the fundamental challenges in HISAC-enabled JCAS lies in understanding its ultimate performance limits under realistic physical constraints. In holographic beamforming, the spatial correlation among channels, the mutual coupling among radiation elements, and the finite-resolution control of the radiation element jointly determine the achievable sensing and communication performance. It is essential to theoretically analyze these limitations and trade-offs to further validate the superiority of HISAC. Such an analysis is non-trivial, as these physical characteristics interact in complex, nonlinear ways that differ significantly from conventional PA architectures. 

Another critical challenge lies in the real-time optimization of digital and holographic beamformers in HISAC systems. Existing approaches typically rely on optimization frameworks such as SDR, SCA, and AO. In particular, SDR- and SCA-based methods are often used for holographic beamforming design. As the computational complexity grows rapidly with the number of RHS elements, these methods are difficult to scale to ultra-massive arrays. To address this issue, HDMA techniques can be employed to reduce the complexity of the holographic beamforming problem~\cite{deng2022hdma}, thereby enabling more scalable implementations. On the other hand, AO-based methods are widely adopted to decouple digital and holographic beamformers. However, their iterative nature can lead to high computational overhead. To improve scalability, low-complexity variants such as closed-form updates for subproblems or learning-based approaches can be explored to reduce the per-iteration complexity and convergence time. Nevertheless, achieving real-time and scalable optimization under dynamic environments remains an open challenge for practical HISAC deployment.

\section{HISAC for Sensing-Assisted Communication}
\label{s_hsac}

Unlike joint communication and sensing, which performs communication and sensing simultaneously using a shared waveform, SAC~(sensing-assisted communication) exploits the outcomes of prior sensing operations to optimize communication performance~\cite{yang2023reconfigurable}. HISAC plays a crucial role in this process by providing accurate spatial perception and high-resolution beam-steering capabilities, thereby facilitating key communication procedures such as beam alignment, channel estimation, and resource allocation. 
In this section, we begin with an overview of the SAC paradigm. We then review two representative RHS-aided SAC scenarios, namely channel estimation and beam training, to illustrate how holographic beamforming can be leveraged to enhance communication performance. The section subsequently presents other related works and concludes with a discussion of the key challenges and opportunities associated with HISAC-enabled SAC.

\subsection{Overview of SAC}

By integrating perception priors into communication design, the communication system can adapt to real-time spatial and environmental dynamics, achieving higher efficiency and reliability. For example, in channel estimation, knowledge of scatterers can refine the wireless channel model, reducing pilot overhead while improving estimation accuracy. In beam training, accurate angular and distance information from sensing can narrow the search space for optimal beam, thereby accelerating link establishment~\cite{zhang2025codebook, cheng2025near}. Furthermore, in resource allocation and scheduling, awareness of user distribution and environmental blockages enables the BS to proactively manage spectrum and power resources, improving energy efficiency and robustness under dynamic channel conditions.

In the above SAC scenarios, there are generally two types of targets involved in the sensing process. The first type includes targets equipped with wireless devices, such as communication users or connected vehicles~\cite{zhang2020towards}. These targets can be detected and localized through device-based techniques, where the BS estimates the user’s position by analyzing the signals transmitted from the user~\cite{zhang2021rss}. Alternatively, they can also be perceived using device-free approaches, in which the BS emits probing signals and infers the user’s location from the reflected echoes. The second type of targets are objects in the surrounding environment, such as walls and trees, which do not carry any wireless transceiver. These targets can only be sensed in a device-free manner by analyzing the propagation, reflection, and scattering characteristics of the transmitted signal. Both types of sensing information are valuable for enhancing the performance of communication functions.

As discussed in Sec.~\ref{ss_cah}, the RHS can achieve high-resolution and flexible beamforming through its large reconfigurable aperture, enabling highly directive beams for precise detection, adjustable beamwidths for wide-area coverage, and tunable antenna position for channel adaptation. These properties allow HISAC to perform accurate and energy-efficient sensing for both types of targets. By dynamically reconfiguring its holographic patterns according to the latest sensed information, communication beams can be updated in real time, thereby realizing a closed-loop sensing-communication interaction.

\subsection{SAC for Channel Estimation}

Accurate channel information is essential for effective beamforming. In conventional channel estimation, the BS directly estimates the full channel matrix $\bm{H}$ through pilot transmission and signal reconstruction. However, in ultra-massive MIMO systems, the extremely large number of radiating elements leads to a significantly larger number of unknown coefficients in $\bm{H}$, which grows linearly with the aperture size. Therefore, if we directly estimate each element within the channel matrix $\bm{H}$, the induced pilot overhead is unacceptable. For example, conventional least-square~(LS) or mean square error~(MMSE)-based estimators require received pilots to have at least
the same dimension as MIMO channels to ensure robust estimation, which is impractical for ultra-massive MIMO systems.
To address this challenge, a sensing-assisted approach can be adopted: instead of estimating every entry of $\bm{H}$ directly, the BS first infers the geometric properties of the dominant propagation paths including their directions, distances, and complex gains, and then reconstructs the full channel~\cite{yue2023channel}. Since the number of unknown coefficients depends on the number of propagation paths rather than the number of RHS elements, and the available propagation paths in wireless environment is limited, this approach substantially reduces pilot overhead while preserving high reconstruction accuracy.

However, in holographic MIMO systems, the extremely large aperture of RHSs significantly extends the Fresnel region, leading to a hybrid propagation environment where both near-field and far-field users or scatterers coexist~\cite{zeng2025hybrid}. Specifically, the Fresnel region is characterized by the Rayleigh distance, which is widely adopted to determine the boundary between near-field and far-field propagation. For typical RHS configurations, this distance can be on the order of tens of meters (e.g., around $80$m for a $256$-element RHS at $30$GHz with quarter-wavelength spacing) or larger, which is comparable to typical link distances. As a result, both near-field and far-field propagation regimes may simultaneously arise in practical systems.

Conventional methods are typically designed under either the far-field or the near-field assumption. When applied to such hybrid environments, these methods suffer from severe performance degradation. A critical reason is the power diffusion (PD) effect, where the energy of one path spreads across multiple steering vectors in the transform domain, creating fake paths that hinder accurate recovery of wireless channels. Thus, new strategies are required to cope with the hybrid near-far field propagation for holographic beamforming-enabled systems.

\subsubsection{System Model}

Consider a channel estimation scenario with an RHS and a single-antenna user.
The multipath channel from the user to the RHS contains both far-field and near-field components, which can be expressed as
\begin{equation}
    \bm{h}_H = \sum_{k \in \mathbb{L}_F} g_k \bm{a}(\theta_k) + \sum_{k \in \mathbb{L}_N} g_k \bm{b}(\theta_k, r_k),
\end{equation}
where $\bm{a}(\theta_k)$ is the far-field steering vector defined as
\begin{equation}
    \bm{a}(\theta_k) = \frac{1}{\sqrt{N}}[1, e^{j2\pi d_e \sin \theta_k/\lambda}, \dots, e^{j 2\pi(N-1)d_e \sin \theta_k/\lambda}]^T.\label{e_a_theta_l}
\end{equation}
Here, $d_e$ is the element spacing in the RHS. $\bm{b}(\theta_k, r_k)$ is the near-field steering vector given by
\begin{equation}
    \bm{b}(\theta_k, r_k) = \frac{1}{\sqrt{N}}[e^{-j2\pi(r_{1,k}-r_k)/\lambda}, \dots, e^{-j2\pi (r_{N,k}-r_k)/\lambda}]^T,
\end{equation}
where $r_{n,k}$ denotes the distance from the $n$-th antenna element to the user via the $k$-th scatterer, and $r_k$ is the distance between the origin and the user. This hybrid model jointly accounts for the planar wavefront in the far-field and the spherical wavefront in the near-field, depending on whether the scatter lies outside or inside the near-far field boundary.

\begin{figure*}[!t]
\centering
\subfloat[]{\includegraphics[height=1.8in]{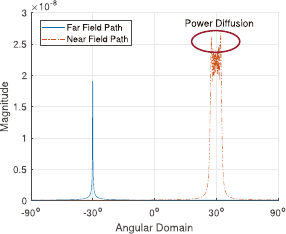}
}
\hspace{1mm}
\subfloat[]{
    \includegraphics[height=1.8in]{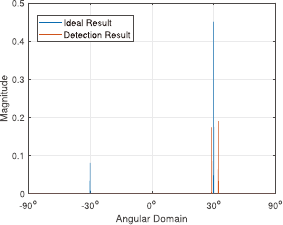}
  }
\hspace{1mm}
\subfloat[]{\includegraphics[height=1.8in]{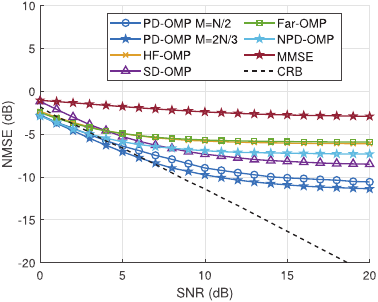}
}
\caption{Sensing-assisted channel estimation for large-scale RHS: (a) The angular-domain transform result of a hybrid-field channel consisting of a far-field path and a near-field path; (b) Inaccurate result of path detection due to power diffusion; (c) The NMSE of channel estimation result versus SNR.}
\label{f_channel_est}
\end{figure*}

\subsubsection{Sensing-Assisted Channel Estimation}

To accurately estimate the unknown channel $\bm{h}_H$, a sensing-assisted channel estimation method can be proposed, where the near and far-field paths are first estimated and the full channel is then reconstructed. 
However, such a hybrid structure introduces the issue of power diffusion which makes the accurate channel estimation challenging~\cite{zeng2025hybrid}. Specifically, existing works first transform the near-field (or far-field) paths to angular or polar domain and then employ sparse-signal-recovery-based method to fully exploit the sparsity of wireless channels. 
When the physical paths are projected onto these transform domains, the energy of a single path component may be dispersed over multiple positions, leading to power diffusion. This effect arises when a near-field path is transformed to the angular domain or when a far-field path is transformed to the polar domain~\cite{yue2024hybrid}, and it can also occur when near-field paths are transformed to the polar domain. Since a hybrid-field XL-MIMO channel contains both near-field and far-field components, power diffusion inevitably appears in both the angular and polar domains. The cause for power diffusion is the strong coherence between the steering vectors in each domain. As a result, transform-domain path components cannot be accurately identified, which ultimately degrades the quality of the reconstructed spatial-domain channel.

To mitigate the impact of power diffusion, a power-diffusion-aware orthogonal matching pursuit (PD-OMP) algorithm is proposed in~\cite{yue2024hybrid}. The key idea is to transform the received pilot signals into a joint angular-polar domain, where both near-field and far-field paths can be effectively separated. In this domain, the algorithm iteratively detects true paths and subtracts their contribution from the pilot signal. Specifically, the strongest path in the signal is identified as a true path, and its contribution contains both the projection of this path and the associated power diffusion. By explicitly accounting for the power diffusion effect, the PD-OMP algorithm can reliably distinguish true paths from power diffusion, leading to more accurate channel reconstruction.

\subsubsection{Simulation Results}

In Fig.~\ref{f_channel_est}(a) and~\ref{f_channel_est}(b), the power diffusion effect for angular-domain is illustrated (details can be found in~\cite{yue2024hybrid}). Specifically, the angular-domain transform of the hybrid-field channel $\bm{h}_F$ can be given by
\begin{equation}
    \bm{h}_F = \bm{F}_A \bm{h}_{A, F},
\end{equation}
where $\bm{h}_{A, F}$ denotes the angular-domain representation of $\bm{h}_F$, and $\bm{F}_A$ is defined as $[\bm{a}(\theta_1, \bm{a}(\theta_2), \dots, \bm{a}(\theta_Q)]$. Here, the whole region is divided into $Q$ angular grids, with $\bm{a}(\theta_q)$ being the steering vector of direction $\theta_q$ and defined in~(\ref{e_a_theta_l}). Suppose the hybrid-field channel $\bm{h}_F$ contains two paths, i.e., a far-field path and a near-field path, and its angular-domain representation is shown in Fig.~\ref{f_channel_est}(a). We can observe that the near-field path spreads over multiple far-field steering vectors, which is different from the far-field path concentrated in one steering vector. When applying the traditional orthogonal matching pursuit~(OMP) algorithm, only two dominant peaks are iteratively detected in the angular domain, which are directly taken as the two channel paths. As a consequence, the true path components are not accurately captured, which can be observed from the results in Fig.~\ref{f_channel_est}(b).

Fig.~\ref{f_channel_est}(c) shows the normalized mean square error~(NMSE) of the sensing-assisted channel estimation result given different SNRs and various estimators. Unlike HF-OMP~\cite{wei2022channel} and SD-OMP~\cite{hu2023hybrid}, which rely on prior knowledge of near-field and far-field path numbers, the proposed PD-OMP achieves higher accuracy without such assumptions. Its NMSE is also lower than that of Far-OMP~\cite{lee2016channel} and NPD-OMP~\cite{cui2022channel} because the former only apply the far-field angular-domain transform, and the latter does not consider the power diffusion range. Furthermore, PD-OMP shows significant gains at higher SNRs (around 20 dB) because the diffusion range can be estimated more precisely. It can also be concluded from Fig.~\ref{f_channel_est}(c) that using more angular grids ($M = 2N/3$) can further enhance the channel estimation accuracy. It is worth noting that PD-OMP uses pilots of length $Q = 20$, which is much smaller than the number of array elements $N = 200$. In conventional schemes that directly estimate the full channel matrix, $N$ pilot symbols are typically required, indicating PD-OMP is highly effective in reducing pilot overhead.

\subsection{SAC for Beam Training}
\label{ss_sbt}

In dynamic environments where user positions and channel conditions vary rapidly, frequent channel estimation still imposes considerable signaling and computational overhead. To enhance communication efficiency, beam training can be employed as an alternative sensing-assisted approach for spatial information acquisition~\cite{chen2025self}. Instead of reconstructing the entire channel, the objective is to determine the optimal transmission beams for all users based on their sensed angular and distance information. In this approach, the RHS first perceives user locations through sensing and then directly aligns its beams toward the identified directions. Furthermore, since holographic patterns on the RHS can be superimposed to form multi-beam codewords, multiple angular and distance locations can be illuminated simultaneously. This property enables simultaneous multi-user beam training, supporting fast and scalable beam alignment for real-time communication in hybrid near-far-field environments.

\subsubsection{System Model}

\begin{figure*}[!t]
\centering
\subfloat[]{\includegraphics[height=1.8in]{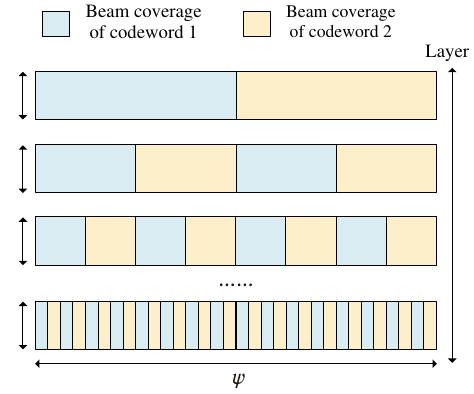}
}
\hspace{4mm}
\subfloat[]{
    \includegraphics[height=1.8in]{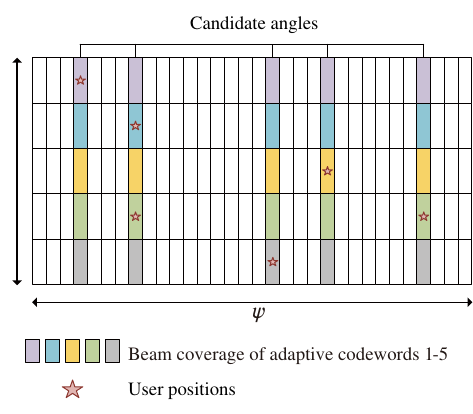}
  }
\hspace{4mm}
\subfloat[]{\includegraphics[height=1.8in]{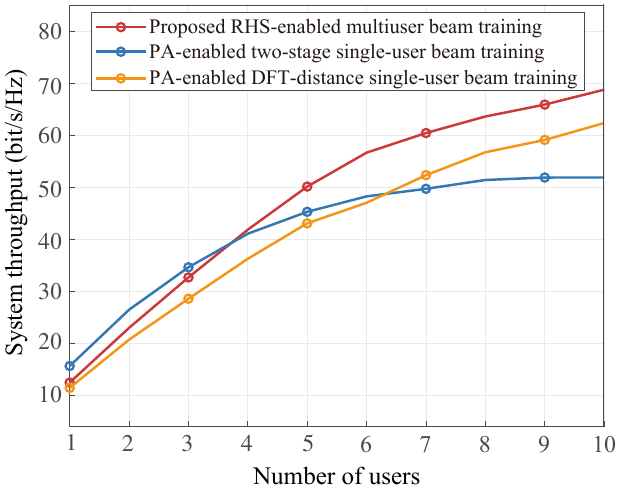}
}
\caption{Sensing-assisted beam training: (a) Multi-user angular codebook structure for holographic beamforming; (b) Multi-user distance-adaptive codebook structure for holographic beamforming; (c) The throughput versus the number of users.}
\label{f_codebook}
\end{figure*}

We consider a downlink multi-user communication system where the BS is equipped with an $N$-element RHS connected to $N_{RF} = L$ RF chains. The $U$ single-antenna users are randomly distributed within the hybrid near-far-field region of the RHS aperture. The signals of the near-field users are modeled as spherical wave, whereas those of far-field users are approximated as planar wave. Note that conventional far-field codebooks based on planar-wave assumptions are no longer applicable in this case. This is because the near-field beam training requires a two-dimensional search over angle and distance, which is different from the one-dimensional angular search in the far-field scenario.

To facilitate efficient codebook design in hybrid near-far-field environments, the spatial channel representation is transformed from the conventional $(r, \theta)$ domain to a new domain $(\varphi, \mu)$, where the angular parameter $\varphi$ is given by
\begin{equation}
    \varphi = \cos \theta,
\end{equation}
and the parameter $\mu$ is given by
\begin{equation}
    \mu = \frac{1-\cos^2 \theta}{r} = \frac{\sin^2 \theta}{r}.
\end{equation}
In this domain, uniformly sampling $(\varphi, \mu)$ ensures that the steering vectors are approximately orthogonal, i.e.,
\begin{align}
    \left| \bm{a}^H(\varphi,\mu)\bm{a}(\varphi + \varphi_{\Delta},\mu) \right| &=0,\\
    \left| \bm{a}^H(\varphi,\mu)\bm{a}(\varphi, \mu + \mu_{\Delta}) \right| &\le \varepsilon,
\end{align}
where $\bm{a}(\varphi,\mu)$ is the steering vector for location $(\varphi,\mu)$, and $\varepsilon$ is a small threshold controlling inter-beam correlation.

\subsubsection{Sensing-Assisted Beam Training}

In the $(\varphi, \mu)$ domain, we design a training process to determine the locations of users, where the angle and distance of users are sequentially searched in the following way:

\begin{itemize}
    \item \textbf{Angle Search:} Different from traditional beam training frameworks where the angles of each users are searched sequentially, a multi-user training process is proposed, where multiple users are served simultaneously. Specifically, a hierarchical codebook shown in Fig.~\ref{f_codebook}(a) is employed to progressively refine angular resolution. Each layer contains two multi-lobe holographic codewords covering different angular sectors. These codewords are carefully optimized to ensure high beamforming gain in both near-field and far-field regions. The beamwidth narrows layer by layer, allowing coarse-to-fine angle determination. Let $\bm{\psi}_{s, p}$ denote the $p$-th codeword in the $s$-th layer, which can be obtained by minimizing the following objective function
    \begin{align}
        \min_{\psi}& \sum^I_{i=1}\sum^J_{j=1}\left(\left|\bm{a}(\varphi_i, \mu_j)\bm{M}\bm{b}\right|-G_{i, j}\right)^2,
    \end{align}
    where $I$ and $J$ are the number of samplings along $\varphi$ and $\mu$ axes, respectively. $G_{i, j}$ is the gain threshold for location $(\varphi_i, \mu_j)$ and can be given by
    \begin{equation}
        G_{i, j} = \begin{cases}
            D, & (\varphi_i, \mu_j) \in \mathcal{C}_{s, p},\\
            0, & \text{otherwise.}
        \end{cases}
    \end{equation}
    Here, $D$ is a predefined positive threshold and $\mathcal{C}_{s, p}$ is a set containing all the sample points for codeword $\bm{\psi}_{s, p}$.
    \item \textbf{Distance Search:} After completing the angle search, the BS obtains the candidate angles for all users and subsequently performs a distance search in a parallel manner. As illustrated in Fig.~\ref{f_codebook}(b), we design distance-adaptive codewords, in which each multi-lobe codeword covers a specific range of the $\mu$ for all angles identified in the previous phase. Let $\bm{\psi}_{k, j}$ denote the predesigned codeword whose energy is concentrated around the direction of the $u$-th user and the $j$-th range bin of $\mu$. To simultaneously determine whether all the users are in the $j$-th range bin of $\mu$, the codeword $\bm{\psi}^{dis}_j$ can be directly generated based on holographic principle, which is given by
    \begin{equation}
        \bm{\psi}^{dis}_j = \dfrac{1}{U}\sum^U_{u=1} \bm{\psi}_{u, j}.
    \end{equation}
    When the BS sequentially transmits the codewords $\bm{\psi}^{dis}_1, \dots, \bm{\psi}^{dis}_J$, each user can measure its received power and reports the index of the codeword that yields the strongest response. In this way, the distance search for all users is completed simultaneously.
\end{itemize}

Compared with existing methods, the proposed approach differs in two key aspects. First, compared to far-field beam training that only performs angular search, the proposed method explicitly incorporates distance search and thus remains effective in near-field regimes. Moreover, compared to existing near-field beam training schemes, which typically perform sequential or user-wise angle-distance search, the proposed scheme enables simultaneous multi-user beam training with an overhead that does not scale with the number of users. This is achieved by exploiting the holographic-pattern superposition property of RHS, which is not available in conventional phased-array or RIS-based systems.

After determining the locations of all the users through the above process, the holographic beamformer can be given by
\begin{equation}
    \bm{\psi} = \dfrac{1}{U}\sum^U_{u=1} \bm{\psi}^*_u,
\end{equation}
where $\bm{\psi}^*_u$ is the codeword whose transmit energy focuses on the location of the $u$-th user. The digital beamformer can then be calculated by using the zero-forcing~(ZF) method and power allocation algorithm, which is omitted here for brevity.

\begin{figure*}[!t]
\centering
{\captionsetup[subfloat]{captionskip=6mm}
\raisebox{4mm}{
\subfloat[]{\includegraphics[height=1.7in]{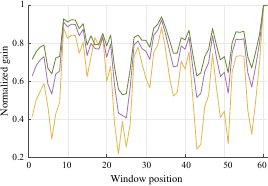}
}}}
\hspace{16mm}
\subfloat[]{\includegraphics[height=2in]{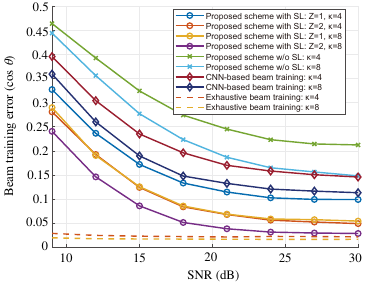}
}
\caption{Sensing-assisted beam training with adjustable aperture position: (a) Channel gain versus the location of the sliding window; (b) Beam training error versus the SNR.}
\label{f_window}
\end{figure*}

\subsubsection{Simulation Results}

To show the impact of low-overhead beam training scheme on system throughput, we compare its performance with existing near-field beam training approaches implemented by PAs with the same aperture size and input power. The system throughput, which jointly accounts for beam training overhead and the achievable sum data rate, serves as an effective indicator of system performance and is defined as
\begin{equation}
    \eta = \left(1 - \frac{t}{T}\right)\sum_{k=1}^{U} R_u,
\end{equation}
where $t$ denotes the number of time slots consumed for beam training, $T$ is the total number of transmission slots, and $R_u$ is the data rate of the $u$-th user. As illustrated in Fig.~\ref{f_codebook}(c), compared with PA-based two-stage~\cite{zhang2024codebook} and DFT-distance~\cite{zhang2022fast} schemes, the SAC-based multi-user beam training scheme for holographic beamforming achieves higher system throughput when the number of users exceeds $3$ (details can be found in~\cite{zhang2025holographic}). As the user number increases, the simultaneous multi-user beam training for holographic beamforming provides higher throughput improvements because it can locate all the users at the same time. In contrast, the training overhead of conventional schemes grows linearly with the number of users.

In the above scheme, each codeword corresponds to a specific array position. When comparing the received signal strengths of different codewords, some may coincide with the peaks of the signal envelope shown in Fig.~\ref{f_window}(a), while others may fall within its troughs due to small-scale fading. This uneven channel response can lead to errors in signal strength comparison, thereby degrading the beam training accuracy. 
To address this issue, we can traverse the candidate sliding windows, i.e., the position of the effective aperture, and the one that yields the maximum received signal strength is selected as the codeword gain. As a result, the probability of the strongest signal appearing at a fading trough is reduced, thereby improving the robustness of beam training.

Fig.~\ref{f_window}(b) illustrates the variation of beam training error with SNR (see~\cite{zhang2025fluid} for details). For comparison, the performances of the CNN-based method~\cite{ma2021deep}, exhaustive search, and the proposed schemes with or without sliding window are presented. The exhaustive method employs a DFT codebook over the full array, containing codewords for all possible angles. Under identical channel conditions (e.g., Rician factor $\kappa$), the proposed sliding window scheme exhibits significantly lower beam training error than the scheme without sliding window, demonstrating that the positional DoF enhances codeword gain and improves training accuracy. The CNN-based approach achieves lower error than the proposed scheme without sliding window owing to its higher prediction capability, but remains less accurate than the proposed scheme because of its sensitivity to channel variations. Furthermore, increasing the number of sliding windows further reduces training error by improving effective channel quality. As SNR increases, the beam training error decreases, and the proposed scheme approaches the lower bound established by exhaustive search performance.

\subsection{Other Related Works}

In addition to detecting environmental scatterers through the uplink between the user and the BS, the BS can also perform monostatic sensing using its own holographic aperture~\cite{zhang2023holographic_o, zhang2022parameter}. For example, the authors in~\cite{zhang2023multi} investigate multi-target detection using two closely deployed RHSs serving as the transmit and receive antennas. To accurately detect multiple unknown targets, the conventional binary hypothesis testing in~(\ref{e_bht}) is extended to a multi-hypothesis testing framework, where each hypothesis represents a possible configuration of target numbers and angular locations. To enhance the discriminability among hypotheses, the paper adopts relative entropy (Kullback-Leibler divergence) as the optimization metric, since it quantitatively measures the separability between the likelihood functions under different hypotheses. A waveform-and-amplitude optimization algorithm is then proposed to jointly design the sensing waveform and the transmit/receive RHS amplitude vectors. Simulation results demonstrate that the proposed sensing system enabled by holographic beamforming improves the probability of detection by approximately $0.13$ compared with a phased-array radar under the same hardware cost, thereby confirming the potential of holographic beamforming for SAC systems.

Moreover, the authors in~\cite{hu2023multi} propose a position-then-transmit scheme for HISAC systems. In this scheme, the user first transmits uplink signals to the BS, during which the BS dynamically adjusts its holographic and digital beamformers. After receiving the uplink signals, the BS performs uplink sensing to estimate the user’s location and subsequently utilizes the estimated position information to guide downlink beamforming for data transmission. Compared with beam training-based methods~\cite{zhang2025holographic, dong2025near}, this approach can directly infer user positions without requiring frequent feedback. To quantify localization performance, the CRB is derived as a theoretical measure of positioning accuracy and further employed as an optimization objective for the joint analog-digital beamforming design. Simulation results show that the proposed scheme achieves a 42\% reduction in positioning mean-square error (MSE) and an 82\% reduction in communication capacity loss compared with benchmark directional beamforming that scans the region of interest (ROI) with focused beams.

The determination of near-far field boundary has also been investigated in the literature. As discussed in Sec.~\ref{ss_cah}, the widely used Rayleigh distance can be employed as the boundary between near-field and far-field~\cite{yue2024hybrid}. This criterion is derived by limiting the phase deviation between the spherical-wave model and its plane-wave approximation within $\pi/8$, such that the far-field model remains sufficiently accurate. Recently, alternative boundaries that enable a more flexible trade-off between computational complexity and system performance have been proposed. For example, the authors in~\cite{zhang2026hybrid} determine the boundary by setting a threshold on the normalized coherence between the near-field and far-field channels. Since a higher coherence indicates a more accurate far-field approximation, a larger threshold can be used to ensure localization performance, while a smaller threshold can be adopted to reduce computational complexity. This provides a practical mechanism to balance localization accuracy and processing complexity in practical implementations. To enable dynamic switching between near-field and far-field processing modes, range-domain discrete Fourier transform (DFT) is first performed to obtain coarse range estimates from the received signals. Based on these estimated ranges, targets are classified as near-field or far-field depending on whether their distances are below or above the boundary. Subsequently, different signal models are adopted, thereby reducing computational complexity while assuring the localization accuracy.

\subsection{Challenges and Opportunities}

One major challenge in SAC lies in achieving real-time sensing while maintaining high communication efficiency. Specifically, the RHS-aided sensing process typically requires multiple time slots, during which the holographic beamformers must be reconfigured to scan different spatial directions. However, this sensing procedure inevitably introduces overhead, which directly reduces the effective communication throughput. To enhance system performance, it is essential to design low-overhead sensing strategies that operate with fewer time slots or even one slot while still ensuring minimal impact on data transmission.

Another significant challenge in SAC arises from sensing moving targets or mobile users in HISAC systems. The motion of objects introduces Doppler shifts that affect the frequency and phase of the received echoes and must be explicitly considered in the optimization of holographic beamformers. Furthermore, the fine-grained sensing resolution provided by large-scale RHSs makes the system highly sensitive to even slight variations in the positions of targets or users, which can easily result in beam misalignment and performance degradation. Addressing this challenge requires novel beamforming mechanisms capable of robustly tracking moving objects, either by dynamically adjusting the beamwidth for enhanced robustness or by performing beam prediction based on estimated motion trajectories.

Moreover, the trade-off between the reconfiguration frequency of holographic patterns and the achievable ISAC performance should also be carefully investigated. On one hand, frequent reconfiguration enables the RHS to closely track time-varying channel and environmental dynamics, thereby improving both communication rate and sensing accuracy. On the other hand, increasing the update frequency incurs higher overhead in control signaling and sensing operations, which may offset the performance gains. This trade-off indicates that a high reconfiguration frequency is not always desirable, and the optimal update rate should be adapted to channel dynamics and user mobility. To this end, adaptive reconfiguration strategies can be developed to avoid unnecessary updates while maintaining reliable performance. For example, sensing and holographic pattern updates can be invoked only when significant channel variations or communication performance degradation are detected. In this way, sensing operations are performed on demand rather than periodically, leading to a more efficient use of system resources.

HISAC systems also face significant challenges under blockage conditions. Although ultra-large antenna arrays extend the near-field region and improve spectral efficiency through beam focusing, the reliability of such links is susceptible to blockage, especially at high frequencies. To address this issue, on the one hand, advanced sensing algorithms are required to accurately detect obstacles and even perceive occluded environments~\cite{zhang2025generative, wang2025ris}. On the other hand, it is also important to develop blockage-aware beamforming strategies that can effectively exploit the sensed information. For example, Airy-beam-based designs, which enable wave propagation along curved trajectories to circumvent obstacles, can be leveraged to enhance spectral efficiency given the knowledge of obstacle locations and sizes~\cite{zhang2026breaking}.

\section{HISAC for Communication-Assisted Sensing}
\label{s_hcas}

Different from the concepts of JCAS and SAC discussed earlier, CAS~(communication-assisted sensing) represents a complementary direction in which communication functionalities are exploited to enhance sensing capability. In this section, we first outline the operating principles of HISAC-enabled CAS. We then describe how communication networks composed of multiple RHS-equipped nodes can be leveraged to support distributed sensing. The section subsequently reviews related works and concludes with a discussion of the key challenges and opportunities.

\subsection{Overview of CAS}

In an HISAC-aided CAS system, each BS equipped with an RHS transmits communication signals toward users, while the echoes reflected from surrounding targets can be captured by the same BS, neighboring BSs, or even user terminals. The received communication signals inherently contain both data and environmental information, from which sensing parameters such as range, velocity, and angle can be extracted through proper signal processing.

Communication can typically assist sensing from two complementary levels in such systems. At the signal level, existing communication waveforms such as wideband orthogonal frequency division multiplexing (OFDM) can be directly reused for high-resolution sensing. The time-frequency structure of these waveforms facilitate accurate delay and Doppler estimation without disrupting communication functionality~\cite{liu2025cp}. At the network level, the interconnections among RHS-equipped BSs enable distributed sensing, where multiple nodes jointly illuminate and observe the same region, exchange sensing-related information, and perform data fusion through backhaul or edge links. This cooperation can significantly enhance sensing coverage, robustness against blockage, and localization accuracy. In the following, a case study of holographic beamforming-enabled distributed sensing is provided.

\subsection{Communication Network for Sensing}

Modern cellular architectures naturally enable distributed sensing because baseband and control functions can be centralized and virtualized across multiple radio nodes. For example, in a cloud radio access network (C-RAN), the conventional BS is decomposed into remote radio heads (RRHs) and a baseband unit (BBU) cloud\cite{farhat2024recent}. The RRHs handle wireless transmission between the BS and user equipments, while the BBU performs encoding, decoding, precoding, and joint signal processing for all connected RRHs via high-capacity fronthaul links. This architecture enables coordinated multipoint transmission, inter-cell interference coordination, and collaborative processing, which can be directly leveraged for distributed sensing. Building upon the foundation of C-RAN, the cell-free network paradigm~\cite{adhikary2024holographic} further eliminates fixed cell boundaries by allowing multiple distributed RRHs to cooperatively serve all users under centralized control, thereby achieving uniform service quality and coordination gains. As a promising implementation of ultra-massive MIMO, RHSs can be incorporated into the cell-free networks and enhance the performance of distributed sensing.

\begin{figure*}[!t]
\centering
{\captionsetup[subfloat]{captionskip=4mm}
\raisebox{4mm}{
\subfloat[]{\includegraphics[height=1.7in]{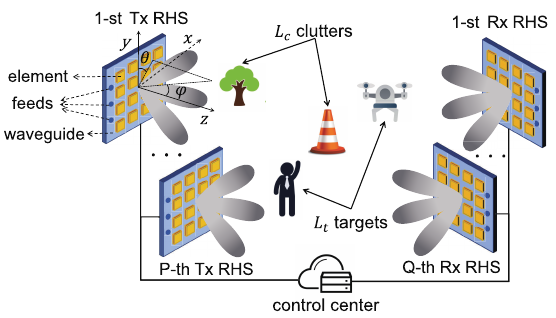}
}}}
\hspace{16mm}
\subfloat[]{\includegraphics[height=2.1in]{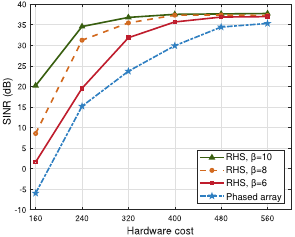}
}
\caption{Holographic beamforming-enabled distributed sensing: (a) A holographic beamforming-enabled distributed sensing network; (b) SINR versus the hardware cost of the RHS and phased array-based schemes.}
\label{f_distri}
\end{figure*}

\subsubsection{System Model}

As shown in Fig.~\ref{f_distri}(a), we consider a distributed sensing system consisting of $P$ transmit and $Q$ receive RHS subarrays connected to a central control center. The waveforms of all the transmit RHSs are orthogonal, and each receive RHS collects echoes emitted by all the transmit RHSs and reflected by $J_t$ targets and $J_c$ clutters. The transmitted signal from the $p$-th RHS can be written as
\begin{equation}
    \bm{X}_p = \bm{M}^t_p\bm{S}_p,
\end{equation}
where $\bm{M}_p$ is the holographic beamfomring matrix of the $p$-th transmit RHS, and $\bm{S}_p$ is the waveform of the $p$-th transmit RHS, satisfying
\begin{equation}
    \bm{S}_p \bm{S}^H_p = \begin{cases}
        \bm{0}_{L_t}, p\ne p',\\
        \bm{I}_{L_t}, p=p'.
    \end{cases}
\end{equation}
Here, $\bm{0}_{L_t}$ represents a zero matrix of size $L_t \times L_t$, and $\bm{I}_{L_t}$ is an identity matrix of size $L_t \times L_t$.

At the receiver side, the signal received by each RHS is first filtered by a matched filter bank to separate the signal from different transmit RHSs. Thus, the signal corresponding to the $p$-th transmit RHS and $q$-th receive RHS can be written by
\begin{equation}
    \bm{Y}_{p, q}
= (\bm{M}^r_q)^T\left(\sum^J_{j=1} \big(\beta^j_{p, q}\bm{A}^j_{p, q}\bm{X}_p\bm{T}^j_{p,q}\big)
	+\mathbf{N}_{p, q}\right),
\end{equation}
where $\beta^l_{p,q}$ denotes the reflection coefficient of the $l$-th target/clutter relative to the $p$-th transmit RHS and $q$-th receive RHS.
$\bm{A}^l_{p, q} = \bm{a}^j_q(\theta^j_q, \phi^j_q)(\bm{a}(\theta^j_p, \phi^j_p))^T$ is the composite transmit-receive steering matrix, with $ \bm{a}^j_q(\theta^j_q, \phi^j_q)$ and $\bm{a}(\theta^j_p, \phi^j_p)$ being the steering vector of the $q$-th receive RHS and $q$-th transmit RHS with respect to the $j$-th target.
$\bm{T}^j_{p, q}$ represents time delay~\cite{zhang2022metaradar}, and $\bm{N}_{p, q}$ is the additive noise.

\subsubsection{Holographic Beamforming for Distributed Sensing}

To optimize the holographic beamformers, the worst-case average SINR can be maximized~\cite{li2025reconfigurable}.
For the $j_t$-th target/clutter and the $(p,q)$-th transceiver pair, the corresponding SINR is given by
\begin{equation}
    SINR^{j_t}_{p, q}(\bm{\psi}^t_p,\bm{\psi}^r_q)
= \frac{\left|\bm{v}^{j_t}_{p,q}(\bm{\psi}^t_p,\bm{\psi}^r_q)\right|^2}
{\sum_{j\neq j_t}\!\!\left|\bm{v}^{j}_{p, q}(\bm{\psi}^t_p,\bm{\psi}^r_q)\right|^2 + |\bm{w}_{p, q}|^2},
\end{equation}
where $v^{j_t}_{p,q}(\bm{\psi}^t_p,\bm{\psi}^r_q) = \text{vec}(\beta^{j_t}_{p, q}\bm{M}^T_q\bm{A}^{j_t}_{p, q}\bm{X}_p\bm{T}^{j_t}_{p, q})$ is the vectorized signal reflected by the $j_t$ target/clutter in $\bm{Y}_{p, q}$, and $\bm{w}_{p, q} = \text{vec}((\bm{M}^r_q)^T\bm{N}_{p, q})$ is the vectorized noise. Thus, the average SINR for the $j_t$-th target across all subarrays is
\begin{equation}
    \overline{SINR}^{j_t}(\bm{\psi}^t,\bm{\psi}^r)
= \frac{1}{PQ}\sum_{p=1}^{P}\!\sum_{q=1}^{Q}
SINR^{j_t}_{p,q}(\bm{\psi}^t_p,\bm{\psi}^r_q),
\end{equation}
where $\bm{\psi}^t = [\bm{\psi}^t_1, \dots, \bm{\psi}^t_P]$ and $\bm{\psi}^r = [\bm{\psi}^r_1, \dots, \bm{\psi}^r_Q]$. Therefore, the worst-case average SINR is given by
\begin{equation}
    \widetilde{SINR}(\bm{\psi}^t,\bm{\psi}^r)
= \min_{j_t} \overline{SINR}^{j_t}(\bm{\psi}^t,\bm{\psi}^r).
\end{equation}

To efficiently solve the holographic beamforming optimization problem for distributed sensing, we can iteratively optimize the holographic patterns for the Tx and Rx RHSs. Moreover, considering the average SINR in the objective function is the sum of multiple fractions, we introduce the slack variables $\bm{\Lambda} = \{\lambda^{j_t}_{p,q}\}^{P, Q, J_t}_{p=1, q=1, j_t=1}$ and use $\lambda^{j_t}_{p,q}$ to approximate $SINR^{j_t}_{p, q}$ in the objective function~\cite{li2025reconfigurable}. 

\subsubsection{Simulation Results}

Fig.~\ref{f_distri}(b) depicts the SINR versus hardware cost for the proposed RHS and the phased array-based schemes with the cost ratio being $\beta = 6, 8, 10$ (see~\cite{li2025reconfigurable} for details). Both the number of transmit and receive RHS subarrays are $2$, and we increase the hardware cost by enlarging the number of elements in each subarray. It can be observed that, under the same power consumption and hardware cost, the SINR of the proposed scheme is at least $4.98$dB higher than that of the phased array scheme on average, indicating a better multi-target detection performance in the distributed sensing scenario.

\subsection{Other Related Works}

Several existing works have investigated HISAC enabled by OFDM or other communication signals. For example, the authors in~\cite{wei2025wideband} consider a near-field scenario that accounts for spherical wavefront and frequency-spatial coupling.
By jointly optimizing the digital and holographic beamformers across wideband subcarriers, this approach enables accurate multi-target detection and robust beam control when the array aperture becomes electromagnetically large. 
Another extension further incorporates a RIS into the system architecture to enhance both communication coverage and sensing performance in non-line-of-sight (NLoS) environments~\cite{wei2024ris}. In this design, the digital, holographic, and RIS beamformers, together with the receive filter, are jointly optimized to maximize the radar SINR under user SINR constraints. 

In~\cite{hu2023holofed}, a HISAC framework that leverages the inherently multi-band nature of cellular networks is investigated. Since different frequency bands exhibit distinct propagation and reflection characteristics, exploiting multi-band information allows the system to jointly utilize communication signals across multiple bands for high-precision environmental sensing and user localization. In the proposed design, the BS performs uplink sensing using signals transmitted by users over several frequency bands. Simulation results demonstrate that this multi-band system enabled by holographic beamforming achieves a $57\%$ lower positioning error variance compared to a beam-scanning baseline.

\subsection{Challenges and Opportunities}

A critical challenge arises in ultra-dense HISAC networks, where a large number of users, targets, and RHS elements coexist. Most existing CAS schemes assume that all communication and sensing tasks are jointly processed by a centralized CPU, which aggregates data and performs global optimization. However, in ultra-dense deployments, this approach leads to prohibitive computational and communication overhead, as the complexity scales with both the number of nodes and the aperture size of each RHS. To address this scalability issue, future HISAC architectures will need to incorporate hierarchical or distributed processing frameworks, where part of the sensing and beamforming computations are executed locally at the edge. This paradigm shift requires new communication-sensing coordination protocols, information exchange mechanisms, and cooperative processing methods that can efficiently balance global coordination with local autonomy. 

\section{Implementation and Experiments}
\label{s_ie}

\begin{figure*}[!t]
\centering
\subfloat[]{\includegraphics[height=2.2in]{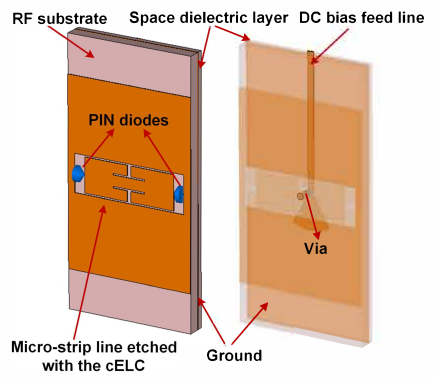}
}
\hspace{10mm}
\subfloat[]{\includegraphics[height=2.2in]{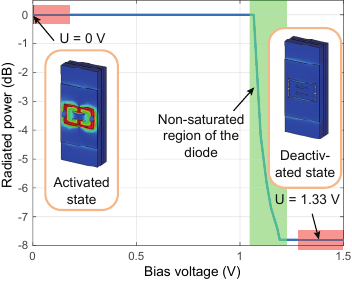}
}
\caption{RHS element: (a) Architecture; (b) Radiated power of an RHS element versus the applied bias voltage.}
\label{f_element}
\end{figure*}

This section presents the implementation and experimental validation of HISAC. We first describe the hardware design of RHSs. Then, prototypes in the literature are introduced to show the feasibility and performance of HISAC, followed by discussions on key implementation challenges and future opportunities.

\subsection{RHS Hardware}

The hardware design of RHSs is crucial for translating holographic principle into practical systems. An RHS typically consists of numerous sub-wavelength radiating elements, a feeding network that excites these elements, and control circuits that adjust their radiation amplitudes. Together, these components enable low-cost, low-power, and highly reconfigurable beam generation. This subsection outlines the key hardware architectures of RHSs, including the design of individual elements and the overall array configuration.

\subsubsection{RHS Element}

The RHS element serves as the core component that controls the radiation amplitude of electromagnetic waves. As illustrated in Fig.~\ref{f_element}(a), an RHS element with five layers is designed~\cite{di2025reconfigurable, deng2023reconfigurable2}. The top layer is a micro-strip line patterned with a complementary electric-inductive-capacitive~(cELC) resonator. Active components such as PIN diodes can be attached to enable dynamic control of radiation characteristics. Beneath it lies an RF dielectric substrate that guides the propagation of the reference wave, followed by a copper ground plane providing electromagnetic isolation. A space dielectric layer separates the ground from the DC bias network on the bottom layer, where the bias voltage is supplied through vias that penetrate the entire structure to reach the PIN diodes mounted on the top layer. Amplitude control is realized by adjusting the effective inductance and capacitance of the cELC resonator through the bias voltage applied to the diodes. 

The electromagnetic response of the element can be tailored by modifying key physical parameters such as the thickness of the substrate, and the geometric dimensions of the cELC resonator~\cite{deng2023reconfigurable, deng2022holographic2}. Specifically, a thicker substrate lowers the resonant frequency and affects the S-parameter response, while the resonator size determines the coupling strength and bandwidth. These parameters are optimized through full-wave simulations and parametric sweeps in CST Studio Suite to achieve the desired resonance and amplitude characteristics, ensuring efficient and stable radiation for the RHS array. The radiation power of the optimized RHS element is illustrated in Fig.~\ref{f_element}(b), where bias voltages of $0$V and $1.33$V correspond to the activated and deactivated states of the element, respectively.

\begin{figure}[!t]
\centering
\subfloat[]{\includegraphics[width=3.2in]{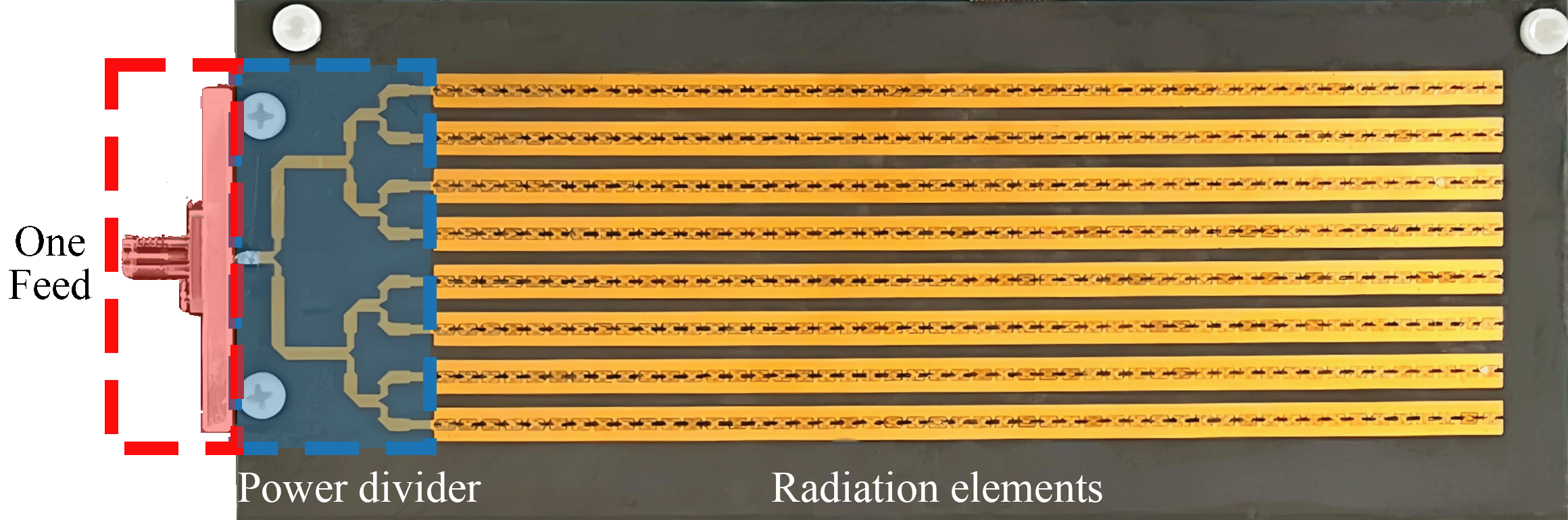}
}\\
\subfloat[]{\includegraphics[width=3.2in]{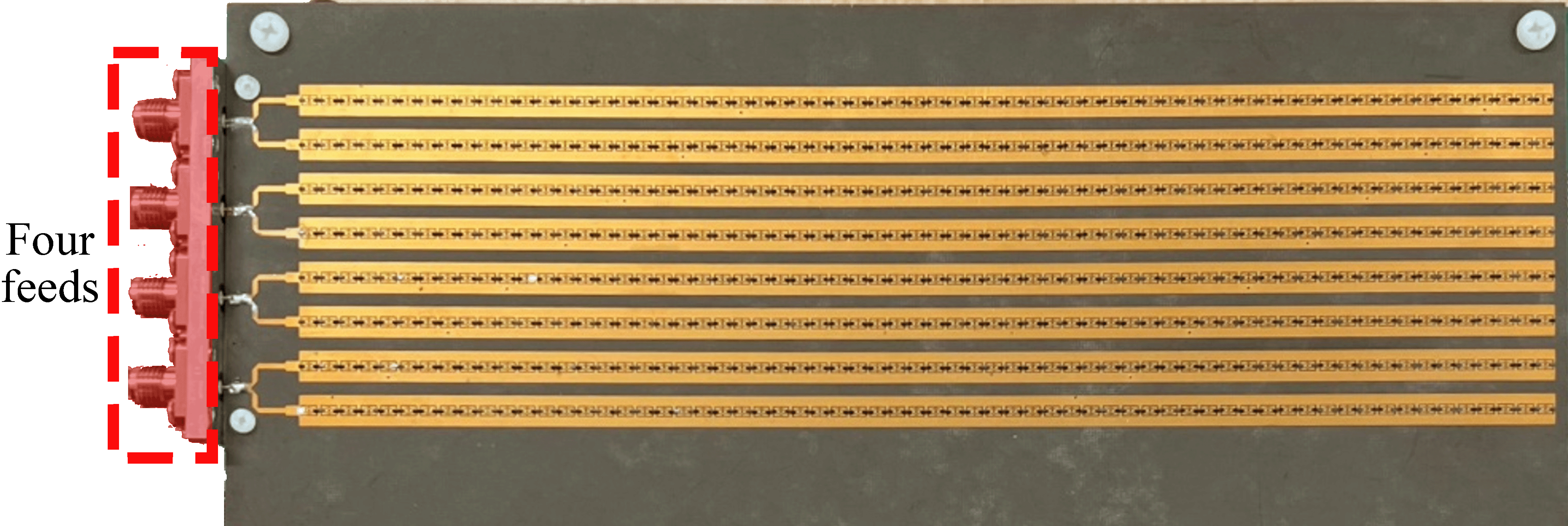}
}
\caption{RHS array with: (a) One feed; (b) Four feeds.}
\label{f_array}
\end{figure}

\subsubsection{RHS Array}

Based on the optimized RHS element, two-dimensional~(2D) RHS arrays can be constructed to enable flexible three-dimensional~(3D) holographic beamforming~\cite{hu2025reconfigurable}. Fig.~\ref{f_array}(a) illustrate an array with an edge-mounted feed, and a power divider, and 384 cELC-based radiation elements. The power divider distributes the input RF signal into multiple outputs, with each output exciting one row of the RHS. Here, one output port of the feeding network in the hardware implementation corresponds to a ``feed'' in the RHS model in Sec.~\ref{s_fr}. Moreover, after the power division, each row still satisfies the leakage power constraint, since isolation is introduced between different rows in the hardware implementation. 
The bottom layer of the array incorporates control lines for supplying external bias voltages, enabling real-time reconfiguration of the holographic pattern. The parameters of the RHS array are also tuned through full-wave simulations to improve the overall radiation performance.

It is also worth noting that RHSs with multiple edge-mounted feeds can also be fabricated, which is shown in Fig.~\ref{f_array}(b). It contains $4$ edge-mounted feeds and $8$ rows, where the input signal associated with each feed is divided into two streams, each exciting one row in the RHS. Compared with the architecture in Fig.~\ref{f_array}(a), this implementation supports multiple independent RF signal inputs, while requiring multiple RF chains and hence leading to higher hardware complexity and cost.

\begin{figure*}[!t]
\centering
\includegraphics[width=5.5in]{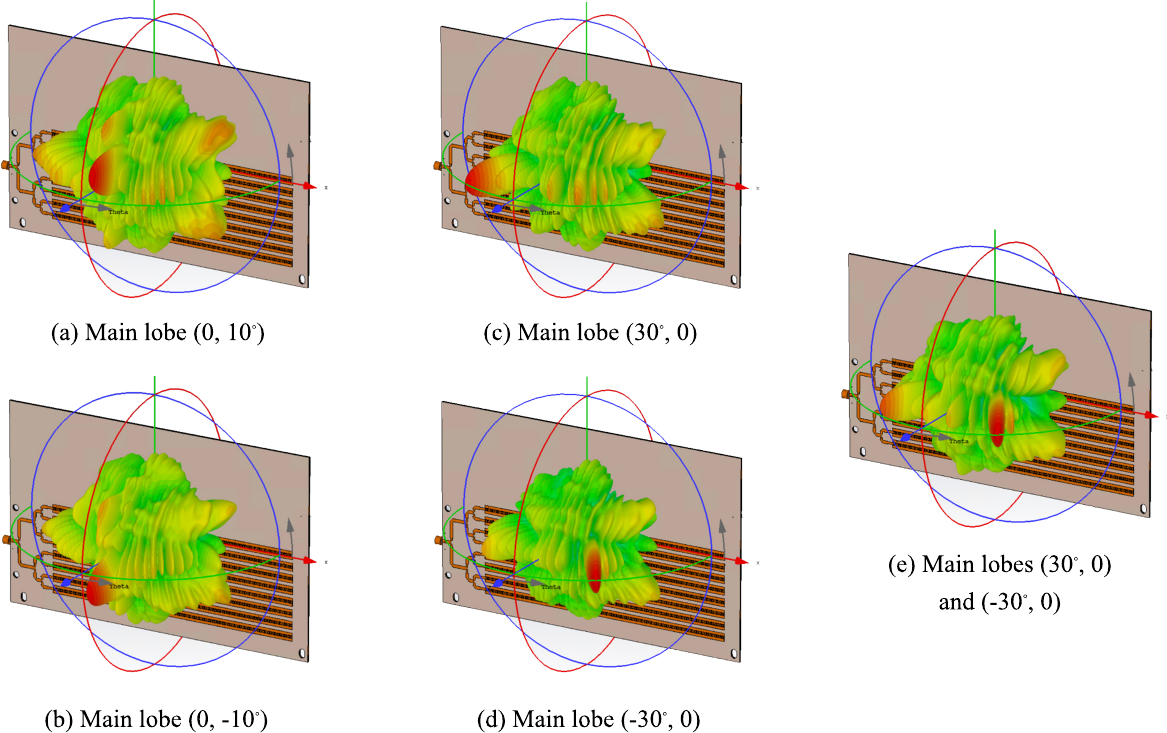}
\caption{Radiation patterns of the RHS array.}
\label{f_patterns}
\end{figure*}

The 2D RHS array can generate single or multiple beams in different directions. Specifically, Fig.~\ref{f_patterns}(a)-(d) demonstrate the simulation results of beam steering toward elevation angles of $\pm 10^\circ$ and horizontal angles of $\pm 30^\circ$, whose holographic patterns are calculated based on holographic principle. Moreover, by superimposing multiple holographic patterns, the array can also form radiation patterns with multiple beams, as shown in Fig.~\ref{f_patterns}(e). 

\begin{figure*}[!t]
\centering
\subfloat[]{\includegraphics[height=1.5in]{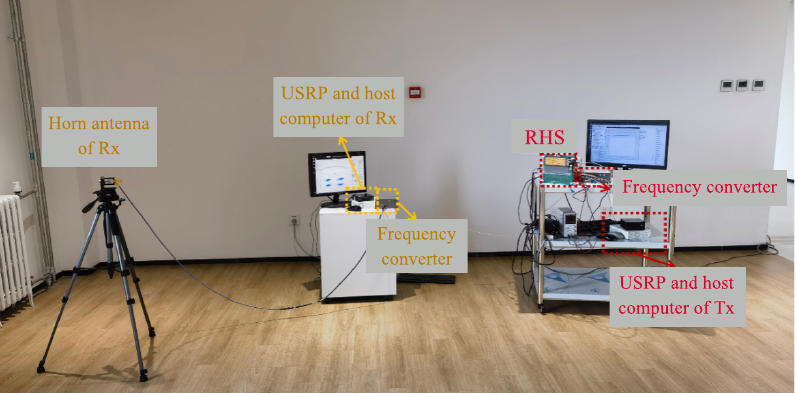}
}
\hspace{4mm}
{\captionsetup[subfloat]{captionskip=4mm}
\raisebox{4mm}{
\subfloat[]{\includegraphics[height=1.2in]{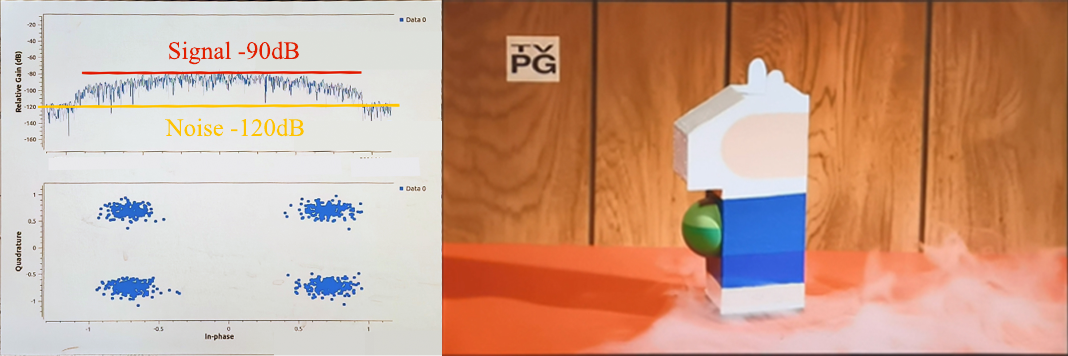}
}}}
\caption{Holographic beamforming-enabled communication: (a) Prototype; (b) Spectrum, constellation, and video snapshot of the Rx.}
\label{f_prot_comm}
\end{figure*}

\subsection{Prototypes}

To verify the practical feasibility of holographic beamforming, four prototypes are developed for different application scenarios, including communication, sensing, JCAS, and SAC.

\subsubsection{Communication Prototype}

Based on the developed RHS array, a point-to-point wireless communication system is established to demonstrate the capability of holographic beamforming for data transmission. As illustrated in Fig.~\ref{f_prot_comm}(a), the prototype consists of a transmitter and a receiver. The transmitter comprises a host computer, a universal software radio peripheral (USRP), and the RHS antenna array, while the receiver includes a standard horn antenna, a USRP, and a receiving host. Specifically, the transmitter employs an LW-N210 USRP, which processes baseband signals using the GNU Radio software framework and performs RF modulation for wireless transmission. The operating frequency of the USRP is set to $2.3$GHz, and its output is up-converted to $26.2$GHz through a frequency converter before being fed into the RHS array. On the receiver side, the horn antenna captures the signal radiated from the RHS and delivers it to a frequency converter, which down-converts the signal to $2.3$GHz. The down-converted signal is then processed by the receiving USRP for demodulation and recovery of the original baseband data.

The system is deployed in an indoor environment to evaluate data transmission performance under typical communication conditions. The horn antenna is positioned at a $30^\circ$ direction relative to the RHS array. Following the digital video broadcasting-terrestrial (DVB-T) standard, real-time high-definition $(1920\times 1080)$ video is transmitted using quadrature phase-shift keying (QPSK) modulation. As shown in Fig.~\ref{f_prot_comm}(b), the received intermediate-frequency signal exhibits a SNR of approximately $30$dB, and the recovered constellation is clearly distinguishable. The reconstructed video at the Rx also matches the transmitted content, verifying that the holographic beamforming-enabled communication prototype can achieve reliable high-definition video transmission. Moreover, the total power consumption of the RHS array is only $5$W under this configuration, demonstrating its potential for low-power and high-throughput wireless communication.

\subsubsection{Sensing Prototype}

\begin{figure*}[!t]
\centering
\subfloat[]{\includegraphics[height=1.6in]{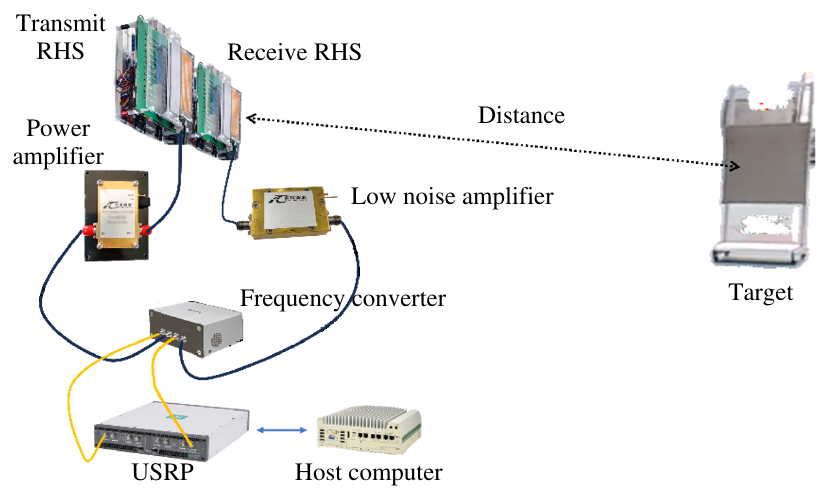}
}
\hspace{4mm}
{\captionsetup[subfloat]{captionskip=2mm}
\raisebox{2mm}{
\subfloat[]{\includegraphics[height=1.3in]{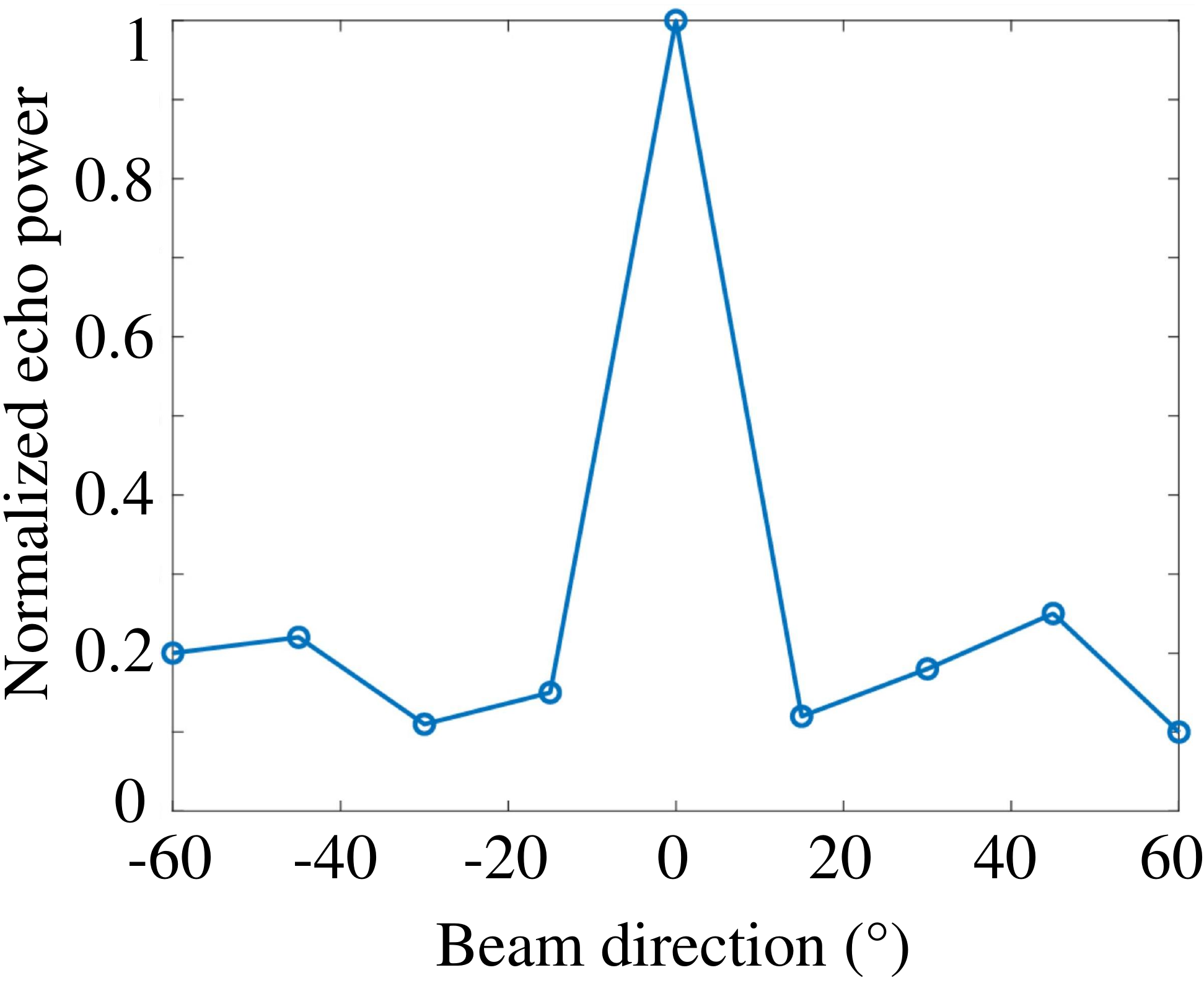}
}}}
\hspace{4mm}
{\captionsetup[subfloat]{captionskip=2mm}
\raisebox{2mm}{
\subfloat[]{\includegraphics[height=1.3in]{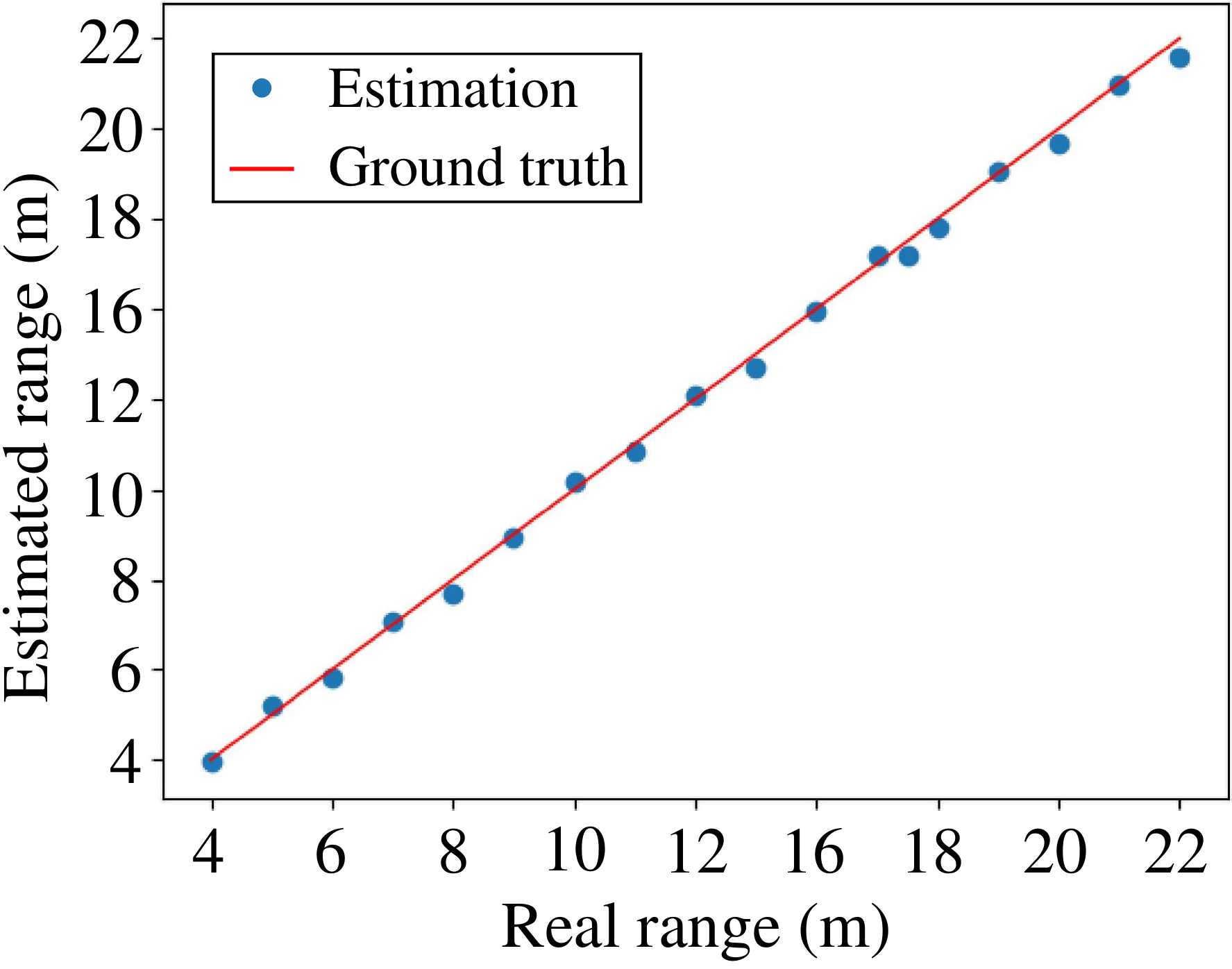}
}}}
\caption{Holographic beamforming-enabled sensing: (a) Prototype; (b) Angle estimation result; (c) Range estimation result.}
\label{f_prot_sense}
\end{figure*}

A holographic beamforming-enabled sensing system is developed to experimentally demonstrate the radar sensing capability. As shown in Fig.~\ref{f_prot_sense}(a), the setup consists of a host computer, a USRP, amplifiers, and two RHS arrays. The computer controls the X410 USRP to generate a 2 GHz intermediate-frequency (IF) signal. This signal is up-converted to $26.2$GHz by a frequency converter, amplified by a power amplifier, and radiated through the transmit RHS array. By applying a specific holographic pattern, the RHS forms a directive beam toward the target, and the reflected echoes are captured by the receive RHS array. The received signal is then amplified and down-converted from $26.2$GHz to $2$GHz using the same frequency converter before being processed by the USRP. Finally, the baseband signal is transferred to the host computer, where radar signal processing algorithms estimate the target’s range and angle.

To evaluate the effectiveness of holographic beamforming-enabled sensing, experiments are conducted with the target positioned at $(0^\circ, 0^\circ)$ relative to the RHS array. During the detection process, the RHS beam sequentially scans from $-60^\circ$ to $60^\circ$ in the horizontal plane, and the echo intensity at each direction is recorded. After normalization, the maximum echo occurs at $0^\circ$, as shown in Fig.~\ref{f_prot_sense}(b), indicating the target’s angular position is accurately recovered. Subsequently, range measurements are carried out with the beam fixed at $0^\circ$. The true distance $D_0$, obtained using a laser rangefinder, is compared with the estimate $D_1$. As illustrated in Fig.~\ref{f_prot_sense}(c), the system achieves a detection range of $20$m and a ranging accuracy better than $10$cm, verifying the practicality of the proposed sensing approach.

\subsubsection{JCAS Prototype}

\begin{figure*}[!t]
\centering
\subfloat[]{\includegraphics[height=1.8in]{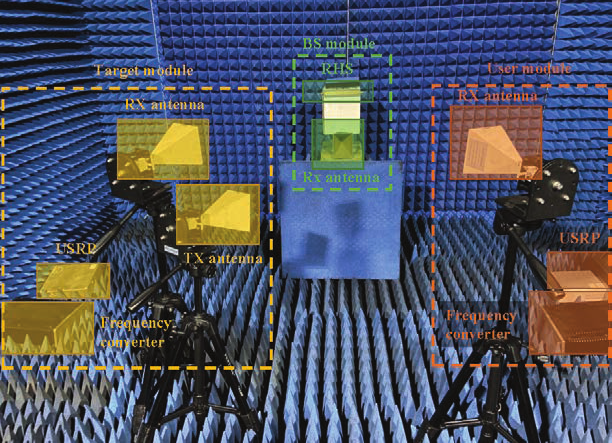}
}
\hspace{4mm}
\subfloat[]{\includegraphics[height=1.8in]{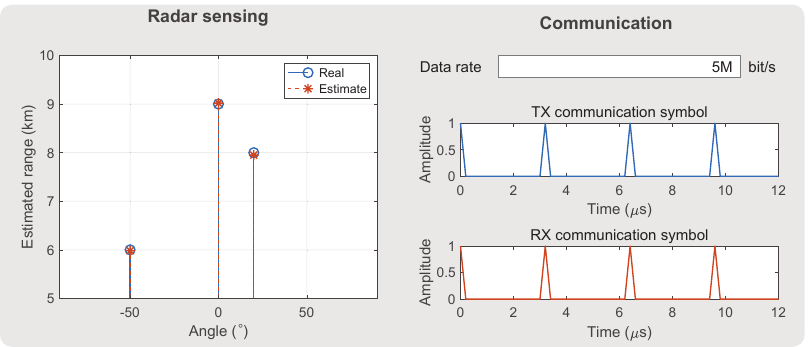}
}
\caption{Holographic beamforming-enabled JCAS: (a) Prototype; (b) Experimental results.}
\label{f_prot_jcas}
\end{figure*}

In this part, we introduce the developed HISAC prototype for simultaneous data transmission and radar sensing. The prototype consists of three functional modules, i.e., the ISAC transceiver module, the user module, and the target module, as illustrated in Fig.~\ref{f_prot_jcas}(a). The transceiver module acts as the BS that transmits ISAC signals and receives radar echoes. An Intel NUC controls the RHS element amplitudes via an FPGA and connects to a USRP for simultaneous transmission and reception. A frequency converter is used to up- and down-convert signals, while a horn antenna captures the echoes. The user module receives the ISAC signal through a horn antenna, down-converts it via a frequency converter, and processes it with a USRP and NUC for data decoding. The target module emulates radar reflections using two antennas, a frequency converter, a USRP, and a NUC. Upon receiving an ISAC waveform, it introduces a programmable delay before retransmission to simulate targets at different distances. This prototype is deployed in an anechoic chamber measuring $4\times3\times2.5,\text{m}^3$. 

Fig.~\ref{f_prot_jcas}(b) presents the experimental results. The target module is positioned at $-50^\circ$, $0^\circ$, and $20^\circ$ with added signal delays of $20,\mu\text{s}$, $30,\mu\text{s}$, and $26.7,\mu\text{s}$, corresponding to simulated ranges of 6 km, 9 km, and 8 km, respectively. During operation, one main lobe of the RHS beam is directed toward each target direction in different cycles, while another main lobe continuously points to the user at $60^\circ$ for communication. The estimated target ranges closely match the ground-truth distances, validating the accuracy of radar sensing. Meanwhile, the received communication symbols at the user module are identical to those transmitted by the BS, achieving a data rate of $5$Mbit/s. These results verify that the proposed HISAC prototype can simultaneously perform reliable communication and sensing within a shared hardware and spectral framework.

\subsubsection{SAC}

\begin{figure*}[!t]
\centering

\subfloat[]{\includegraphics[height=1.9in]{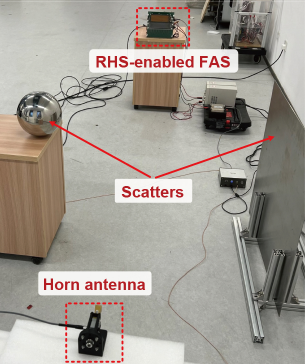}
}
\hspace{4mm}
\subfloat[]{\includegraphics[height=1.9in]{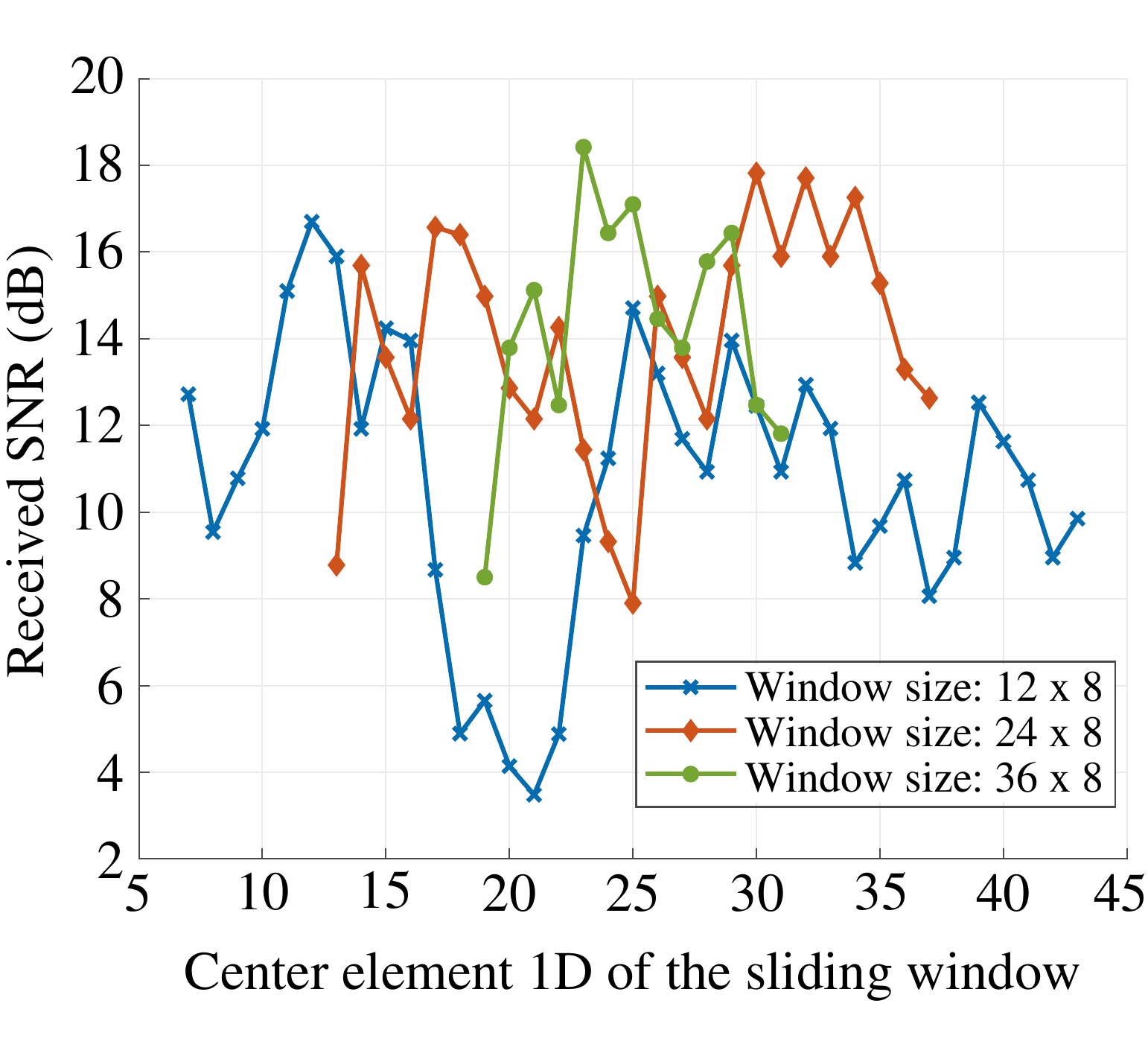}
}
\hspace{4mm}
\subfloat[]{\includegraphics[height=1.9in]{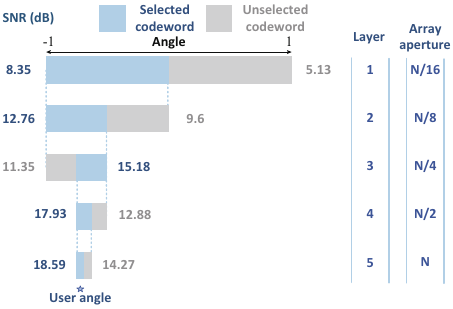}
}
\caption{Holographic beamforming-enabled SAC: (a) Prototype; (b) Received SNR versus position of sliding window. The received SINR significantly varies with the position of the sliding window, indicating a strong small-scale fading effect in this experimental setup; (c) Beam training process when user is at $-35^\circ$.}
\label{f_prot_sac}
\end{figure*}

We further developed a holographic beamforming-enabled SAC platform, as shown in Fig.~\ref{f_prot_sac}(a). The system consists of two modules, i.e., a Tx and a Rx. In addition to the LOS path, several metal scatterers are placed to create NLOS components. The Tx module comprises a host computer, a USRP, a frequency converter, and the RHS-enabled BS. The Rx module consists of a computer, a USRP, a frequency converter, and a standard horn antenna.

To examine the effect of the sliding window position (i.e., the equivalent array position) on the received signal strength, the following setup is created, where the RHS and horn antenna are placed at the same height and separated by $3.5$m. The horn antenna directly faces the RHS at a horizontal angle of $0^\circ$. Three window sizes ($36 \times 8$, $24 \times 8$, and $12 \times 8$) are tested, corresponding to $13$, $25$, and $37$ possible window positions, respectively. Fig.~\ref{f_prot_sac}(b) shows the variation of received SNR across different window positions and aperture sizes. The received SNR fluctuates with the window position for all three configurations, with differences between the maximum and minimum SNRs being $9.92$dB, $10.12$dB, and $13.22$dB for window lengths of $36$, $24$, and $12$, respectively. The average received SNRs for window sizes of $36$, $24$, and $12$ are $14.35$dB, $13.98$dB, and $10.75$dB, respectively, indicating that larger apertures yield stronger received signals due to increased beam directivity.

Following the procedure described in Sec.~\ref{ss_sbt}, a five-layer fluid beam training experiment is carried out using the designed platform. Two horn antennas are placed at azimuth angles of $-35^\circ$. As shown in Fig.~\ref{f_prot_sac}(c), the RHS transmits signals starting from the first layer of the codebook. During beam training, each codeword utilizes two sliding windows for signal transmission, and the receiver selects the one that maximizes the received SNR. During the training process, the window size increases progressively, generating beams with finer angular resolution and eventually locking onto the user’s angle. The blue codewords in Fig.~\ref{f_prot_sac}(c) represent the selected beams at each layer. The results show that the user direction is accurately identified at the final layer, thereby validating the feasibility and effectiveness of the proposed beam training scheme.

\subsection{Challenges and Opportunities}

The hardware design of RHSs involves extensive parameter optimization through full-wave simulations, which are often computationally intensive and time-consuming. The geometry of each element, substrate property, and element spacing must be iteratively tuned to meet the desired electromagnetic response. To accelerate this process, data-driven surrogate modeling and machine-learning-based optimization can be investigated as an alternative to full-wave evaluations. Techniques such as neural network-assisted parameter prediction and transfer learning between frequency bands can be developed to further reduce the design cycle while maintaining high accuracy.

Another challenge lies in evaluating holographic beamforming performance in real-world ISAC conditions. Existing tests are often conducted in controlled environments, while practical deployment typically introduces challenges such as multipath propagation, interference, mobility, and dynamic targets. A unified experimental framework that covers diverse sensing and communication scenarios can be developed for comprehensive validation, thus bridging the gap between laboratory demonstrations and practical applications.

In addition, practical hardware impairments of RHSs also affect the communication and sensing performance of HISAC systems, which should be carefully tackled. Specifically, hardware non-idealities such as calibration errors of PIN diodes, inconsistent element responses, and residual mutual coupling introduce element-wise amplitude and phase perturbations. These impairments result in a mismatch between the intended and the actually realized holographic pattern, thereby distorting the synthesized wavefront. As a result, such distortions degrade the beam focusing capability, leading to increased beamwidth, beam pointing errors, reduced peak gain, and elevated sidelobe levels, which in turn degrades ISAC performance. For instance, an increased beamwidth may cause two closely spaced targets that are originally resolvable to become indistinguishable. To mitigate these effects, both hardware- and algorithm-level solutions are required. On the hardware side, accurate calibration and consistent manufacturing are essential to ensure accurate and reliable element responses. On the algorithmic side, impairment-aware beamforming~\cite{guo2024transmit} and robust optimization techniques~\cite{shen2026robust} should be developed to preserve sensing performance under practical hardware constraints.

Moreover, since most existing prototypes rely on finite-resolution control, its impact should also be investigated. From a performance perspective, finite-resolution control introduces amplitude quantization errors at each element, which lead to deviations from the ideal holographic pattern. In~\cite{hu2022holographic}, the effect of quantization on sum rate is analyzed, and simulation results show that the performance gap compared to continuous control decreases as the number of quantization bits increases. In addition, these results indicate that a small number of quantization bits is often sufficient to approach the performance of continuous amplitude control, especially for large-scale RHS systems. In the context of HISAC systems, finite-resolution control may also affect sensing performance by distorting the synthesized wavefront, and a systematic characterization of this impact remains an open problem.

\section{Future Directions}
\label{s_e}

This section explores the future research directions of HISAC toward broader application scenarios and emerging technologies. Specifically, HISAC can be implemented in diverse environments such as satellite communications, autonomous driving, and robotics, where the combination of high-resolution sensing and adaptive communication provides unique advantages. Furthermore, integrating HISAC with advanced techniques including deep learning, wireless power transfer, and non-orthogonal multiple access (NOMA) can further enhance its intelligence, flexibility, and energy efficiency. Finally, the security and privacy aspects of HISAC are discussed.

\subsection{Application in Various Scenarios}

\subsubsection{Satellite}

Extending HISAC to satellite communication offers new opportunities for spaceborne joint communication and sensing~\cite{yin2024integrated, lukito2025learning}. In satellite systems, payload size and power consumption are tightly constrained~\cite{gu2024isac}, making the compact and power-efficient nature of RHSs particularly attractive. Integrating RHSs into satellite payloads enables adaptive beamforming without mechanical steering, allowing dynamic coverage control and multi-beam operation for both communication and remote sensing tasks. Moreover, by adopting advanced multiplexing schemes, HISAC can support various sensing tasks such as high-resolution imaging and atmospheric observation, while simultaneously maintaining broadband downlink connectivity.

Initial works have investigated the use of holographic beamforming for satellite communication. In~\cite{hu2023holographic}, a holographic beamforming-aided LEO satellite broadcasting system was considered, where the holographic pattern was optimized to maximize the downlink sum-rate. The minimum number of RHS elements required for the system to outperform phased arrays was derived, showing that holographic beamforming can achieve comparable throughput with significantly lower hardware cost and power consumption. In addition,  \cite{peng2025on} extended the study to network-level coverage performance. A closed-form expression and a tractable approximation for the coverage probability in holographic beamforming-aided LEO networks were derived, revealing how the number of RHS elements, satellite density, and user distribution jointly affect coverage. Simulation results indicate the existence of an optimal satellite density and highlight the energy efficiency advantage of RHSs over phased arrays. These works highlight that integrating holographic beamforming into satellites is promoting to enable lightweight and power-efficient architectures for future space networks.

\subsubsection{Autonomous Driving}

Autonomous driving requires both precise environmental perception and reliable low-latency communication among vehicles and roadside units (RSUs)~\cite{du2025toward, li2025integrated}. HISAC provides a unified platform to jointly fulfill these requirements by enabling shared use of spectrum, hardware, and signal processing resources. In HISAC-assisted vehicular networks, RHSs can be deployed on vehicles~\cite{zhang2023reconfigurable} or RSUs to generate adaptive beams for vehicle-to-everything (V2X) communication and high-resolution sensing. Leveraging their fine-grained control over radiation amplitude and phase, RHSs can dynamically adjust beamwidth and direction to track fast-moving vehicles while simultaneously maintaining high-throughput links.

For example, in vehicle-to-infrastructure (V2I) networks, holographic beamforming can be used to perform SAC, where radar echoes are exploited to estimate vehicle motion parameters such as angle, velocity, and distance, thus supporting predictive beam tracking and reducing feedback overhead. Meanwhile, CAS can use V2X signals to improve radar detection and localization accuracy. HISAC systems can also benefit from advanced waveform designs such as orthogonal time frequency space~(OTFS) modulation, which offers robustness to Doppler effects in high-mobility scenarios. Furthermore, by jointly optimizing communication rate and sensing accuracy, HISAC can support multi-vehicle coordination and cooperative perception under complex traffic conditions. 

In the literature, the authors in~\cite{zhang2024reconfigurable} have demonstrated the feasibility via a holographic beamforming-aided wireless simultaneous localization and mapping~(SLAM) framework, a key component of autonomous driving. In this framework, vehicles equipped with multiple RHS-based radars perform full $360^\circ$ environmental scanning using amplitude-controlled, series-fed elements and optimized SLAM algorithms. Simulation results show that the proposed holographic beamforming-aided SLAM achieves over threefold improvement in localization accuracy compared with phased-array-based baselines under the same hardware cost, confirming the potential of HISAC for autonomous driving applications.

\subsubsection{Robotics}

HISAC can play a transformative role in next-generation robotic systems by enabling robots to simultaneously perceive and communicate within their environment using shared electromagnetic resources. Compared with traditional multi-sensor fusion architectures, which rely on separate radar, vision, and communication modules, the integration of HISAC allows a single RHS array to provide high-resolution environmental sensing and reliable data transmission. This capability is particularly valuable in dynamic and confined environments such as smart warehouses, manufacturing floors, and healthcare facilities, where multiple robots operate in close proximity and require precise localization~\cite{zhang2022metaradar2} and coordination~\cite{singh2025empowering, ni2025reconfigurable}.

Specifically, by embedding RHSs onto robotic platforms, holographic beamforming can dynamically adjust the radiation pattern to focus on objects or robot partners, supporting real-time obstacle detection, map reconstruction, and cooperative task execution. The holographic beamforming-enabled robot swarm can also perform collaborative perception by exchanging compressed sensing data over low-latency communication links, forming a unified 3D environmental map. The large-aperture, near-field characteristics of the RHS can improve short-range imaging accuracy and motion awareness~\cite{torcolacci2024holographic}.

Despite these advantages, implementing HISAC in robotic systems presents several challenges. First, latency is a critical constraint, as robots must perform real-time decision-making and precise motion control. The shared use of RHS for both sensing and communication introduces significant processing and coordination overhead, particularly when managing a large number of radiation elements. Second, the conformal design of RHSs poses another major challenge. Unlike BSs or planar platforms, the surfaces of robots are typically curved and continuously moving. Implementing RHSs on non-planar and dynamically deformable surfaces requires new electromagnetic designs that can maintain stable amplitude control under mechanical distortion. Addressing these challenges will enable HISAC to serve as the foundation for intelligent, cooperative, and self-aware robotic systems in 6G and beyond.

\subsection{Integration with Emerging Techniques}

\subsubsection{Deep Learning for HISAC}

Deep learning has emerged as a powerful tool to enhance the adaptability, intelligence, and efficiency of HISAC systems~\cite{luong2025advanced}. Traditional optimization-based approaches for joint sensing and communication often suffer from high computational complexity, limited scalability, and dependence on accurate channel or environmental models. In contrast, deep learning provides a data-driven paradigm that can learn the complex nonlinear relationships among communication, sensing, and propagation characteristics directly from data~\cite{temiz2025deep}, enabling real-time optimization and environment-aware RHSs.

In HISAC, deep neural networks (DNNs) can be employed for a wide range of tasks, including beamforming optimization, waveform co-design, channel estimation, and target recognition. For instance, convolutional neural networks (CNNs) can extract spatial features from channel or sensing data for holographic beam control, while recurrent networks such as long short-term memory (LSTM) networks can exploit temporal correlations for predictive beamforming and dynamic target tracking. Unsupervised and reinforcement learning approaches further enable adaptive holographic optimization without explicit labels or prior models, allowing HISAC systems to operate efficiently in unknown or time-varying environments.

In the literature, the unrolling algorithm has been introduced to reduce the complexity of holographic beamforming~\cite{he2025reconfigurable}. By embedding traditional iterative optimization algorithms within neural networks, this algorithm can not only achieve fast convergence, but also improve sum rate performance compared to conventional fractional programming or phased-array methods. In addition, federated learning frameworks are also emerging to support collaborative HISAC systems. In~\cite{zhang2023reconfigurable}, multiple vehicles equipped with RHSs train local models using their own sensing data, sharing only the model parameters rather than raw information. Simulation results validates that this approach outperforms non-cooperative schemes while preserving data privacy.

Despite these promising advances, several challenges remain for learning-assisted holographic beamforming in HISAC systems. First, the high-dimensional control space of RHSs, especially for ultra-large apertures, leads to significant training complexity and data requirements. Second, the integration of learning with near-field beamforming remains largely underexplored, where the coupling between angle and distance introduces additional complexity in both data representation and network design.

Another promising direction is to explore learning-based strategies for joint hardware and holographic beamforming design. While most existing works focus on optimizing holographic patterns under a fixed RHS aperture, the joint design problem considering both the effective aperture size and holographic pattern has not been investigated. Such a joint design further complicates learning-based approaches, as it involves both discrete and continuous variables, and leads to variable-length outputs since the dimension of the holographic pattern depends on the selected aperture configuration.

\subsubsection{Wireless Power Transfer}

As the demand for sustainable and maintenance-free wireless networks continues to grow, the ability to to support wireless energy transfer has become increasingly important. Wireless power transfer enables the transmission of energy from a source to a receiver through electromagnetic waves, eliminating the need for wired connections. By radiating controlled RF signals, transmitters can charge or power distant devices such as sensors, IoT nodes, or mobile terminals. This technology is particularly valuable in large-scale or hard-to-reach deployments, where replacing or recharging batteries is costly or impractical.

Integrating wireless power transfer into the HISAC framework extends its functionality from dual-purpose operation to triple-function integration, where sensing, communication, and power transfer share the same spectrum, waveform, and hardware resources. Holographic beamforming offers a unique advantage in this setting due to the large apertures and precise electromagnetic control. By dynamically adjusting the amplitude and phase of each radiation element, holographic beamforming can generate narrow, high-gain beams for efficient near-field power focusing~\cite{li2024performance} and energy transmission~\cite{wang2025reconfigurable}.

\subsubsection{NOMA}

Unlike traditional orthogonal multiple access (OMA), which allocates distinct time or frequency resources to different users, non-orthogonal multiple access~(NOMA) allows multiple users to share the same time-frequency resource blocks by separating their signals in the power or code domain~\cite{ahmed2025unveiling}. In power-domain NOMA, the transmitter superimposes multiple users’ signals with different power levels, and the receivers apply successive interference cancellation (SIC) to decode their intended information~\cite{sultana2025joint}. This approach improves spectral utilization and supports a larger number of simultaneous connections, which is especially attractive for 6G networks~\cite{nasser2024rendezvous}.

Integrating NOMA into the HISAC framework provides an effective way to manage the shared spectrum between sensing and communication while enhancing multi-user connectivity. Specifically, NOMA offers additional flexibility for resource allocation in HISAC, and SIC-assisted receivers can mitigate inter-user and sensing interference. On the other hand, the fine-grained electromagnetic control of holographic beamforming allows precise power allocation among users and sensing targets, improving the trade-off between communication throughput and sensing accuracy. The feasibility of combining holographic beamforming with NOMA to enhance energy efficiency in UAV-assisted systems has been demonstrated in~\cite{song2025miniature}.

\subsection{Security and Privacy}

With the increasing integration of communication and sensing functions in wireless systems, security and privacy have become critical concerns in ISAC. 
The shared spectrum and dual-functional nature of ISAC expose wireless systems to new vulnerabilities~\cite{su2023security, wang2025simultaneous}. On the one hand, the transmitted ISAC waveforms carrying user data may be intercepted, leading to the leakage of confidential communication information. On the other hand, the sensing functionality may inadvertently reveal sensitive parameters such as user or radar locations. 
As a result, protecting both communication confidentiality and sensing privacy is essential for future HISAC networks. 
In the following, we discuss potential solutions from the communication and sensing perspectives, respectively.

\begin{itemize}
    \item To avoid the leakage of confidential communication data, physical-layer security (PLS) offers a promising method beyond conventional cryptography. By exploiting the capability of holographic beamforming, HISAC systems can shape electromagnetic fields with high precision to promote the signal strength at users and suppress that at eavesdroppers, thus enhancing the secrecy rate while minimizing information leakage. This has been demonstrated in recent work~\cite{xu2024reconfigurable}, in which the holographic beamforming-assisted PLS can achieve better secrecy rates than conventional phased arrays.
    \item As for sensing functions, unauthorized inference of sensitive parameters such as radar locations needs to be prevented to ensure privacy preservation. Future HISAC networks may adopt privacy-aware beamforming, where the gradient of the holographic radiation pattern is controlled to mislead adversarial sensing or to mask the presence of the sensing node. In addition, sensing-assisted security can be achieved by using the radar perception capability of HISAC to localize potential eavesdroppers and adaptively redesign secure beams in real time.
\end{itemize}

\section{Conclusion}
\label{s_c}

In this paper, a comprehensive overview of HISAC has been presented. We first introduced the working principles, hardware structure, and unique leakage power constraint of RHSs, followed by an overview of the HISAC architecture. We then explored three basic models of HISAC, i.e., joint communication and sensing, sensing-assisted communication, and communication-assisted sensing. We also described the implementation of RHS hardware and presented communication, sensing, and HISAC prototypes to experimentally verify their feasibility and efficiency. Finally, future directions of HISAC were discussed, including applications in satellite, vehicular, and robotic systems, integration with deep learning, wireless power transfer, and NOMA techniques, and the security and privacy issues. We hope this paper serves as a useful reference for future research on HISAC, paving the way for cost- and power-efficient 6G and beyond.

\bibliographystyle{IEEEtran}
\bibliography{myReference.bib}

\vfill

\end{document}